\pdfoutput=1
\documentclass{article}
\usepackage{amsmath}
\usepackage{amssymb}
\usepackage{amsthm}

\usepackage{mathtools} 

\DeclarePairedDelimiter\floor{\lfloor}{\rfloor}

\usepackage{fancyhdr}
\usepackage{geometry}
\usepackage{sectsty}
\usepackage{graphicx}
\usepackage[numbers]{natbib}
\sectionfont{\bfseries\Large\raggedright}
\usepackage{caption}
\usepackage{subcaption}
\usepackage{wrapfig}
\usepackage[acronym]{glossaries}
\usepackage{xcolor}
\usepackage{booktabs}       
\usepackage{placeins}
\usepackage{hyperref}
\usepackage{url}
\usepackage[capitalise,nameinlink]{cleveref}

\newacronym{nn}{NN}{Neural Network}
\newacronym{gpu}{GPU}{Graphics Processing Unit}
\newacronym{tf}{TF}{TensorFlow}
\newacronym{cnn}{CNN}{Convolutional Neural Network}
\newacronym{dns}{DNS}{Direct Numerical Simulation}
\newacronym{cfl}{CFL}{Courant-Friedrichs-Lewy}
\newacronym{rans}{RANS}{Reynolds-Averaged-Navier-Stokes}
\newacronym{les}{LES}{Large Eddy Simulation}
\newacronym{sgs}{SGS}{subgrid-scale}
\newacronym{pde}{PDE}{partial differential equation}
\newacronym{ks}{KS}{Kuramoto-Sivashinsky}
\newacronym{wake}{WAKE}{Wake Flow}
\newacronym{kolm}{KOLM}{Kolmogorov Flow}
\newacronym{aero}{AERO}{Compressible Aerofoil flow}
\newacronym{one}{ONE}{one-step}
\newacronym{nog}{NOG}{no-gradient}
\newacronym{wig}{WIG}{with-gradient}

\definecolor{onecol}{HTML}{B59190}
\definecolor{nogcol}{HTML}{5581B0}
\definecolor{wigcol}{HTML}{F09235}
\colorlet{onecoldark}{black!16!onecol}
\colorlet{nogcoldark}{black!16!nogcol}
\colorlet{wigcoldark}{black!16!wigcol}
\newcommand{\one}[1]{\textcolor{onecoldark}{\gls{one}#1}} %
\newcommand{\nog}[1]{\textcolor{nogcoldark}{\gls{nog}#1}} %
\newcommand{\wig}[1]{\textcolor{wigcoldark}{\gls{wig}#1}} %

\begin{document}
\title{Differentiability in Unrolled Training of Neural Physics Simulators on Transient Dynamics}
\author{Bjoern List, Li-Wei Chen, Kartik Bali, Nils Thuerey 
\\
School of Computation, Information and Technology\\
Technical University Munich\\
Boltzmannstraße 3, 85748 Garching, Germany \\
\texttt{\{bjoern.list, ge35mav, kartik.bali, nils.thuerey\}@tum.de} 
}
\maketitle

\begin{abstract}
Unrolling training trajectories over time strongly influences the inference accuracy of neural network-augmented physics simulators. We analyze this in three variants of training neural time-steppers. In addition to one-step setups and fully differentiable unrolling, we include a third, less widely used variant: unrolling without temporal gradients. Comparing networks trained with these three modalities disentangles the two dominant effects of unrolling, training distribution shift and long-term gradients. We present detailed study across physical systems, network sizes and architectures, training setups, and test scenarios. It also encompasses two simulation modes: In prediction setups, we rely solely on neural networks to compute a trajectory. In contrast, correction setups include a numerical solver that is supported by a neural network. Spanning these variations, our study provides the empirical basis for our main findings: Non-differentiable but unrolled training with a numerical solver in a correction setup can yield substantial improvements over a fully differentiable prediction setup not utilizing this solver. The accuracy of models trained in a fully differentiable setup differs compared to their non-differentiable counterparts. Differentiable ones perform best in a comparison among correction networks as well as among prediction setups. For both, the accuracy of non-differentiable unrolling comes close. Furthermore, we show that these behaviors are invariant to the physical system, the network architecture and size, and the numerical scheme. These results motivate integrating non-differentiable numerical simulators into training setups even if full differentiability is unavailable. We show the convergence rate of common architectures to be low compared to numerical algorithms. This motivates correction setups combining neural and numerical parts which utilize benefits of both.
\end{abstract}

\section{Introduction}\label{sec:introduction}
Our understanding of physical systems relies on capturing their dynamics in mathematical models, often representing them with a \gls{pde}. Forecasting the behavior with these models thus involves the notoriously costly and difficult task of solving the \gls{pde}. By aiming to increase simulator efficiencies, machine learning was successfully deployed to augment traditional numerical methods for solving these equations. Common areas of research are network architectures \citep{sanchez2020learning, li2020fourier, geneva2020modeling, ummenhofer2019lagrangian, tripura2023wavelet}, reduced order representations \citep{lusch2018deep,wiewel2019latent,ATKINSON2019166, mou2021data, brunton2021modern, wu2022learning}, and training methods \citep{Um2020Solver-in-the-loop, Sirignano2020, MacArt2021, brandstetter2022message}. Previous studies were further motivated by performance and accuracy boosts offered by these methods, especially on GPU architectures. Machine learning based simulators aim to address the limitations of classical \gls{pde} solvers in challenging scenarios with complex dynamics \citep{frank2020machine, von2020combining, brunton2020machine}. Computational fluid dynamics represents an application of machine learning augmented solvers~\citep{beck2021perspective, Duraisamy2019}. This especially includes turbulence closure modeling in Reynolds-averaged Navier-Stokes modeling~\cite{ling2016reynolds, zhou2021learning, zhou2022frame, zhang2022ensemble, zhang2023combining} as well as closure models for large eddy simulation \cite{Novati2021AutomatingLearning, bae2022scientific,list_chen_thuerey_2022, Sirignano2020, van2023energy}.
In tackling these challenges, various architectural advancements have recently been made.
On the one hand, physics-informed networks have gained popularity in continuous \gls{pde} modeling 
\citep{raissi2019physics, Duraisamy2019, sun2020surrogate}, and extensions of this have been made to discrete representations \citep{ranade2021discretizationnet, chiu2022can}. However, many learned approaches work in a purely data-driven manner on discrete trajectories. Several of these use advanced network architectures to model the time-evolution of the \gls{pde}, such as graph networks \citep{pfaff2021learning, sanchez2020learning} and particle simulators \citep{wessels2020neural, prantl2022guaranteed}, 
problem-tailored architectures \citep{wang2020towards,stachenfeld2021learned, de2019deep}, bayesian networks \citep{yang2021b}, generative adverserial networks \cite{wu2020enforcing}, transformer models \citep{han2021_Predicting,geneva2022transformers,li2022transformer}, or lately diffusion models \citep{lienen2023generative,lippe2023pde,kohl2023turbulent}. Variations of these approaches compute the evolution in an encoded latent space \citep{wiewel2019latent, geneva2020modeling, geneva2022transformers,brunton2021modern, wu2022learning, vlachas2022multiscale}. 
In an effort to achieve discretization-independence of the learned simulators, \textit{neural operators} have shifted the learning task towards finding a mapping between the function spaces of the numerical solutions. The architectural changes that become necessary when moving to this paradigm have been made for convolutional architectures \cite{raonic2024convolutional}, graph networks \cite{li2020neural} or Fourier-based features \cite{li2020fourier}, among others. 
Additionally, physical knowledge is often integrated into the neural solver by keeping some classical numerical component as part of the solver architecture \citep{oishi2017computational, yin2021augmenting}. In this context, the practice of training neural networks in combination with numerical solvers emerged. The resulting simulator is sometimes referred to as \textit{hybrid} solver or \textit{augmented} solver in literature \cite{Kochkov, list_chen_thuerey_2022, lippe2023pde}. Since neural network training is based on automatic differentiation, differentiable solvers are required to calculate network updates. In his setting, network optimization is conceptually equivalent to adjoint optimization. Here, the adjoint equation provides "sensitivities", which are referred to as "gradients" in the discrete automatic differentiation setting \cite{sanderse2024scientific}. Due to this similarity, early works utilized adjoint solvers into which networks were embedded \cite{Sirignano2020, MacArt2021}.

Out of all physical systems of interest in simulations, those exhibiting transient dynamics are particularly challenging to model. Solutions to these systems are typically based on a time-discrete evolution of the underlying dynamics. Consequently, the simulator of such dynamics is subject to many auto-regressive invocations. Designing simulators that achieve stability and accuracy over long auto-regressive horizons can be challenging in numerical and neural settings. However, unlike numerical approaches, where unstable components such as dispersion errors are well studied and effective stabilizing approaches exist, little is known about the origin of instabilities and mitigating strategies for neural networks. As an interesting development, a stability analysis of echo-state networks was presented recently \cite{margazoglou2023stability}, allowing deeper insights into the behavior of recurrent networks. Nevertheless, stabilizing autoregressive neural simulators remains particularly difficult.

One promising strategy is \textit{unrolling}, where multi-step trajectories are evolved at training time.
Um et al. \cite{Um2020Solver-in-the-loop} used an auto-differentiable implementation of an incompressible Navier-Stokes solver to train correction networks to accelerate numerical simulations of a cylinder-wake flow.
Kochkov et al. \cite{Kochkov} used a very similar technique also employing unrolled correction learning on the two-dimensional Kolmogorov flow. Both of these setups used neural network architectures based on discrete convolutions. An adjoint solver was used by Sirignano et al. \cite{Sirignano2020} on challenging three-dimensional turbulence scenarios to train a point-wise closure model for Large-Eddy simulation. These previous works all successfully deployed unrolling. In this practice, one update to the network weights is based on an unrolled trajectory, where the neural simulator is applied multiple times in an auto-regressive fashion. By design, these unrolled training trajectories are closer to the inference scenario than simple one-step training. This effectively reduces the \textit{data shift}, which describes the difference in the distribution of training and inference data. By exposing autoregressive trajectories at training time, unrolling moves the observed training data closer toward realistic inference scenarios \citep{wiles2021fine, wang2022koopman}.
A measurable gain in long-term stability of networks trained with this method was found in previous work \cite{Um2020Solver-in-the-loop, list_chen_thuerey_2022, melchers2023comparison}.
By now, unrolling represents a widely used approach when training neural networks that predict \gls{pde} trajectories on their own \citep{geneva2020modeling}, or support numerical solvers in doing so, for instance, in LES closure modeling \cite{Sirignano2020, MacArt2021, list_chen_thuerey_2022}. An important observation is that these previous works utilized the full backpropagation gradient for network optimization. In an unrolled setup including a numerical solver, this can only be calculated using differentiable or adjoint solvers. Since differentiability is rarely satisfied in existing code bases, this is a first limitation of differentiable unrolling.
A resulting open question is how much the numerical solver's differentiability assists in training accurate networks.

Additionally, possible downsides of differentiable unrolling concerning gradient stability were investigated in recent work. While a vanishing/exploding effect is well known for recurrent networks \cite{goodfellow2016deep}, it is only sparsely studied for correction approaches featuring numerical and neural components. The stability of the backpropagation gradients was found to align with the Lyapunov time of the physical system for recurrent prediction networks \cite{mikhaeil2022difficulty}. Furthermore, unrolling for too many steps was associated with diminishing returns \cite{Um2020Solver-in-the-loop, list_chen_thuerey_2022} and even showed negative effects in previous work \citep{melchers2023comparison}. This indicates that the \textit{gradient stability} of unrolled setups dependent on the unrolled horizon marks an important limiting factor. In practical considerations, unrolled setups need to balance the benefits of data shift reduction with sufficient gradient stability. The reported success of some methods used in previous work can be related to this balancing act. For instance, recent studies \cite{brandstetter2022message, prantl2022guaranteed} proposed a variant of classic truncated backpropagation in time \citep{sutskever2013}. These recent works reported a positive effect on the network performance when "warm-up" steps were introduced. A warm-up step increases the unrolled horizon without contributing to the learning signal since backpropagation gradients are discarded for these steps. This effect was reported to stabilize the trained networks. 
As a similar potential remedy, it was proposed to cut the backpropagation chain into individual sequences \cite{list_chen_thuerey_2022}.

Within this paper, we explore unrolling from a dynamical systems perspective. A first fundamental insight is that no machine-learned approximation \textit{identically} fits the underlying target.
For physics simulators, this means that the dynamics of the learned system will always differ from the target dynamics. As such, the attractors for the target and the learned system do not coincide entirely. Unrolling can be seen as an exploration of the machine-learned dynamics' attractor at training time. In contrast to training without unrolling, this approach thus allows sampling from a different distribution in the solution space. An unrolled training procedure samples from the learned dynamics and exposes (more of) the inference attractor at training time. The difference between training and inference, which we referred to as \textit{data shift}, is reduced. For longer horizons, more of the learned dynamics' attractor is exposed, stabilizing the neural simulator. 

\begin{figure}
    \centering
    \includegraphics[height=5.2cm]{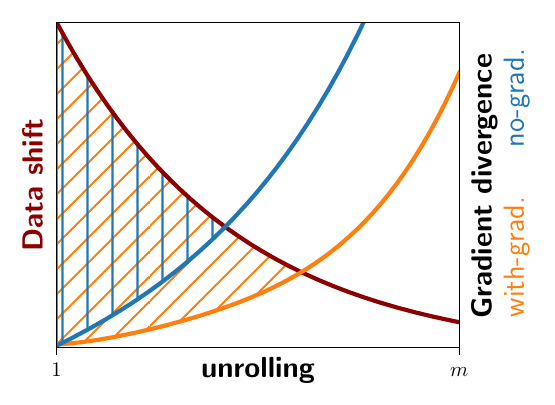} 
    \caption{Illustration of data shift and gradient divergence over unrolling; gradients of non-differentiable unrolling diverge from the true gradient landscape (blue), gradients of differentiable simulators are prone to explosions over long horizons (orange); shaded areas mark potential benefits over one-step learning; full gradient calculation performs best.}
    \label{fig:schematic}
\end{figure}

In principle, this would motivate ever longer horizons, as illustrated in Figure \ref{fig:schematic}. However, obtaining useful gradients through unrolled dynamical systems limits this horizon in practice \cite{chung2022optimization}. The differentiability of numerical solvers is crucial to backpropagate through solver-network chains. However, if the learned system exhibits chaotic behavior, backpropagating long unrolled chains can lead to instabilities in the gradient calculation itself \cite{mikhaeil2022difficulty}. Due to the strong non-linearities of deep-learning architectures, the auto-regressive dynamics of the learned simulator can show chaotic behavior in early training stages, regardless of the proper\-ties of the target system. A limitation to the effective unrolled horizon in fully differentiable setups was thus observed in previous studies~\cite{Um2020Solver-in-the-loop, list_chen_thuerey_2022}. 

A central and rarely used variant employs unrolling \emph{without backpropagating gradients} through the simulator.
We study this training variant in detail because of its high practical relevance, as it can be realized by integrating existing, non-differentiable solvers into a deep learning framework. The forward pass of a non-differentiable unrolling setup is identical to the fully differentiable setup and thus features the same attractor-fitting property. In the backward pass, the non-differentiability of the solver is equivalent to an operator returning a zero gradient. 
This effectively truncates the backpropagation and avoids long gradient propagation chains. While cutting gradients potentially has a stabilizing effect, we find that it also limits the effective unrolling horizon, albeit for different reasons: the computed gradient approximations diverge from the true gradient as the unrolling horizon increases, making the learning process less reliable. 
Figure \ref{fig:schematic} illustrates these gradient inaccuracies for both training modalities. To the best of our knowledge, extensive studies of non-differentiable unrolled training do not yet exist. This is despite its high relevance to the scientific computing community as an approachable variant integrating numerical solvers in network training.

We analyze these two unrolling variants using standardized solver and network architectures while contrasting them with one-step training.
This allows disentangling the effects of data shift introduced by forward unrolling and accurate long-term gradients calculated by backpropagation. 
Our study targets four open questions in the context of unrolling:

\paragraph{(I) How effective is non-differentiable unrolling?}
The vast majority of existing numerical solvers are not differentiable. We investigate the effectiveness of non-differentiable unrolling and whether integrating such a solver into the training loop yields measurable benefits. While previous studies have provided a good intuition on the effectiveness of differentiable unrolling, there are no prior results on non-differentiable unrolling.

\paragraph{(II) How do the unrolling variants compare?} Differentiable unrolling, non-differentiable unrolling, and one-step learning require different levels of software engineering efforts. A comparison of these training setups is necessary to choose the most suitable setup for a given application.

\paragraph{(III) Which benefits does correction training yield?}
Whether machine learning architectures should be trained to augment numerical codes in a corrective setup or rather constitute surrogates in a predictive setup, is up to scientific debate. The integration of numerical solvers as indicated in (I) and (II) has to be motivated by distinct benefits of such correction approaches.

\paragraph{(IV) Do the effects of unrolling show in different \gls{pde}s?}
The concepts of unrolling and differentiablility may arise in various applications governed by different physical laws. As such, we want to investigate how the effects of unrolling translate between physical systems, solvers and network architectures.
\newline

We test our findings on various physical systems with multiple network architectures, including convolutional and graph networks. To achieve this, we develop differentiable solvers that could serve as a benchmark in future studies. In this context, it is worth noting that several benchmarks and datasets have been published to increase comparability and promote standardization of machine-learned \gls{pde} simulators, especially in fluid mechanics. 
Datasets for advection-diffusion type PDEs \citep{takamoto2022pdebench}, and measured real-world smoke clouds \citep{eckert2019scalarflow} have been proposed.
A high-fidelity Reynolds-averaged Navier-Stokes dataset is provided in \cite{bonnet2022airfrans}. Furthermore, more specialized datasets focus on fluid manipulation \citep{xian2022fluidlab}, and navigation for aerial vehicles \cite{janny2023eagle}. 
We choose three widely used differentiable scenarios and an additional non-differentiable case as the basis for the following investigation.

\section{Unrolling PDE Evolutions}\label{sec:unrolling_theory}
Unrolling is a common strategy for learning time sequences \cite{Um2020Solver-in-the-loop, MacArt2021, Kochkov, list_chen_thuerey_2022}. A first important distinction is whether the task at hand is a \textit{prediction} or a \textit{correction} task. Additionally, we formally introduce the differences between \textit{non-differentiable} and \textit{differentiable} unrolling in gradient calculation. These are then used to derive an intuition about the gradient effects of unrolling.

\subsection{Evolving partial differential equations with neural networks}
Let us consider the general formulation of a \gls{pde} in the form of 
\begin{equation}\label{ground_truth_pde}
    \frac{\partial \mathbf{u}}{\partial t} = \mathcal{F}(\mathbf{u}, \nabla\mathbf{u},     \nabla^2\mathbf{u},\dots), 
\end{equation}
with $\mathbf{u}$ representing the field variables of the physical system. From the perspective of dynamical systems, the state $\mathbf{u}$ is said to evolve on the \textit{attractor} $A_\mathcal{F}$ of the underlying \gls{pde} system.
For the following analysis, we define an attractor $A$ to satisfy the following conditions: 
\begin{enumerate}
    \item The attractor $A$ is forward invariant such that for any state $\mathbf{u}_{t=t_0}\in A$, the temporal evolution of this state is also in $A$ such that $\mathbf{u}_{t>t_0}\in A$. Thus, trajectories on $A$ cannot "leave" this attractor and will further evolve within $A$.
    \item The attractor has a \textit{basin of attraction} $\mathcal{B}(A)$ in the neighborhood of $A$ for which all states $\mathbf{u}\in \mathcal{B}(A)$ converge to $A$ as $t\rightarrow\infty$. By this property, $A$ attracts neighboring states.
    \item There is no subset of $A$ that satisfies the previous conditions.
\end{enumerate}
The attractor can be seen as a subspace of the phase space of $\mathbf{u}$ containing only those realizations of $\mathbf{u}$ that are physical under Equation \eqref{ground_truth_pde}.
For an initial condition in the \textit{basin} of the attractor $\mathcal{B}(A)$, the solution will converge to states on $A_\mathcal{F}$ and further evolve within this attractor.

The learning tasks presented in the following sections assume that we have access to a highly-resolved ground truth solver that produces solution trajectories that are physical under Equation \eqref{ground_truth_pde}.
Its trajectories follow the dynamics of a \gls{pde} $\mathcal{F}$ and thus converge to $A_\mathcal{F}$ for a reasonably chosen initial condition in $\mathcal{B}(A_\mathcal{F})$ and an adequate burn-in time. The resulting dataset of spatiotemporally discrete snapshots $\bar U=[\mathbf{\bar u}^0\dots\mathbf{\bar u}^N]$ lies strictly on this attractor such that $\bar{\mathbf{u}}^t \in A_\mathcal{F}$ for all $t$. Additionally, the physical system has a governing characteristic timescale $t_c$, where $t_c$ is dependent on the underlying \gls{pde} and its characteristic hyperparameters. The dataset size $N$ is typically large in comparison to this timescale. 

\subsection{Training Neural Simulators}
We study \textit{Neural Simulator} architectures that use neural networks for evolving discretized forms of \gls{pde}s.
Simulating a system's fully resolved dynamics becomes prohibitively expensive in many applications. A classic example is turbulent flows, where modeling approaches like large-eddy simulations often replace direct numerical simulations. In this context, Neural Simulators can applied on a reduced representation of the dynamical state. A mapping between the ground truth and its reduced representation leads to
\begin{equation}
    \tilde{\mathbf{u}}^t = R(\bar{\mathbf{u}}^t), \quad \forall \bar{\mathbf{u}}^t \in \bar U.
\end{equation}
It is important to note that the reduction is not strictly necessary. If one is interested in the full dynamics, the reduction $R$ becomes the identity. In our studies, the reduction depends on the physical system and will be specified for each system at a later stage.

The task of a Neural Simulator is to provide a time-stepping operation on discrete data frames in accordance with the \gls{pde} in \eqref{ground_truth_pde}. We denote the trajectory on which the Neural Simulator operates as $\mathbf{u}^t$. Due to modeling errors, $\mathbf{u}^t$ may not exactly fit $\tilde{\mathbf{u}}^t$. The Neural Simulator operation $g_\theta(\mathbf{u})$ on a sample from the dataset can then be denoted as
\begin{equation}\label{eq:trajectory_fit}
    \tilde{\mathbf{u}}^{t+1} \approx  \mathbf{u}^{t+1}=g_\theta(\tilde{\mathbf{u}}^t), \quad 
\end{equation}
on the dataset $\tilde{\mathbf{u}}^t \in \tilde U$, where $\theta$ represents the weights of the neural network. We want to stress that, at this point, we have not limited the structure of $g$ in any way. However, when specifying prediction and correction setups, we will differentiate the construction of $g$, even though both fit this general operation signature.

\textit{Prediction} setups fully rely on a neural network to calculate the next timestep. Denoting the network as $f_\theta$ leads to 
\begin{equation}\label{eq:prediction}
\mathbf{u}^{t+1} = g_{\theta,p}(\mathbf{u}^{t}) = f_\theta(\mathbf{u}^{t}).
\end{equation}
In a prediction configuration, the network fully replaces a numerical solver with the goal of improving its accuracy and performance. They're also known as neural solvers or surrogates.
Various previous works fit this description. It encompasses full state \gls{pde} predictions with a whole range of architectures including graph networks~\cite{sanchez2020learning, pfaff2021learning}, coarse-grained predictions of the full dynamics like in weather prediction \cite{rasp2020weatherbench, lam2023learning} as well as reduced order models or latent evolutions that operate on encoded spaces \cite{wiewel2019latent, EIVAZI2021108816}.
We will focus on networks based on convolutional or graph message-passing operations. Details of the architectures are introduced with the physical system.

\textit{Correction} setups are also concerned with time-evolving a discretized \gls{pde} but additionally include a numerical solver $\mathcal{S}$ that approximates the solution by computing only parts of a timestep. The underlying assumption is that the solver cannot capture the full dynamics of the \gls{pde} in \eqref{ground_truth_pde}. 
From a machine learning perspective, the numerical solver introduces a powerful prior and inductive bias.
A timestep $g_\theta$ thus consists of the solver $\mathcal{S}$ and a neural network model $f_\theta$.
The network complements the numerical solver such that 
\begin{equation}\label{eq:correction_general}
    \mathbf{u}^{t+1} = g_{\theta,c}(\mathbf{u}^t;f_\theta ; \mathcal{S}).
\end{equation} 
These setups represent a combination of numerical and neural architectures, potentially combining the advantages of both domains. Possible applications include closure and turbulence modeling in RANS or LES, coarse-graining, as well as dynamical system control setups. The formulation in \eqref{eq:correction_general} is quite general and fits many previous works. For instance, the correction networks used in previous work \cite{Um2020Solver-in-the-loop, Kochkov} used the following more specific formulation
\begin{align}\label{eq:correction_additive}
\begin{split}
    \mathbf{u}^*&=\mathcal{S}(\mathbf{u}^t)\\
    \mathbf{u}^{t+1}&=f_\theta(\mathbf{u}^*).
\end{split}
\end{align}
We will also use this setup for our corrections. Different approaches exist. For instance in the context of LES, where the subgrid-scale terms are usually embedded into the filtered equation. This leads to
\begin{align}\label{eq:correction_forcing}
\begin{split}
    \mathbf{f}&=f_\theta(\mathbf{u}^t) \approx  \nabla \cdot \tau_{\mathrm{sgs}}\\
    \mathbf{u}^{t+1}&=\mathcal{S}(\mathbf{u}^t, \mathbf{f}).
\end{split}
\end{align}
This closure-inspired formulation is more popular in the computational fluids community \cite{Sirignano2020, list_chen_thuerey_2022}. In this context, training a neural network directly on ground truth values of $\tau_{\mathrm{sgs}}$ is referred to as \textit{a-priori} training while training with respect to the solution state $\mathbf{u}$ is frequently called \textit{a-posteriori} training \cite{sanderse2024scientific}.

\begin{figure}
\centering
\includegraphics[height=3.8cm]{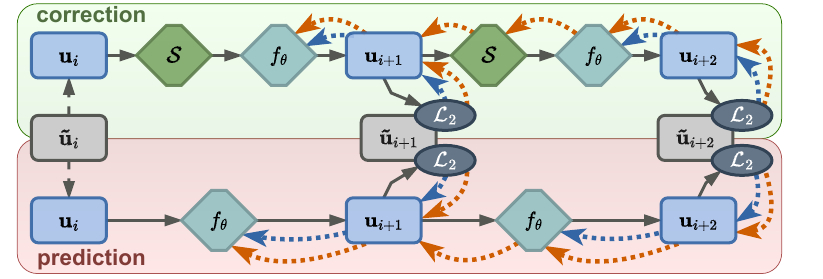}
\hspace{-0.25cm}
\includegraphics[height=3.8cm]{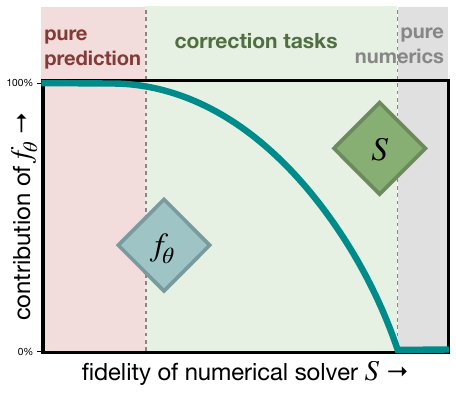}
\caption{Left: Illustration of unrolling for a horizon $m=2$, with numerical solver $\mathcal{S}$ and neural network $f_\theta$, visualized for correction and prediction chains;
importantly, it highlights differences in the gradient flow shown with dotted lines in the backward pass for non-differentiable (blue, \textcolor{nogcoldark}{no-gradient (NOG)}) and differentiable (orange, \textcolor{wigcoldark}{with-gradient (WIG)}) setups: note how the gradients do not flow through a time step of $\mathcal{S}$ for \nog{}, instead they contain two shorter, disconnected backpropagation chains; Right: Contributions of networks and numerical solvers in possible prediction and correction paradigms}
\vspace{-0.2cm}
\label{fig:flowchart}
\end{figure}

For the prediction and correction setups from equation \eqref{eq:correction_additive} and \eqref{eq:prediction}, respectively,
the input and output vectors of the networks share the same solution space. While the learned mapping differs in prediction and correction cases, one does not necessarily need different architectures to train one or the other. Thus, we use different instances of identical neural network architectures in both setups. We use the same notation $f_\theta$ for both applications to stress this property. A representative computational chain for correction and prediction setups using this notation is visualized in Figure \ref{fig:flowchart}.

\subsection{Unrolling dynamical systems at training time}
In practice, Neural Simulators rarely satisfy Equation \eqref{eq:trajectory_fit} exactly and rather form an approximation of the ground truth. Over long auto-regressive rollouts at inference time these small-scale errors accumulate and lead to a substantial shift in the solution, which can cause instabilities and ultimately makes the solution diverge. The \textit{data shift} between snapshots observed at training time and those observed in inference is commonly identified as the cause of these instabilities \citep{quinonero2022dataset}. When simply training on one-step trajectories as in Equation \eqref{eq:trajectory_fit}, only data sampled from the ground truth attractor $A_\mathcal{F}$ is observed, as only data from the ground truth dataset $\tilde{\mathbf{u}}\in \tilde{U}$ is supplied. The same model might then used autoregressively at inference time with $\mathbf{u}^{i+1}=g_\theta(\mathbf{u}^i)$. An inference trajectory is then constructed as $[\mathbf{u}^{i+1}, g_\theta(\mathbf{u}^{i+1}), \dots]$ . Crucially, we supply the model with self-generated data $\mathbf{u}_i$ for the first time when computing this inference trajectory. Based on the fact that learned simulators never perfectly reproduce the ground truth, this can lead to error accumulation and unwanted behavior in the trajectories. This means, that the attractor of the learned dynamics $A_{g_\theta}$ includes additional states in such cases. These states are exposed by the auto-regressive inference rollout which feeds unseen and fundamentally different data to the network. Thus, the learned attractor $A_{g_\theta}\neq A_\mathcal{F}$.

We study unrolling as a possible remedy for this problem. Unrolling describes the process of auto-regressively evolving the learned system in a training iteration. It can be formalized as 
\begin{equation} \label{eq:learned_dynamics}
    \mathbf{u}^{t+s} = g_\theta^s(\tilde{\mathbf{u}}^t),
\end{equation}
where $g^s_\theta$ represents the recurrent application of $s$ steps of $g_\theta$. 
When unrolling at training time from a starting frame $\tilde{\mathbf{u}}^t$, multiple frames $[\mathbf{u}^{t+1}, \mathbf{u}^{t+2},\dots,\mathbf{u}^{t+m}$] are generated within one training iteration, where $m$ denotes the unrolling horizon. A simple one-step training is obtained by setting $m=1$. 
However, one-step training without unrolling does not guarantee that the states $\mathbf{u}$ observed during training sufficiently represent $A_{g_\theta}$. In mathematical terms, 
\begin{equation}
    \forall \mathbf{\tilde{u}}^t \in A_\mathcal{F} \: \exists \: {g_\theta}: \lim_{N\rightarrow\infty}\{g_\theta(\mathbf{\tilde{u}}^0), g_\theta(\mathbf{\tilde{u}}^1), . . . , g_\theta(\mathbf{\tilde{u}}^N)\} \neq A_{g_\theta}.
\end{equation}
This means that we are not guaranteed to explore the attractor of the learned system $A_{g_\theta}$ when only those states are observed that are based on one discrete evolution of $g_\theta$, regardless of the dataset size $N$. Simple examples of this are networks sampled during training. A common data-augmentation approach is to perturb the ground truth states in $\tilde U$ with noise, in the hope that the perturbed states fit $A_{g_\theta}$ more closely \cite{sanchez2020learning}. While this approach was reported to be successful, it still does not guarantee that the observed states are from $A_{g_\theta}$, and thus, those augmented states are not necessarily closer to states observed during inference.

Let us now suppose we unroll to the horizon $m$ with $g_\theta^s(\mathbf{\tilde u})$ and $s\in[1,\dots,m]$ at training time.
Based on the definition of the attractor $A_{g_\theta}$ it can be proven that
\begin{equation}\label{eq:unroll_attractor_fit}
    \lim_{m\rightarrow\infty} \bigcap_{i=1..N}\{g_\theta^0(\mathbf{\tilde{u}}^t), g_\theta^1(\mathbf{\tilde{u}}^t), . . . , g_\theta^m(\mathbf{\tilde{u}}^t)\} \supset A_{g_\theta},
\end{equation}
for sufficiently many $\mathbf{\tilde u}^t$ in the basin $\mathcal{B}(A_{g_\theta})$. Details of the proof are provided in \ref{app:unrolling}.
In contrast to one-step training without unrolling, enabling it thus exposes the inference attractor $A_{g_\theta}$ at training time for sufficiently large datasets and horizons $m$. While the horizon $m$ is practically never chosen long enough to expose $A_{g_\theta}$ with a single initial condition, a sufficiently large dataset ensures that $A_{g_\theta}$ is approached from multiple angles and observed states from this set converge to the attractor of $g_\theta$ for large enough $m$.
Precisely the difference between the observed training set (sampled from $A_\mathcal{F}$) and the inference attractor $A_{g_\theta}$ represents a form of data shift that can impede inference accuracy.
The dark red curve in Figure 1 shows how this shift is thus reduced by choosing larger $m$.

\subsection{Training neural networks to solve PDEs:} \label{sec:training_to_solve_pde}
Since the training procedure is largely similar for correction and prediction, we will use the generalized $g_\theta$ for the following analysis. We will put special emphasis on the differences between correction and prediction setups when they arise. The general training procedure is given as 
\begin{equation}\label{eq:unrolling_loss}
\theta^* =\mathrm{arg}\min_\theta \bigg [ \sum_{t=1}^{N-m} \sum_{s={1}}^{m} 
 \mathcal{L}_2\big ( \tilde{\mathbf{u}}^{t+s}, g_\theta^s(\tilde{\mathbf{u}}^t)
 \big)\bigg]. 
\end{equation}
For one learning iteration, a loss value is accumulated over an unrolled trajectory. We can write this total loss over the unrolled trajectory as
\begin{equation}
    \mathcal{L} = \sum_{s=1}^{m} \mathcal{L}_2(\tilde{\mathbf{u}}^{t + s}, g^s_\theta(\tilde{\mathbf{u}}^t)) = \sum_{s=1}^m  \mathcal{L}^s,
\end{equation}
where $\mathcal{L}^s$ represents the loss evaluated after a step. The simplest realization of the training procedure \eqref{eq:unrolling_loss} utilizes 
the forward process with $m=1$, i.e. no unrolling. We refer to this variant without unrolling as \one{} in the following. This approach is common practice in literature and the resulting training procedure is easily realized with pre-computed data.
In contrast, the full backpropagation through an unrolled chain with $m > 1$ requires a differentiable solver $\mathcal{S}$ for the correction setup.
The differentiable setup can calculate the full optimization gradients by propagating gradients through a simulation step, as illustrated by the orange arrows in Figure \ref{fig:flowchart}. The gradients are thus evaluated as 
\begin{equation} \label{eq:wig_gradient}
    \frac{\partial \mathcal{L}_\mathrm{WIG}^s}{\partial \theta} = \sum_{B=1}^s \bigg[
    \frac{\partial \mathcal{L}^s}{\partial g_\theta^s} 
    \bigg ( \prod_{b=1}^{s-B}
    \frac{\partial g_\theta^{s-b}}%
    {\partial g_\theta^{s-b-1}} 
    \bigg)
    \frac{\partial g_\theta^B}
    {\partial \theta}\bigg].
\end{equation}
We refer to this fully differentiable setup as \wig{}. The loss contribution $\mathcal{L}_\mathrm{WIG}^s$ originates from the $s$th step in the unrolling horizon. Gradients associated with this loss contribution are propagated to all previous network contributions, which results in the outer sum in equation \eqref{eq:wig_gradient}. The individual network contributions require backpropagation through (multiple) simulator operations $g_\theta$, resulting in the chain-rule product in equation \eqref{eq:wig_gradient}. Refer to Figure \ref{fig:flowchart} for a visualization. The gradient flow through a simulator step $g$ is caluclated as $\frac{\partial g_{\theta,c}^{s+1}}{\partial g_{\theta,c}^s}=\frac{\partial f_\theta^s}{\partial \mathcal{S}^s}\frac{\partial \mathcal{S}^s}{\partial g_{\theta,c}^s}$ for correction setups. Here it becomes apparent that gradients propagate through a simulator step $g_{\theta,c}$. The numerical solver $\mathcal{S}$ is part of this operation in correction setups and needs to be differentiable with respect to the input state, i.e. $\frac{\partial \mathcal{S}(u)}{\partial u}$ has to be available. The propagated gradient values carry information for all network contributions lying earlier in the unrolled chain. Since there is no numerical solver in the prediction setup, the gradient propagation solely depends on the network and evaluates to $\frac{\partial g_{\theta,p}^{s+1}}{\partial g_{\theta,p}^s}=\frac{\partial f_\theta^s}{\partial g_{\theta,p}^s}$.

To test the effect of differentiability, we introduce a second strategy for propagating the gradients $\frac{\partial \mathcal{L}^s}{\partial \theta}$.
If no differentiable solver is available, optimization gradients can only propagate to the network application, not through the solver. The blue arrows in Figure \ref{fig:flowchart} do not pass through a simulation step. In mathematical terms, the gradients are thus evaluated as 
\begin{equation}
    \frac{\partial \mathcal{L}_\mathrm{NOG}^s}{\partial \theta} = 
    \frac{\partial \mathcal{L}^s}{\partial g_\theta^s} 
    \frac{\partial g_\theta^s}
    {\partial \theta}.
\end{equation}
This setup is referred to as \nog{}. While gradients can still flow into the parameter space $\theta$ of an operation $g_\theta$, they cannot \textit{pass through} it. Hence, $\frac{\partial g_\theta^{s+1}}{\partial g_\theta^s}:=0$ holds for correction and prediction setups. Refer to Figure \ref{fig:flowchart} again for a visualization. In short, a simulator step in the \nog{} setup is not differentiable and yields no gradients with respect to previous steps. Most existing code bases in engineering and science are not fully differentiable. Consequentially, this \nog{} setup is particularly relevant, as it can be implemented using existing traditional numerical solvers.
It is important to note a potential limitation of \nog{} unrolling. It only works without implementing additional differential operations when the ground truth information is available directly at the network output. For the correction formulation used within this study from equation \eqref{eq:correction_additive}, this is trivially given since the output of the network constitutes the new solution. If equation \eqref{eq:correction_forcing} is used without a differentiable solver instead, only one option exists - Moving the ground truth information to the output of the network and employing \nog{} unrolling. In the example of closure modeling, this could be achieved by explicitly filtering the high-resolution ground truth to derive subgrid-scale stresses \cite{park2021toward}. However, deriving a filter is reliant on strong assumptions for the actual filter operation. The behavior of the resulting subgrid-scale quantities depends on the choice of filter \cite{piomelli1991subgrid}. Due to these approximations, a-priori derived ground-truth values for the subgrid-scale quantities generally are not ideal targets to maximize a-posteriori performance. This leads to a mismatch between the a-priori trained networks and filtered LES equations, which can make these networks ineffective \cite{shankar2024differentiableturbulenceclosurepartial}. This problem is circumvented by a-posteriori training of the subgrid-scale model, which in turn requires differentiable solvers. Utilizing the non-differentiable \nog{} strategy for closure models is thus only achievable with the correction formulation of equation \eqref{eq:correction_additive}.

\begin{table}
\caption{Calculated gradients with unrolled step $s$;
for correction: 
$\frac{\partial g^{s+1}}{\partial g^s}=\frac{\partial f_\theta^s}{\partial \mathcal{S}^s}\frac{\partial \mathcal{S}^s}{\partial g^s}$;
for prediction: 
$\frac{\partial g^{s+1}}{\partial g^s}=\frac{\partial f_\theta^s}{\partial g^s}$}
\centering
\begin{tabular}{c|c|c|c|c}
      & $\mathcal{L}$ & $\partial \mathcal{L}/\partial\theta$  & Requires diff. $\mathcal{S}$ & Realization \\
      \hline & & & &\\[-8pt]
     ONE & $\mathcal{L}_2^1$&$\dfrac{\partial \mathcal{L}_2^1}{\partial f^1_\theta}\dfrac{\partial f^1_\theta}{\partial \theta}$ & no & precompute\\[12pt]
     NOG & $\sum\limits_s\mathcal{L}_2^s$&$\sum\limits_s\dfrac{\partial \mathcal{L}_2^s}{\partial f^s_\theta}\dfrac{\partial f^s_\theta}{\partial \theta}$ & no & interface\\[12pt]
     WIG & $\sum\limits_s\mathcal{L}_2^s$& 

     $    \sum\limits_s \sum\limits_{B=1}^s 
    \dfrac{\partial \mathcal{L}^s_2}{\partial {g}^s} 
     \dfrac{\partial g^s}{\partial g^B}
     \dfrac{\partial g^B}
     {\partial \theta}
     $
     & yes & re-implement
\end{tabular}%
\label{tab:gradients}%
\end{table}
Figure \ref{fig:flowchart} contains a visualization including the backpropagation flow through an unrolled chain. 
It highlights the fundamentally different nature of accumulated Jacobian terms for \wig{} and \nog{}: the former grows quadratically in $m$, i.e., inducing three terms in the example of the figure, while \nog{} grows linearly and induces only two Jacobian terms.  
The gradient calculations for our three setups, additionally including the \one{} variant, are further denoted in Table \ref{tab:gradients}. Further detail of the gradient calculations in correction and prediction setups is found in \ref{app:gradients}.

\subsection{Gradient behavior in unrolled systems} 
As \wig{} unrolling provides the true gradient, we can thus derive the gradient inaccuracy of the \nog{} setup as 
\begin{equation} \label{eq:nog_gradient_mismatch}
\frac{\partial \mathcal{L}_\mathrm{WIG}}{\partial \theta} - \frac{\partial \mathcal{L}_\mathrm{NOG}}{\partial \theta} =
\sum_{s=1}^m  \sum_{B=1}^{s-1} 
\bigg[\frac{\partial \mathcal{L}^s}{\partial g_\theta^s}
\bigg ( \prod_{b=1}^{s-B}
\frac{\partial g_\theta^{s-b}}{\partial g_\theta^{s-b-1}}\bigg)
\frac{\partial g_\theta^B}{\partial \theta}\bigg]
\end{equation}
The long backpropagation chains used in \wig{} are missing in \nog{} unrolling and thus lead to the innermost product in Equation \ref{eq:nog_gradient_mismatch}. At this stage, we can integrate a property of chaotic dynamics derived by \cite{mikhaeil2022difficulty} in this analysis. Therein, the authors show how the Jacobians 
$\mathbf{J_s}=\frac{\partial g_\theta^{s}}{\partial g_\theta^{s-1}}$ have eigenvalues larger than 1 in the geometric mean:
\begin{equation}
\bigg|\bigg|\prod_{b=1}^{s-B}\frac{\partial g_\theta^{s-b}}{\partial g_\theta^{s-b-1}}\bigg|\bigg| > 1  .
\end{equation}
This means that the scale of the backpropagated gradients increases with unrolling length in the \wig{} setup. Since the gradient flow through $g_\theta$ is cut in the \nog{} approach, this also means that more information is omitted for long unrolling horizons, which increases the gradient inaccuracy of \nog{}. The largest contribution cut from the \nog{} gradients stems from the longest chains in equation \eqref{eq:nog_gradient_mismatch}. These long chains scale with $m^2$ due to the forward and backward pass (equivalent to the maximum values of the iterants $s$ and $B$ in equation \eqref{eq:nog_gradient_mismatch}). We can thus conclude that the \nog{} gradient inaccuracy grows with $\propto m^2$, while the \nog{} gradients only linearly depend on $m$. For increasing unrolling lengths, the error in this surrogate gradient will outgrow the gradient itself. As a consequence, this approximation of the gradient computed in the \nog{} system diverges from the true (\wig{}) gradients for increasing $m$, as shown with the blue line in Figure \ref{fig:schematic}. Due to the fact that numerically covering a characteristic timescale $t_c$ usually requires many timesteps, the \nog{} gradient divergence will be fast compared to this timescale. Using the gradient approximations used in \nog{} as a learning signal does not accurately match the loss used in training. This hinders the network optimization, which is visualized by the gradient divergence curve in Figure \ref{fig:schematic}.

Let us finally consider the full gradients of the \wig{} setup. We can use Theorem 2 from \cite{mikhaeil2022difficulty} stating
\begin{equation}
    \lim_{m\rightarrow\infty}\big|\big|
    \frac{\partial g_\theta^m}{\partial g_\theta^0}\big|\big|=\infty
\end{equation}
for almost all points on $A_{g_\theta}$. The gradients of the \wig{} setup explode for long horizons $m$. As a direct consequence, the training of the \wig{} setup becomes unstable for chaotic systems when $m$ grows. In contrast to the \nog{} gradients discussed above, it is not an inaccuracy in the computation that makes the training become unstable, but rather this intrinsic property of the Jacobian of chaotic dynamics. As such, the gradient divergence of the \wig{} system does not depend on the number of discrete steps like the \nog{} ones did, but is rather governed by the characteristic physical timescale $t_c$ of the learned system. Due to the fact that one physical timescale of a \gls{pde} system is usually discretized by numerous timesteps, the \nog{} gradients diverge earlier compared to \wig{} gradients for increasing $m$. This is indicated by the orange line below the blue one in Figure \ref{fig:schematic}. 

It is important to note that stopping the gradient as done by the \nog{} approach limits some aspects of neural network design. While exact gradients are available for the network outputs with respect to the network parameters, the same is not true for the unrolled trajectory. Consequently, a sensitivity analysis with respect to individual network parameters or design choices  \cite{pizarroso2020neuralsens, zhang2022bilateral} would not be possible for the entire trajectory, but rather only with respect to the direct network output. Similarly, interactions between initial variables, parameters and the output like in \cite{gevrey2006two, tsang2017detecting} is only possible in a fully differentiable setting.

\subsection{Curriculum learning}\label{sec:curriculum}
As illustrated above, unrolling can have negative effects on the gradient stability. This effect can even be amplified for newly initialized networks with random weights, where erroneous outputs make even the forward process unstable. As such, it is common practice to limit the unrolling horizon in the initial phases of the training run \cite{Um2020Solver-in-the-loop, list_chen_thuerey_2022}. The number of unrolled steps is then increased incrementally, which we term \textit{curriculum} in this study. Curriculums are necessary when training with long unrolling horizons. A simple approach is to start with one-step learning, with longer horizons following afterward, e.g., $m=[1,2,4]$.

\subsection{Summary}
We can summarize the hypotheses stated above as follows:
Unrolling training trajectories for data-driven learning of chaotic systems can be expected to reduce the data-shift, as the observed training samples converge to the learned attractor. At the same time, long unrollings can lead to unfavorable gradients for both \nog{} and \wig{} setups. 
This means that intermediate ranges of unrolling horizons can exist, where the benefits of reducing the data-shift outweigh instabilities in the gradient. 
Based on this trade-off, our goal is to identify preferred training regimes in our experiments where models can be trained to higher accuracy without destabilizing the learning process.
We also expose the unrolling horizons and quantify the expected benefits of training with \nog{} and \wig{} modalities.

\section{Physical Systems and Architectures}
\label{sec:systems_architectures}

\begin{figure}
\centering
\includegraphics[height=3.5cm]{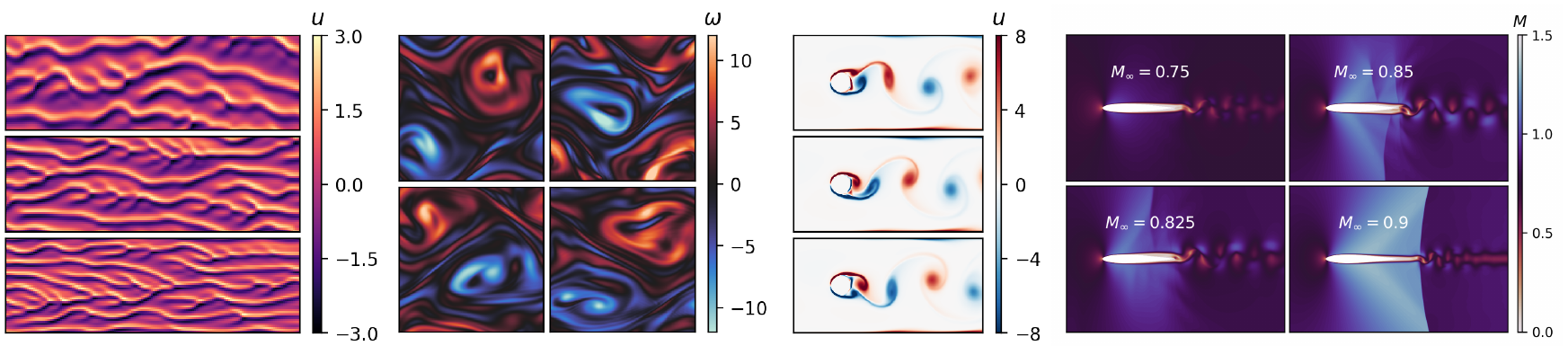} 
\caption{Visualizations of our physical systems, from left to right: KS equation state over time, KOLM vorticity field, WAKE vorticity field, AERO local Mach number field visualization for different upstream Mach numbers M$_\infty$}
\label{fig:dataset_vis}
\end{figure}

The four physical systems shown in Figure \ref{fig:dataset_vis} were used for our learning tests. We focus on several widely used and well-understood test-cases from the scientific machine learning community.
All individual systems are parameterized to exhibit varying behavior within the dataset, and each test set contains unseen values inside and outside the range of the training data set.  
\subsection{Kuramoto-Sivashinsky}
This equation is a fourth-order chaotic \gls{pde}. It is a popular testbed for machine-learning based studies of chaotic systems. Its time-evolution was targeted by a series of previous works \cite{vlachas2018data,geneva2020modeling,stachenfeld2021learned,brandstetter2022message,qu2022learning,melchers2023comparison,lippe2023pde,maddu2023stencil}.
The domain size, which leads to more chaotic behavior and shorter Lyapunov times for larger values \citep{edson_bunder_mattner_roberts_2019}, was varied across training and test data sets. 
The ground truth solver uses an exponential RK2 integrator \citep{cox2002exponential}. In order to create a challenging learning scenario, the base solver for correction resorts to a first-order version that diverges within 14 steps on average. 
We base most of our empirical tests on this \gls{ks} case since it combines challenging learning tasks with a small computational footprint.
The \gls{ks} equation is a fourth-order stiff \gls{pde} governed by 
\begin{equation}
    \frac{\partial u}{\partial t} + u\frac{\partial u}{\partial x} + \frac{\partial^2 u}{\partial x^2} +\frac{\partial^4 u}{\partial x^4} =0,
\end{equation}
with simulation state $u$, time $t$, and space $x$ living in the periodic domain of size $\mathcal{X}$. The dynamics exhibit a highly chaotic behavior. 
The domain was discretized with $N_x=48$ grid points such that the first computational node lies at $x_0=0$ and the last one at $x_{N_x}=\mathcal{X}*(1-1/N_x)$. The timesteps were set to $1$. The physical domain length $\mathcal{X}$ is the critical parameter in the \gls{ks} equation. The training dataset was computed for a range of $\mathcal{X}=2\pi * [5.6, 6.4, 7.2, 8, 8.8, 9.6, 10.4]$.
To calculate the dataset, a randomized initial condition $u_0 = cos(x) + \alpha sin(x)$, with $\alpha$ being sampled randomly from $[-0.5,0.5]$. A sequence of $5000$ steps was then recorded for each domain length. The numerical solves and network training were performed in PyTorch \citep{NEURIPS2019_9015}.
To calculate the dataset, a similarly randomized initial condition $u_0 = cos(x) + \alpha sin(x)$ was used. After an initial burn-in time of 100 steps, a sequence of $5000$ steps was recorded for each domain size. The numerical solves and network training were performed in PyTorch \citep{NEURIPS2019_9015}.
The characteristic timescale of the \gls{ks} system is based on the Lyapunov timescale of chaotic systems. This timescale $t_{s,\text{KS}}= \frac{1}{\lambda_1}$ requires knowledge of the largest Lyapunov exponent $\lambda_1$. We calculate this exponent using the Jacobian of the discretized \gls{pde} \citep{Pikovsky_Politi_2016}, which is readily available due to the differentiability of the solver. Let us denote the \gls{ks} solver and its Jacobian as $u^{t+1}=\mathcal{S}_\text{KS}(u^t)$, $J^t = \frac{\partial u^{t+1}}{\partial u^t} = \frac{\partial \mathcal{S}_\text{KS}(u^{t})}{\partial u^t}$.
An initial perturbation $\epsilon^0$ is then propagated alongside the tangent of the trajectory such that
\begin{align}
\tilde\epsilon^{t+1}&= J^t\epsilon^t, \\
\epsilon^{t+1} &= \frac{\tilde\epsilon^{t+1}}{||\tilde\epsilon^{t+1}||}.
\end{align}
The largest Lyapunov exponent can then be calculated by estimating the exponential growth rate of the perturbation, which is achieved by averaging the normalization factor
\begin{equation}
    \lambda_1=\frac{1}{T}\sum_t^T\ln||\tilde{\epsilon}^t|| ,
\end{equation}
for a sufficiently long horizon $T$. We use this method to calculate the \textit{average} Lyapunov exponents for the range $\mathcal{X}$ used in our training dataset. Setting $T=1000$ after discarding the first transient period of 100 steps for the calculations of $u_0$ and $\epsilon_0$, the average exponent evaluates to $\lambda_1= 0.0804$. The timescale of our discretized system is thus $t_{c,\text{KS}}=12.4$, which corresponds to $\approx 12$ timesteps of the neural simulator.

\paragraph{Neural Network Architectures:}
We use two architectures for learning tasks with the KS case: a graph-based (GCN) and a convolutional network (CNN). Similar networks were previously used on the \gls{ks} equation in the case of GCNs \cite{brandstetter2022message, dulny2023dynabench}, while CNN based architectures were used in several other works  \cite{geneva2020modeling,stachenfeld2021learned,qu2022learning,melchers2023comparison, maddu2023stencil}. 
Stachenfeld et al. \cite{stachenfeld2021learned} compared several types of convolutional networks, including ResNets and U-Nets,
on the \gls{ks} as well as Navier-Stokes systems. More complex convolutional architectures on the \gls{ks} system were also investigated by Geneva et al. \cite{geneva2020modeling}, who proposed an encoder-decoder based CNN, or Qu et al. \cite{qu2022learning}, who split the CNN into linear and nonlinear parts for modeling of (non-)linear \gls{pde}s.
In our study, both networks types follow a ResNet-like structure \citep{he2016deep}, have an additive skip connection from input to output, and are parameterized in terms of their number of \textit{features} per message-passing step or convolutional layer, respectively. They receive 2 channels as input, a normalized domain size $\mathcal{X}$ and $u$, and produce a single channel, the updated $u$, as output.
The GCN follows the hierarchical structure of directional Edge-Conv graph nets \cite{wang2019dynamic,li2019deepgcns} where each Edge-Conv block is comprised of two message-passing layers with an additive skip connection. The message-passing concatenates node, edge, and direction features, which are fed to an MLP that returns output node features. At a high level, the GCN can be seen as the message-passing equivalent of a ResNet.
These message-passing layers in the GCN and convolutional layers in the CNN are scaled in terms of whole ResBlocks, i.e., two layers with a skip connection and leaky ReLU activation.
The networks feature additional linear encoding and decoding layers for input and output.
The architectures for GCN and CNN were chosen such that the parameter count matches, as listed in Table \ref{tab-app:ks-arch}.
\begin{table}[tbh]
    \caption{KS Network Architectures and Parameter Counts; we denote the number of blocks as network \textit{depth}, and specify the network architectures with a tuple containing (features, depth); For both GCN and CNN the smallest network size with a depth indicated by ''-'' is a special case using a single non-linear layer with the listed number of features.}
    \label{tab-app:ks-arch}
    \centering
    \begin{tabular}{l c c c c c c c}
        \toprule
        \textit{Architecture}: &   .5k*  & .5k & 2k & 33k & .2M & 1M  \\

        \midrule
        GCN (features, depth):
        & (46,-) & (9,1)  & (14,2) & (31,8) &  (63,12) &  (126,16) \\ 
        GCN Parameters:   
        & 511 & 518 & 2007 & 33,578 & 198,770 & 1,037,615  \\ 
        \midrule
        CNN (features, depth):
        & (48,-) & (8,1) & (12,2) & (26,8) & (52,12) & (104,16) \\ 
        CNN Parameters: &
        481 & 481 & 1897 & 33125 & 196457 & 1042705 \\ 
        \bottomrule
    \end{tabular}
\end{table}

\paragraph{Network Training:}
Every network was trained for a total of $50000$ iterations of batchsize $16$. The learning rate was initially set to $1\times10^{-4}$, and a learning rate decay of factor $0.9$ was deployed after every $2000$ iterations during training. The \gls{ks} systems was set up to quickly provide substantial changes in terms of simulated states. As a consequence, short
unrollments of $m=3$ were sufficient during training unless otherwise mentioned. For this unrollment number, no curriculum was necessary. 

For fairness across the different training variants the KS setup additionally keeps the number of training data samples constant for \one{} versus \nog{} and \wig{}. For the latter two, unrolling means that more than one state over time is used for computing the learning gradient. Effectively, these two methods see $m$ times more samples than \one{} for each training iteration. Hence, we increased the batch size of \one{} training by a factor of $m$. However, we found that this modification does not result in an improved performance for \one{} in practice.

For the tests shown in Figure \ref{fig:unrolling_horizon} we varied the training data set size, reducing the number of time steps in the training data set to $1000$, while keeping the number of training iterations constant. The largely unchanged overall performance indicates that the original data set could be reduced in size without impeding the performance of the trained models.

\paragraph{Network Evaluation:}
We computed extra- and interpolative test sets for the domain length $\mathcal{X}$. Similarly to the training dataset, the testset is based the same parameterization for the initial conditions. Again, the simulations feature a burn-in time until their dynamics are fully developed. The extrapolation uses $\mathcal{X}=2\pi*[4.8, 11.2]$, whilst interpolation is done for $\mathcal{X}=2\pi*[6.8, 9.2]$.
For our $\mathcal{L}_2$ results, we selected 20 fully developed stated the test dataset and started autoregressive runs with the learned simulators on each $\mathcal{X}$. We then accumulated the relative errors for sequences of $12$ steps, corresponding to $1 t_{c,\text{KS}}$. 
We averaged the losses over all extra- and interpolative cases and computed the standard deviation over the total set of tests. 

\subsection{Wake Flow} \label{sec:systems_wake}
Our second system is a two-dimensional flow around a cylinder governed by the incompressible Navier-Stokes equations. This classical fluid mechanics setups also frequently used in a scientific machine learning context. See \cite{miyanawala2017efficient,lee2019data,Um2020Solver-in-the-loop,peng2020unsteady, ranade2021discretizationnet, morimoto2022generalization} for some examples.
Our \gls{wake} dataset consists of Karman-Vortex streets with varying Reynolds numbers,  uses numerical solutions of the incompressible Navier-Stokes equations 
\begin{align}\label{eq:navier-stokes}\begin{split}
    \frac{\partial \mathbf{u}}{\partial t} + (\mathbf{u}\cdot\nabla) \mathbf{u} &= - \nabla p + \frac{1}{Re}\nabla^2\mathbf{u} + \mathbf{f}, \\
    \nabla \cdot \mathbf{u}&=0,
\end{split}\end{align}
with the two-dimensional velocity field $\mathbf{u}$ and pressure $p$, and an external forcing $\mathbf{f}$ which is set to zero in this case.

\paragraph{Numerical Data}
The ground truth data was simulated with operator splitting using a Chorin-projection for pressure and second-order semi-Lagrangian advection \cite{bridson2015book, ferziger2019computational}. The training data is generated with second-order advection, while learning tasks use a truncated spatial resolution ($4 \times$), and first-order advection as a correction base solver. Simulations were run for a rectangular domain with domain size $N_x, N_Y = [2, 1]$. A cylindrical obstacle with a diameter of $L_D=0.1$ placed at $(0.5, 0.5)$. The streamwise boundaries use a constant bulk inflow velocity of $U=1$ at the inlet and Neumann conditions $\frac{\partial u}{\partial x}=0$ at the outlet. Inviscid, impermeable walls were used for both spanwise boundaries. The Reynolds number is defined as $Re=\frac{UL_D}{\nu}$. The physical behavior is varied with the Reynolds number by changing the viscosity of the fluid. The training data set contains 300 time-steps with $\Delta t=1$ for six Reynolds numbers $Re = [97.5, 195, 390, 781.25, 1562.5, 3125]$ at resolution $[256,128]$. For learning tasks, these solutions are down-sampled by $4\times$ to $[64,32]$.
The solver $\mathcal{S}$ for the correction setup replaces the second-order advection with first-order semi-lagrangian advection, which introduces more numerical dissipation \cite{Stam1999}. The numerical solves and network training were performed in PyTorch \citep{NEURIPS2019_9015}.
The characteristic timescale of the cylinder flow $t_{c,\text{WAKE}}$ is based on the vortex shedding frequency $f_s$. This frequency is calculated using the Fourier transform of a sample point downstream of the cylinder at [32,16] in the downsampled dataset. A time history of $K=500$ steps is collected, and the dominant vortex shedding frequency is extracted by
\begin{equation}
 f_s = \arg\max \hat u_y[f], \quad \text{where}\quad \hat u_y[f] = \sum_{k=0}^K u_y[k]e^{2\pi\iota kf/K}.
\end{equation}
Following this analysis, the average vortex shedding frequency in our dataset is $f_s=0.01$ which gives a characteristic timescale of $t_{c,\text{WAKE}}=\frac{1}{f_s}=100$.

\paragraph{Neural Network Architectures:}
This test case employs fully-convolutional residual networks \citep{he2016deep}. Our setup is an extension of the networks used in Um et al. \cite{Um2020Solver-in-the-loop} with respect to parameter count. Similar networks are a popular choice for the \gls{wake} flow and are used in \cite{miyanawala2017efficient, morimoto2022generalization}. Further development by Lee et al. \cite{lee2019data} looked into multiscale CNNs on a range of physical systems including \gls{wake} flow. CNN-based encoder-decoder architectures for similar scenarios were studied by Peng et al. \cite{peng2020unsteady} and Ranande et al. \cite{ranade2021discretizationnet}.
Our CNNs can also be seen as a two-dimensional extensions of those used the KS setup: the neural networks have a ResNet structure, contain an additive skip connection for the output, and use a certain number of ResBlocks with a fixed number of features. 
The \textit{smallest} network uses 1 ResBlock with 10 features, resulting in 6282 trainable parameters.
The \textit{medium}-sized network has 10 ResBlocks with 16 features and 66,178 parameters, while the \textit{large} network has 20 blocks with 45 features resulting in more than 1 million (1,019,072) parameters.

\paragraph{Network Training and Evaluation:}
The networks were trained for three steps with 30000 iterations each, using a batch size of 3 and a learning rate of $1\times10^{-4}$. The training curriculum increases the number of unrolled steps (parameter $m$ of the main text) from $m=1$, to $m=4$ and then $m=16$. Each stage applies learning rate decay with a factor of $1/10$.
Metrics are computed and accumulated over 32 steps, i.e. ca.
$1/3 t_{c,\text{WAKE}}$,
for 8 test cases with previously unseen Reynolds numbers: three interpolative ones $Re = [2868.5, 2930, 2990]$ and five extrapolative Reynolds numbers $Re = [3235, 3296, 3906, 4516.5, 4577.5]$.

\subsection{Kolmogorov Turbulence}
Thirdly, we study a periodic two-dimensional \gls{kolm} \citep{Arnold2009} also following the incompressible Navier-Stokes equations. 
This setup is a widely used fluid flow case in scientific machine learning
\cite{Kochkov,sun2023neural,shu2023physics,lippe2023pde,li2024physics,li2024scalable,azizzadenesheli2024neural}.

In contrast to the \gls{wake} flow, \gls{kolm} introduces a forcing $\mathbf{f}$ in equation \eqref{eq:navier-stokes}. The additive forcing causes the formation of a shear layer, whose instability onset develops into turbulence \citep{Arnold2009}.

\paragraph{Numerical data:}
The two-dimensional \gls{kolm} system is also governed by the incompressible Navier-Stokes equations, with the addition of a forcing term $\mathbf{f}$. The forcing was set to $\mathbf{f} = [a_x*\sin(k_x*y),\: 0]^T$ with wavenumber $k_x=6$ and amplitude $a_x=1$. The domain has periodic boundary conditions in each dimension.
We use a more involved semi-implicit numerical scheme (PISO by \citep{Issa1986}) for this setup. Similarly to the \gls{wake} system, the \gls{kolm} learning cases are based on a spatiotemporal resolution truncation with a ratio of $4\times$ in both space and time, and the Reynolds number is varied across training and test cases. The physical domain size was set to $L_x=L_y=2\pi$ and discretized by a $128\times128$ grid. The timestep was set to $\Delta \tilde t=0.005$ for the high-resolution ground truth dataset, which maintained a Courant number smaller than $0.5$. Following the definition by \citep{Chandler2013}, we calculate the Reynolds number as 
\begin{equation}
    Re=\frac{\sqrt{a_x}}{\nu} \Big(\frac{L_y}{2\pi}\Big)^{\frac{3}{2}}.
\end{equation}
The training dataset is based on a variation of the viscosity $\nu$, which results in a Reynolds number variation of $Re= [300, 400, 500, 700, 800, 900]$. After initializing the simulations with a profile matching $\mathbf{f}*0.1$ across the entire domain, a burn-in time of $15000 \Delta\tilde t$ was run for the high-res simulations. Afterward, $6000$ frames with $\Delta t = 4\times\Delta \tilde{t}$ were added to the dataset. The numerical solves and network training were performed in Tensorflow \citep{199317}.
To choose a characteristic timescale for the \gls{kolm} system that is consistent with the Reynolds number definition, we set 
\begin{equation}
    t_{c,\text{KOLM}}=\sqrt{\frac{L_y}{2\pi*a_x}}.
\end{equation}
This definition leads to $t_{c,\text{KOLM}}=1$, which corresponds to 50 timesteps of the neural simulator since it uses a timestep of $4\times\Delta \tilde{t}$.

\paragraph{Neural Network Architectures:}
The \gls{kolm} networks are also based on ResNets. Similar architectures were also used on a \gls{kolm} system in \cite{Kochkov}, and in \cite{sun2023neural,li2024scalable} for baseline comparison. Kochkov et al. \cite{Kochkov} used it to study a range of unrolled systems, including one that is equivalent to our correction cases. Lippe et al. \cite{lippe2023pde} used the KOLM flow as a two-dimensional testbed for their method based on convolutional U-Nets, and compared it to standard U-Nets, as well as CNNs in corrections setups. Azzizadenesheli et al. \cite{azizzadenesheli2024neural} based their study of neural operators on the KOLM system, and their range of networks includes operator formulations of graph networks as well as convolutional U-Nets. ResNets are structured into blocks, where data is processed through convolutions and added to a skip connection. In contrast to \gls{ks} and \gls{wake} architectures, the number of features varies throughout the network. Four different network sizes were deployed, details of which are listed in Table \ref{tab-app:kolm-arch}. The network sizes in the \gls{kolm} case range from $32$ thousand to $1$ million. 

\begin{table}[tbh]
    \caption{KOLM Network Architectures and Parameter Counts}
    \label{tab-app:kolm-arch}
    \centering
    \begin{tabular}{l| r c l }
        \toprule
        \textit{Architecture} &   \textit{\# Parameters} & \textit{\# ResNet Blocks} & \textit{Block-Features} \\
        \midrule
        CNN, 32k & 32369 & 5 & [8, 20, 30, 20, 8] \\ 
        CNN, 0.1M & 115803 & 7 & [8, 16, 32, 64, 32, 16, 8] \\
        CNN, 0.5M & 461235 & 7 & [16, 32, 64, 128, 64, 32, 16] \\
        CNN, 1M & 1084595 & 9 & [16, 32, 64, 128, 128, 128, 64, 32, 16]\\
        \bottomrule
    \end{tabular}
\end{table}

\paragraph{Network Training and Evaluation:}
A spatiotemporal downsampling of the ground truth data formed the basis of the training trajectories. Thus, training operated on a $4\times$ downsampled resolution with $4$ times larger time-steps, i.e. $32\times32$ grid with $\Delta t=0.02$.
The networks were trained with a curriculum that incrementally increased the number of unrolled steps such that $m=[1,2,4]$, with an accompanying learning rate schedule of $[10^{-4}, 10^{-5}, 10^{-6}]$. Additionally, a learning rate decay with factor $0.9$ after an epoch of $36$ thousand iterations. For each $m$, 144 thousand iterations were performed. The batch size was set to $1$. 
$\mathcal{L}_2$ errors are computed and accumulated over 250 steps, corresponding to $5t_{c,\text{KOLM}}$. The procedure is repeated for 5 test cases each and previously unseen Reynolds numbers: $Re = 600$ for interpolation and $Re = 1000$ for extrapolation. 

\subsection{KS in 3D:} \label{sec:ks3d} High-dimensional  problems arise in many applications, such as in fluid mechanics in the form of LES. As a representative of higher-dimensional chaotic problems, we use a 3D extension of the \gls{ks} equation. Once again, we vary the domain size in training and testing datasets.
The \gls{ks} equation in 3D describes a scalar-valued field $u$ in three dimensions. Similarly to the one-dimensional counterpart, it is highly chaotic. We use a cubic domain of size $\mathcal{X}\times\mathcal{X}\times\mathcal{X}$ with periodic boundaries in all directions. The equation is given by
\begin{equation}
\frac{\partial u}{\partial t} + \big(\nabla u\big)^2 + \nabla^2u +\nabla^4u = 0.
\end{equation}
\paragraph{Numerical Data:}
The ground truth solver uses an exponential RK2 integrator \cite{cox2002exponential}. The domain was resolved by 32 cells in each direction. The correction is based on a first-order integrator that diverges over 14 timesteps. The networks are thus tasked to stabilize the solver and close the gap to the higher-order solution.
\paragraph{Network Architecture:}
We focus on training correction networks for this setup, and use a ResNet \cite{he2016deep} architecture that represent a 3D extension of the previously used 1D and 2D variants.
The network features 4 ResNet blocks with 16 latent features, amounting to 55k parameters.
\paragraph{Network Training and Evaluation:}
For the training dataset, the domain sizes were varied such that $\mathcal{X}=[12, 13, 15, 17, 19, 21, 22]$, and trajectories of 200 timesteps based on 5 initial conditions were run for each domain size. To construct an initial condition, we first draw a random superposition parameterized $\sin$ and $\cos$ functions, in line with the 1D case. Then, a burn-in simulation of $500$ steps was run to ensure de-correlation from these superpositions. The resulting frame is used as an initial condition for the dataset recording. Across the training dataset, we estimate the characteristic timescale based on the first Lyapunov exponent to be $t_{c, \text{KS}_\text{3D}}= 1.8$. This is equivalent to 18 timesteps of size $\Delta t=0.1$.
We trained three networks for each training variant, where each of the three used a different random seed to generate the initial network weights. All networks were trained for 50,000 steps. The unrolled versions used a horizon of $m=4$.
The data-generation procedure is repeated for interpolative and extrapolative test sets, which operate on $\mathcal{X}= [14, 16, 18, 20]$ and $\mathcal{X}=[10, 11, 23, 24]$ respectively. The temporal horizon for testing was chosen to be 20 timesteps, which is equivalent to $1.1t_{c, \text{KS}_\text{3D}}$. The test losses are computed across 10 random initial conditions for each domain size.

\subsection{AERO:}
As a more complex physical system, we investigate compressible flow around an aerofoil, a task centered on pure prediction. We utilize the open-source structured-grid code CFL3D \citep{Rumsey1997, Rumsey2010} to solve the compressible Navier-Stokes equations, thus generating our ground truth dataset. Here, we vary the Mach number $\text{Ma}$ while keeping the Reynolds number constant. Details of this system are found in \ref{app:aero_case}.
\section{Results}
Our results compare the three training methods \one{}, \nog{}, and \wig{} in various scenarios. The figure titles mark the physical system, whilst subscripts represent the network architecture (\textit{graph}: graph network, \textit{conv}: convolutional ResNet), and superscripts differentiate between correction (\textit{corr}) and prediction (\textit{pred}) tasks. In the following sections, we will report on convolutional networks for the \gls{ks}, \gls{wake}, \gls{kolm}, and 3D-KS systems, graph networks on the \gls{ks} system as well as U-nets in the \gls{aero} case. Additional U-net results for \gls{ks} are presented in \ref{sec:appendix_additional_results}.

For each test, we train multiple models (typically 8 to 20) for each setup that differ only in terms of initialization (i.e. random seed). 
When reporting relative errors, we use the following definition for the relative error of a single prediction
\begin{equation}
    \mathcal{L}_\mathrm{rel}^t = \frac{\mathrm{avg}\big [k(\mathbf{u}^t-\tilde{\mathbf{u}}^t)\big ]}{|k(\mathbf{u}^t)|_\infty}, \quad \text{with} \quad k(\mathbf{u}^t)=\sum_c^D(\mathbf{u}^t_c)^2,
\end{equation}
where $D$ denotes the number of spatial dimensions. The $k$ operation is equivalent to calculating the kinetic energy for each grid cell and thus reduces the channel dimension of the data. The stepwise relative error is then averaged across a testing trajectory for plotting such that
\begin{equation}
    \mathcal{L}_\mathrm{rel} = \frac{1}{N_T} \sum_t^{N_T}\mathcal{L}_\mathrm{rel}^t
\end{equation}
The evaluation metrics were applied independently for outputs generated by each of the models, which are displayed in terms of mean and standard deviations below.
This indicates the expected performance of a training setup and the reliability of obtaining this performance. A Welch's test for statistical significance was performed for the resulting test distributions, and p-values can be found alongside the tabulated data in \ref{app:evaluation_data}.
\begin{figure}[b]
\centering
\includegraphics[height=4.49cm]{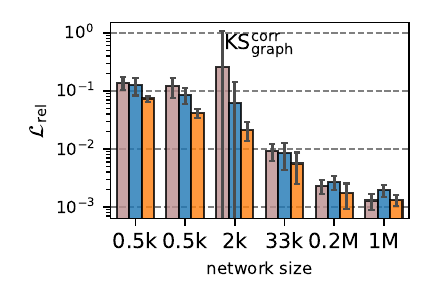} \ 
\includegraphics[height=4.49cm]{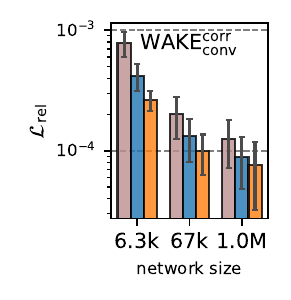} \
\includegraphics[height=4.49cm]{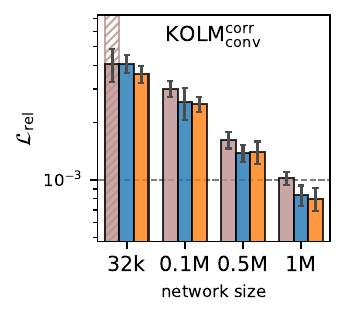} 
\caption{Inference accuracy measured in $\mathcal{L}_\mathrm{rel}$ for correction setups on \gls{ks}, \gls{kolm}, and \gls{wake} systems; displayed models were trained with \one{}(brown), \nog{}(blue), \wig{}(orange); across network architectures (graph networks for KS, conv-nets otherwise), and network sizes \wig{} has lowest errors; one 32k \one{} model diverged in the \gls{kolm} case that \nog{} and \wig{} kept stable}
\label{fig:corr_agnostic}%
\end{figure}%
\subsection{Distribution Shift and Long-term Gradients}
We first focus on establishing a common ground between the different variants by focusing on correction tasks for the physical systems. Training over 400 models with different architectures, initialization, and parameter counts, as shown in Figure \ref{fig:corr_agnostic}, reveals a first set of fundamental properties of training via unrolling: 
it reduces inference errors for graph- and convolutional networks, despite changes in dimensionality, the order of the \gls{pde}, and the baseline solver architecture.
For the vast majority of tests, models trained with unrolling outperform the corresponding one-step baselines. Unrolling is also versatile concerning the type of the modeled correction, as networks were tasked to either learn truncations due to reduced
convergence orders  with graph networks (\gls{ks}), or 
spatial grid coarsening errors via CNNs (\gls{kolm}, \gls{wake}). 

These findings are not new \citep{bar2019learning,Um2020Solver-in-the-loop}, but confirm that our setup matches previous work. 
They already relate to question (IV), i.e. that testing unrolled architectures can be performed efficiently on cheaper low-dimensional problems, such as the \gls{ks} system.
Figure \ref{fig:kolm_inference_trajectory} visualizes exemplary inference rollouts of the \gls{kolm} system, where differences in the learning systems emerge.

\begin{figure}[t]
    \centering
    \includegraphics[height=5.9cm]{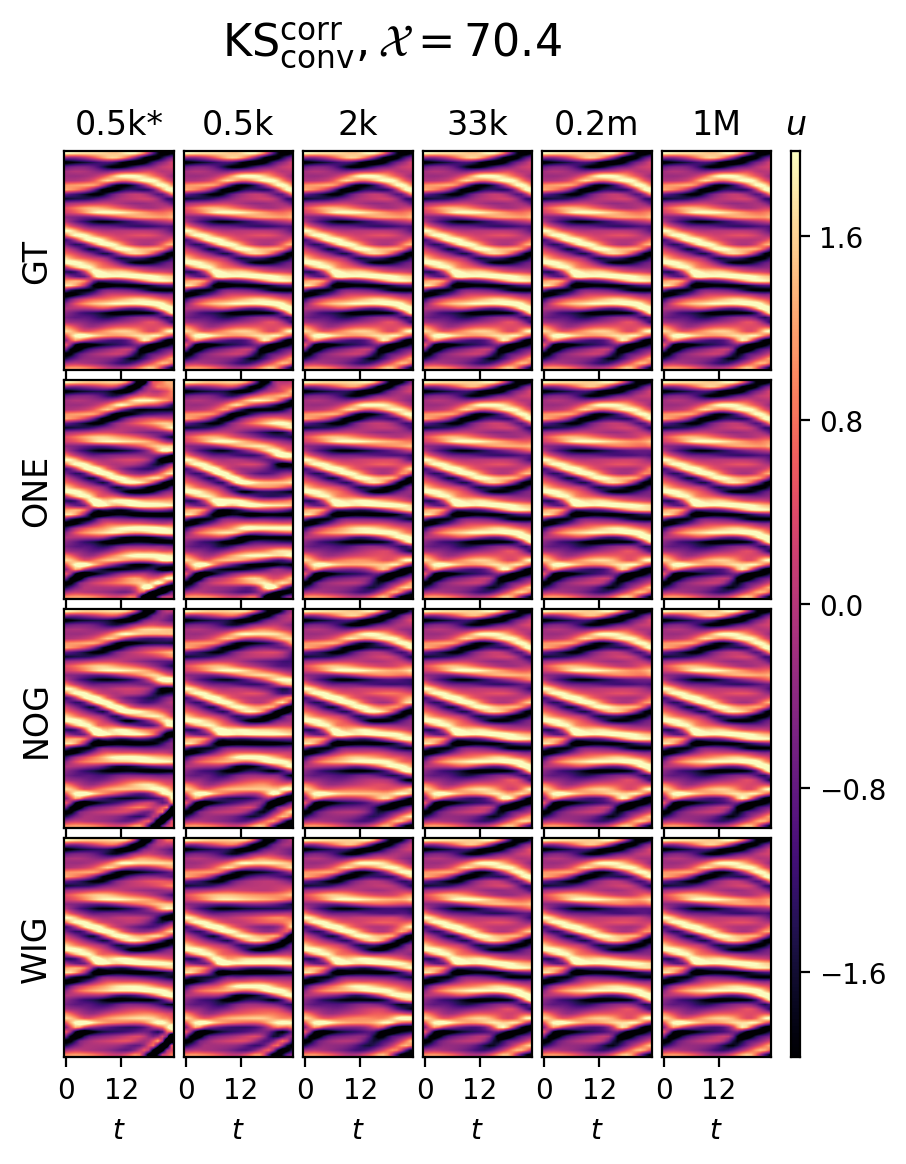}
    \includegraphics[height=5.8cm]{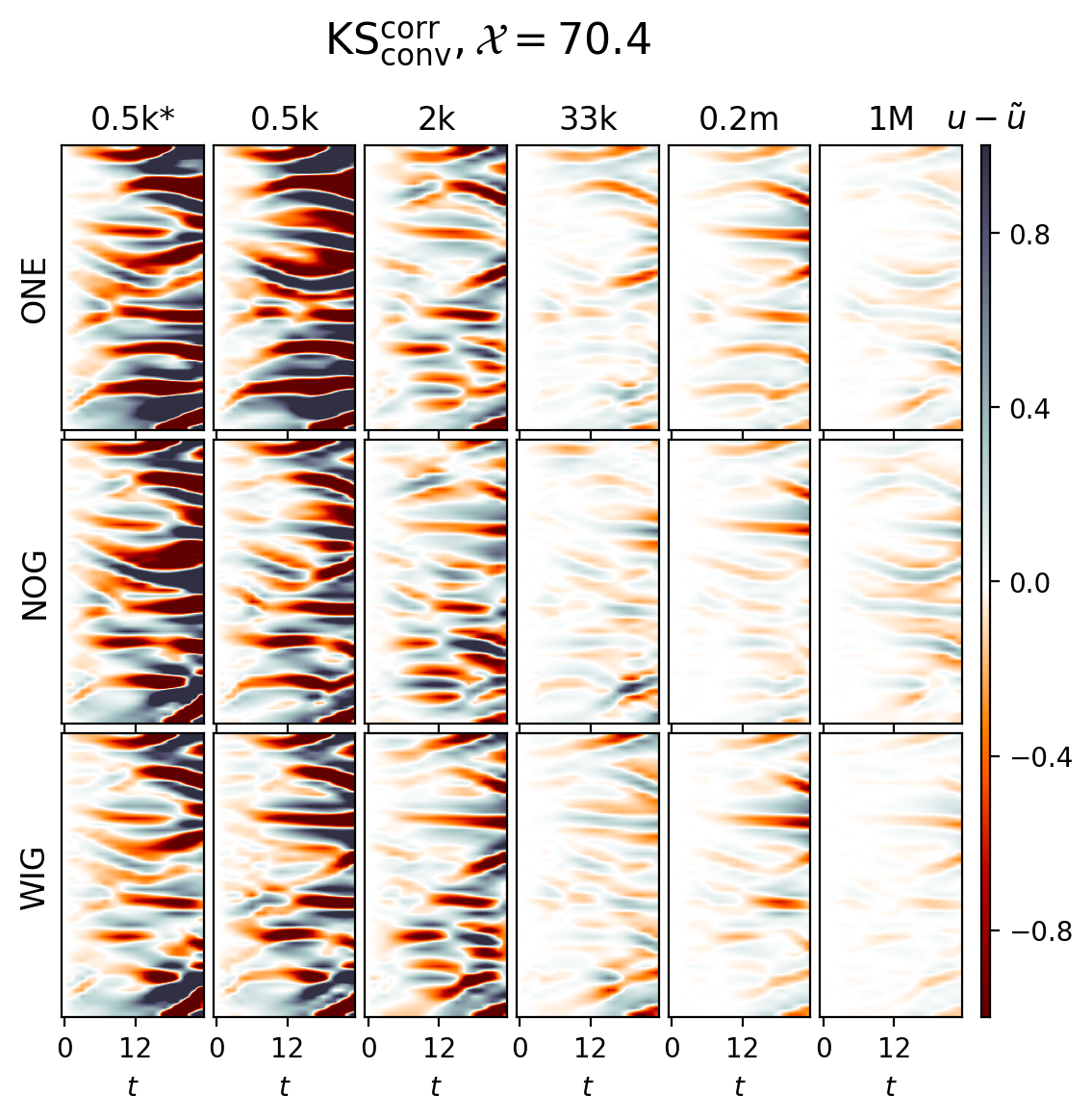}
    \includegraphics[height=5.8cm]{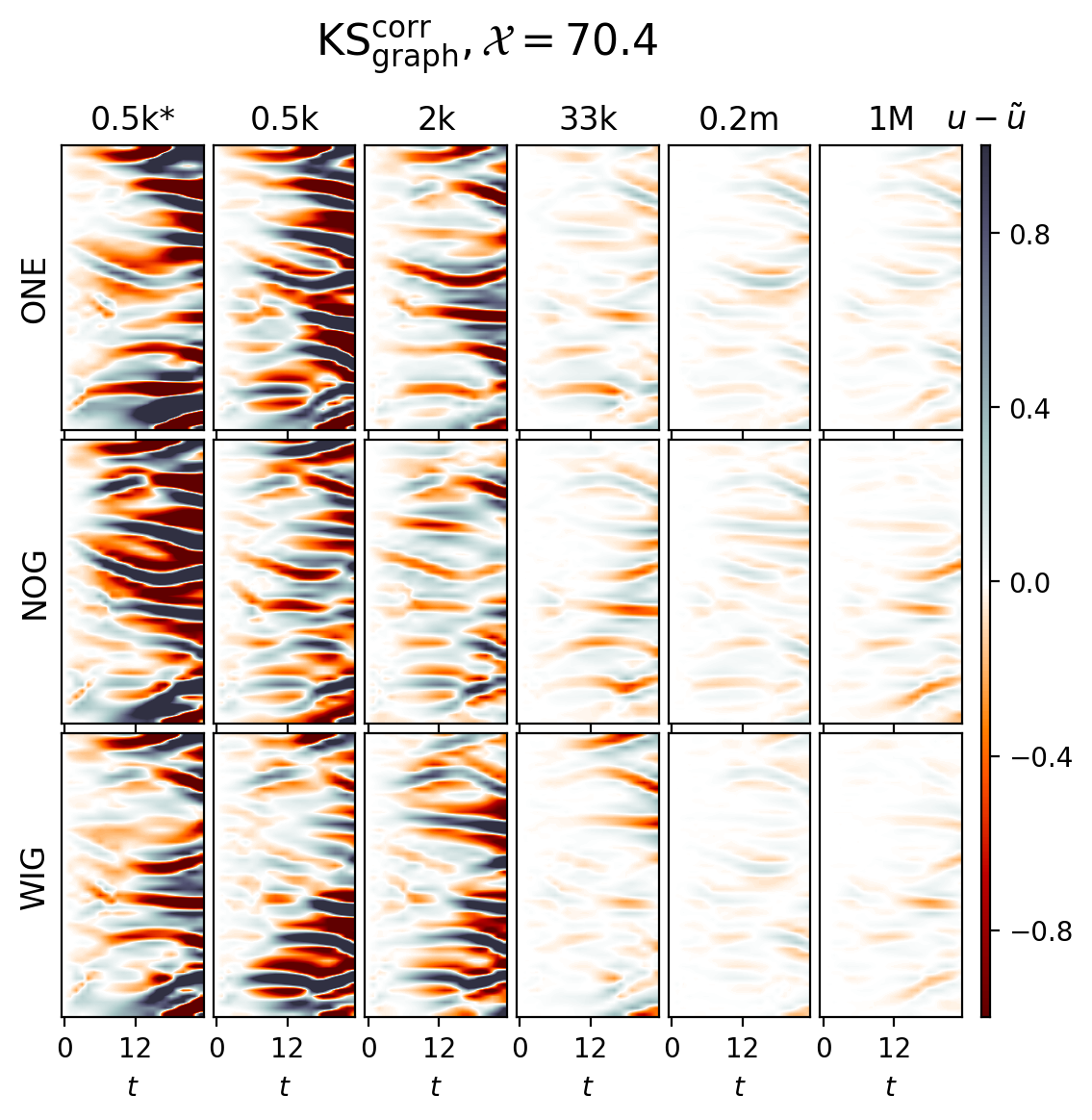}
    \caption{Exemplary inference trajectories for  the \gls{ks} system on extrapolation data for $\mathcal{X}=70.4$; trajectories of CNNs on the left, errors of CNNs and GCN in the middle, and on the right respectively for all training variants}
    \label{fig:ks_inference_short}
\end{figure}%

\begin{figure}[t]
    \centering
    \includegraphics[height=3.5cm]{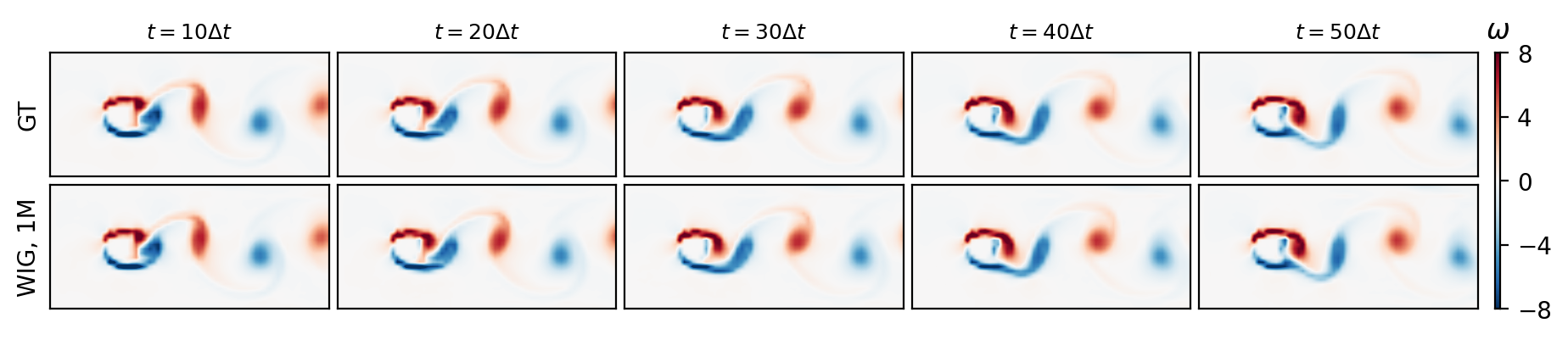}
    \caption{Exemplary inference trajectories for  the \gls{wake} system for the ground truth and \wig{} training over 50 timesteps}
    \label{fig:wake_inference}
\end{figure}%

\begin{figure}[ht]
    \centering
    \includegraphics[height=6cm]{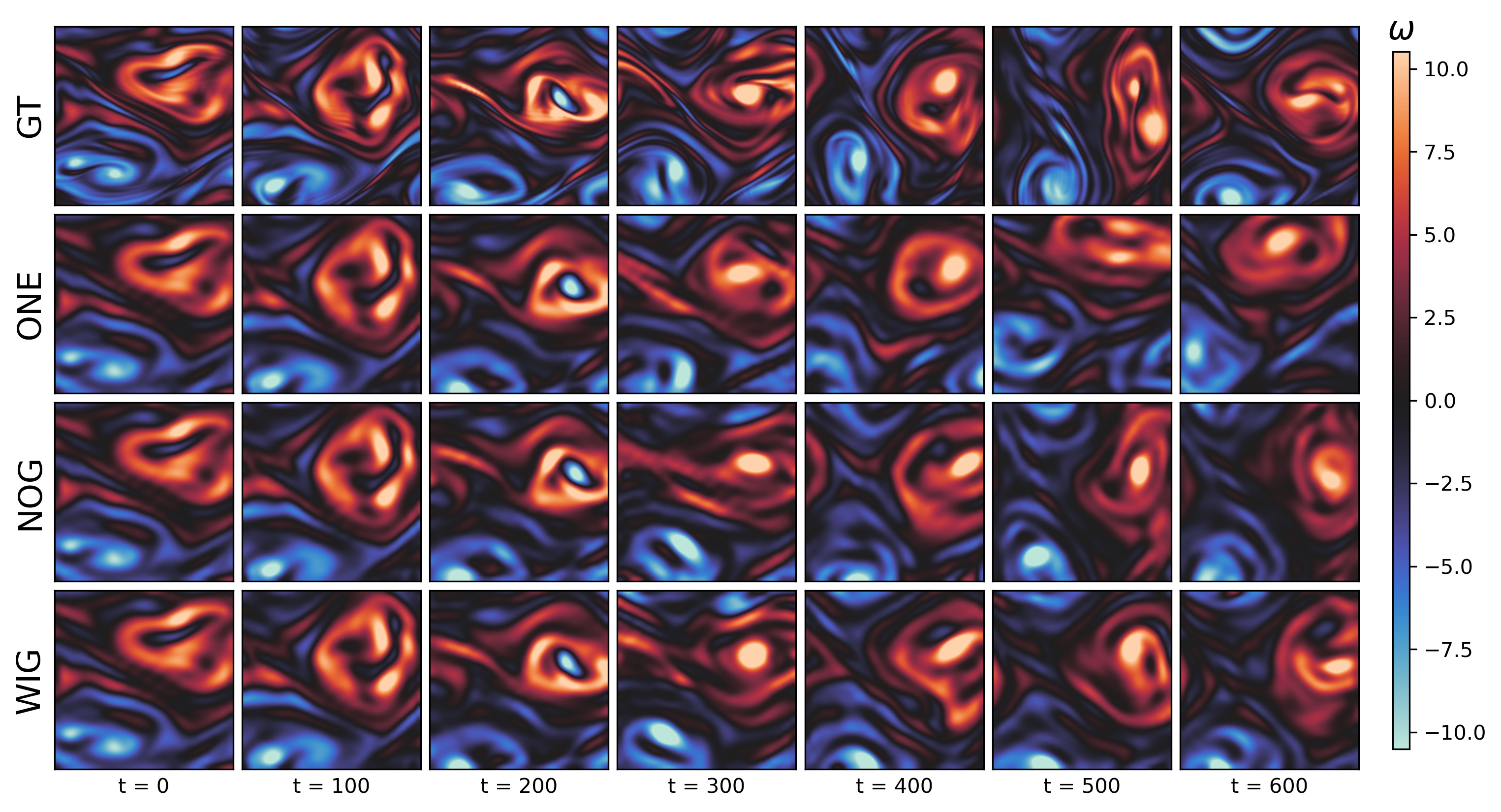}
    \caption{Exemplary inference trajectories over long inference horizons for $Re=1000$, comparing vorticity values for the ground truth simulation, and corrective setups trained with \one{}, \nog{}, and \wig{}; corrective setups used the 1M network initialized with the same seed}
    \label{fig:kolm_inference_trajectory}
\end{figure}%
\paragraph{Disentangling Contributions} 
Next, we investigate the effect of backpropagating gradients in the unrolled chain of network and simulator operations. 
The non-differentiable \nog{} setup already addresses the data shift problem, as it exposes the learned dynamics' attractor at training time. The inference errors are visualized in Figure \ref{fig:corr_agnostic}. The graph-network setup on the \gls{ks} is equivalent to the one studied in \cite{brandstetter2022message,dulny2023dynabench}, and the 1M model matches the one used by Brandstetter et al. \cite{brandstetter2022message} in terms of size and overall architecture. We present correction results in this section, while previous works have used these graph networks for prediction. The \gls{wake} experiments match the architecture of \cite{miyanawala2017efficient} and span the network size use by Um et al. \cite{Um2020Solver-in-the-loop}, who also studied correction setups. The closest relative to our \gls{kolm} setup can be found in the "learned correction" by Kochkov et al. \cite{Kochkov}, where a convolutional ResNet with $\sim 0.2$M parameters was trained in a very similar correction setup.

On average, training with \nog{} over \one{} training yields an error reduction of 33\%, which answers question (II). For large architectures, inference accuracy increases and learned and ground truth systems become more alike. We believe that in these cases, their attractors are similar, and unrolling is less crucial to expose the learned attractor. However, \nog{} training still uses crude gradient approximations, which negatively influences training regardless of how accurate the current model is. We believe this is why the \nog{} variant falls behind for large sizes.
While \nog{} models remain closer to the target than \one{} in all other cases, the results likewise show that the differentiable \wig{} setup further improves the performance.
These networks reliably produce the best inference accuracy. Due to the stochastic nature of the non-linear learning processes, outliers exist, such as the $0.5$m \nog{} model of the \gls{kolm} system. Nonetheless, the \wig{} models consistently perform best and yield an average improvement of 92\% over the \one{} baseline. The conclusions from the error average can also be drawn from the behavior of the best-performing networks of each training variation (see \ref{sec:appendix_additional_results}).
\begin{figure}[h]
\centering
\includegraphics[height=4.5cm]{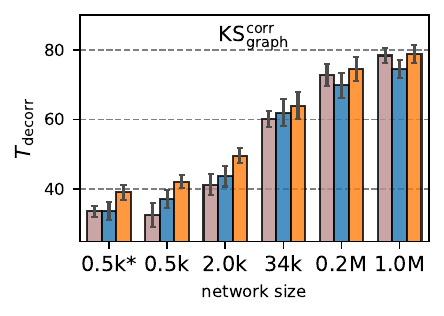} 
\includegraphics[height=4.5cm]{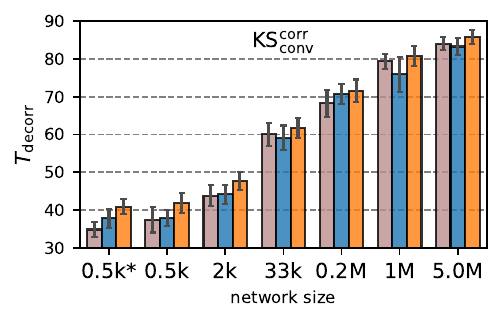} 
\caption{Number of timesteps until the threshold is reached for de-correlation }
\label{fig:ks_corr_cnn_correlation}
\end{figure}%

\paragraph{Long-term Evaluations}
We evaluate a cross-correlation metric to measure the similarity of the data in the inferred trajectories to the ground truth data. We first run an autoregressive inference sequence and calculate the correlation as 
\begin{equation}
    C^t = \mathrm{corr}(\mathbf{u}^t, \tilde{\mathbf{u}}^t) 
\end{equation}
After calculating the correlation of inference frames to the ground truth for each timestep, we measure the time until the two trajectories are decorrelated. For the trajectories to be decorrelated, we define a threshold of $C<0.8$, and measure the timestep until at least 5 frames in the trajectory exceed this threshold, such that the decorrelation time becomes
\begin{equation}
    T_\mathrm{decorr} = \min_T \big\{T : \sum_{t=0}^T\mathbf{1}(C^t<0.8) \geq 5\big\}.
\end{equation}
The statistics of the reached step-counts were gathered over all test cases, as for the $\mathcal{L}_2$ errors.
Figure \ref{fig:ks_corr_cnn_correlation} depicts the time until de-correlation between inference runs and the ground truth for the \gls{ks} system. For all model sizes and both network architectures, \wig{} achieves the longest inference rollouts until the de-correlation threshold is reached. Similarly to the observations on the $\mathcal{L}_2$ metric in Figure \ref{fig:corr_agnostic}, \nog{} is the second best option for small and medium-sized networks when it comes to correlation. This observation again translates between the network architectures.  

\begin{figure}[h]
\centering
\includegraphics[height=4.5cm]{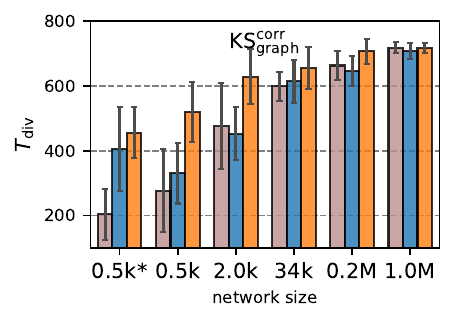} 
\includegraphics[height=4.5cm]{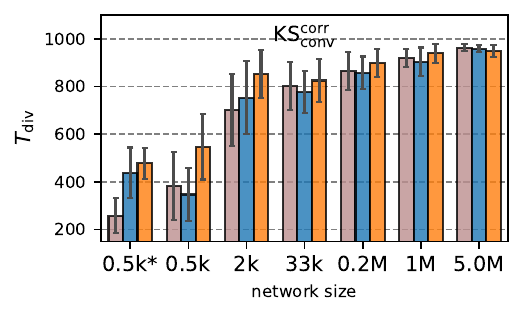} 
\caption{Number of timesteps until the threshold is reached for divergence}
\label{fig:ks_corr_cnn_divergence}
\end{figure}

In addition, we studied the time until divergence for the \gls{ks} system. This metric sets a high $\mathcal{L}_2$ threshold, which would correspond to a hundred-fold increase in the state of a simulation, and again measures the timesteps until this value is reached. It can be formalized as
\begin{equation}
    T_\mathrm{div} = \min_T \big\{T : \mathcal{L}_2(\mathbf{u}^t,\tilde{\mathbf{u}}^t) \geq 500\big\}.
\end{equation}
Note that this metric does not measure whether there is a particular similarity to the ground truth trajectory. Rather, a large number of timesteps can be seen as a metric for the autoregressive stability of the network. Figure \ref{fig:ks_corr_cnn_divergence} shows this evaluation for the \gls{ks} system. Here, we also show a correction networks with convolutional networks for the correction of the \gls{ks} system. These are similar to the ones trained in \cite{melchers2023comparison}, where convolutional networks of size 0.5k and smaller were used in a correction setting. The unrolled setups excel at this metric. Unsurprisingly, mitigating the data shift by unrolling the training trajectory stabilizes the inference runs. Through unrolling, the networks were trained on inference-like states. This also leads to better alignment of the ground truth and the learned attractor. Figure \ref{fig:ks_corr_cnn_divergence} shows how models trained with \one{} are most likely to experience instabilities. The fact that the unrolled setups are more stable indicates that the exploration of the learned attractor at training time improves inference performance.
The most stable models were trained with the \wig{} setup, whose long-term gradients further discourage unphysical outputs that could lead to instabilities in long inference runs. The large networks' divergence time comes close to the upper threshold we evaluated, i.e. 1000 steps.

\begin{figure}
\hspace{-0.2cm}
    \includegraphics[height=2.6cm]{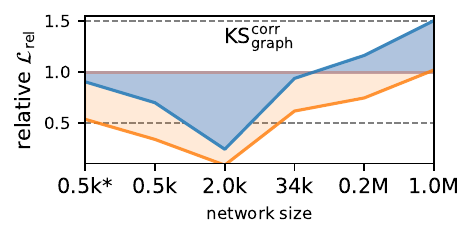}\hspace{-0.3cm}
    \includegraphics[height=2.6cm]{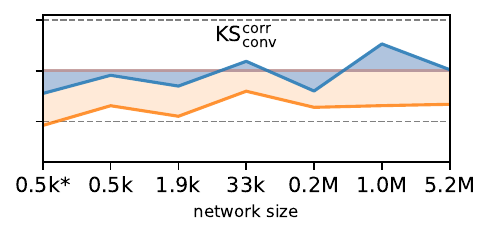}\hspace{-0.3cm}
    \includegraphics[height=2.6cm]{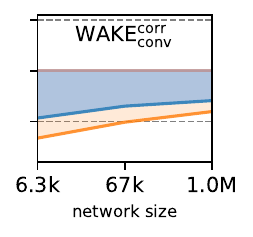}\hspace{-0.3cm}
    \includegraphics[height=2.6cm]{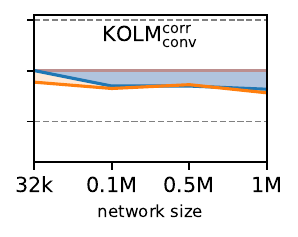}\hspace{-0.3cm}
\caption{$\mathcal{L}_\mathrm{rel}$ error of \nog{} and \wig{} setups relative to the error achieved by \one{} (i.e. ${\mathcal{L}^\mathrm{NOG}_\mathrm{rel}}/{\mathcal{L}^\mathrm{ONE}_{\mathrm{rel}}}$  and ${\mathcal{L}^\mathrm{WIG}_\mathrm{rel}}/{\mathcal{L}^\mathrm{ONE}_{\mathrm{rel}}}$); \wig{} training reliably produces more accurate results than \nog{} and \one{}; shading added for better identification of positive/negative factors}

\label{fig:corr_relative}%
\end{figure}

To conclude, our results allow for disentangling the influence of data shift and gradient divergence. 
As all training modalities, from data sets to random seeds, were kept constant, the only difference between \nog{} and \wig{} is the full gradient information provided to the latter.
As such, we can deduce from our measurements that reducing the data shift contributes to the aforementioned improvement of 33\%, while long-term gradient information yields another improvement of 15\% (question II).
Figure \ref{fig:corr_relative} highlights this by depicting the accuracy of the unrolled setups normalized by the respective \one{} setup for different model sizes and physical systems. \wig{} yields the best $L_2$ reductions. \nog{} in certain cases even performs worse with factors larger than one, due to its mismatch between loss landscape and gradient information, but nonetheless outperforms \one{} on average. A crucial takeaway is that unrolling, even without differentiable solvers, provides measurable benefits to network training. As such, interfacing existing numerical codes with machine learning presents an attractive option that does not require the re-implementation of established solvers. Our results provide estimates for the expected gains with each of these options. 

\begin{figure}
\centering
\includegraphics[height=3.75cm]{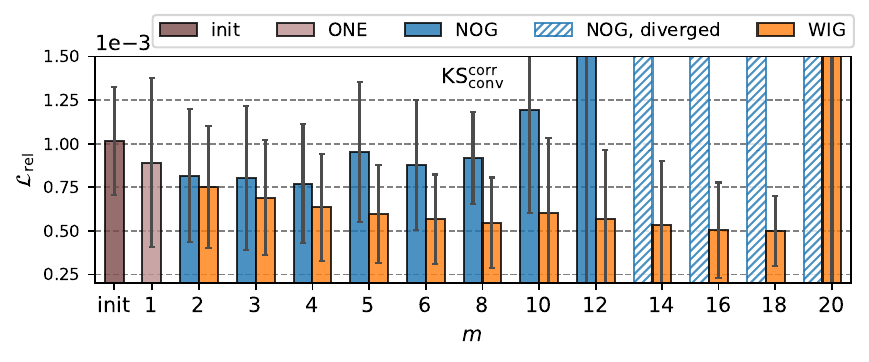} \hspace{-0.4cm}
\includegraphics[height=3.70cm]{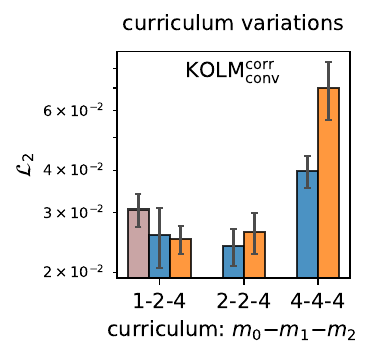} \hspace{-0.4cm}
\includegraphics[height=3.70cm]{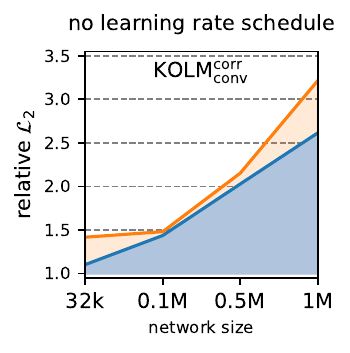}
\caption{Left: variation of unrolling horizons $m$ trained for the \gls{ks} system; all models were pre-trained with the leftmost one-step setup marked as \texttt{init} and then finished training in their specific $m$ parameterizations; 
with \nog{} training, the error increases and explodes for $m>6$, which corresponds to $\frac{1}{2}t_{c, \text{KS}}$ characteristic timescales in our discretization; \wig{} remains stable for longer horizons, diverges at $m>18$, i.e. for $m>t_{c, \text{KS}}$; Middle: different curriculums for the \gls{kolm} system; starting the training with large $m$ leads to unfavorable network states; Right: Inference accuracy without learning rate schedule relative to with scheduling, i.e. $\mathcal{L}_2^\mathrm{no-schedule}/\mathcal{L}_2^\mathrm{schedule}$; All models benefit from learning rate scheduling, indicated by values are larger than one, and especially for larger models the schedule is crucial}
\label{fig:unrolling_horizon}
\end{figure}

\paragraph{Unrolling horizons}
Based on the theoretical analysis in Section \ref{sec:unrolling_theory}, the learned simulations should perform best for moderate unrolling horizons and deteriorate when these horizons are further increased. As such, the effects of unrolling on data shift and gradient divergence depend on the unrolled length $m$, as summarized in Figure \ref{fig:schematic}.
A long horizon diminishes the data shift, as the learned dynamics can be systematically explored by unrolling. At the same time, gradient inaccuracy impairs the learning signal in the \nog{} case, while exploding gradients can have a similar effect for long backpropagation chains in the \wig{} setup. We observe these effects in an experimental setup where the unrolled training horizon is varied for networks 
where training is continued from an identical starting point obtained via pre-training.
The results are visualized in Figure \ref{fig:unrolling_horizon}. 
For increasing horizons, the inference performance of \nog{} variants improves until a turning point is reached. After that, \nog{} performance deteriorates quickly. In the following regime, the stopped gradients cannot map the information gained by further reducing the data shift to an effective parameter update. The quality of the learning signal deteriorates.
As outlined in Section~\ref{sec:unrolling_theory}, this divergence does not only depend on the underlying timescale, but is also influenced by the number of timesteps. With an increasing number of timesteps, more gradients are cut in the \nog{} system and the gradients become inaccurate. Nevertheless, it is worth mentioning that the divergence occurs at  6 steps, which is roughly equivalent to $t_{c,\text{KS}}/2=12.4/2$.
This inaccuracy effect is mitigated by allowing gradient backpropagation through the unrolled chain in the \wig{} setup. Herein, the inference accuracy benefits from even longer unrollings and only diverges for substantially larger $m$ when instabilities from recurrent evaluations start to distort the direction of learning updates. As the unrolling horizon $m$ significantly surpasses the timescale $t_{c,\text{KS}}=12.4$, the network optimization produces 
diminishing returns.
These empirical observations confirm the theoretical analysis from Section \ref{sec:training_to_solve_pde}: From this theoretical perspective, the chaotic timescale gives an estimate for the unrolling horizon for which gradients explode in the target dynamical system. However, in the early stages of the optimization when training losses are still high, the neural simulator is not a good representation of the target dynamics. In this case, the chaotic timescale of the target is not very meaningful. Thus, the training horizons can not only be determined based on the target dynamics, which is also reflected in our study. Here, we recommend using a curriculum instead as described in Section~\ref{sec:curriculum}.
%
The curriculum-based approach features an incremental increase of the unrolling length $m$ at training time \citep{Um2020Solver-in-the-loop,lam2023learning}. Figure \ref{fig:unrolling_horizon} shows how long unrollings in the initial training phases can hurt accuracy. Untrained networks are particularly sensitive since autoregressive applications can quickly lead to divergence of the simulation state or the propagated gradients. At the same time, learning rate scheduling is necessary to stabilize gradients (question IV). 

While designing unrolling horizons is not trivial, the fundamental takeaway is that there exist such horizons for which training performance is improved for both \nog{} and \wig{}, while this effective horizon is longer for \wig{}, as illustrated in Figure~\ref{fig:schematic}.

\paragraph{Gradient Stopping}
Cutting long chains of gradients was previously proposed as a remedy for training instabilities of unrolling \citep{list_chen_thuerey_2022, brandstetter2022message, pmlr-v162-suh22b}.
A popular option is to discard gradients for a number of initial steps. This technique is referred to as "pushforward trick" in \citep{brandstetter2022message} and as "warm-up" steps in \citep{prantl2022guaranteed}. This procedure is similar to the unrolling variants we study. Like unrolling, it time-advances the system, thus drawing data from a more inference-like distribution. However, there are significant differences in the treatment of the losses on intermediate steps and their backpropagation gradients. In the warm-up approach, a number $w$ of initial steps are discarded for loss calculation, and their gradients are stopped. Only steps after the warm-up horizon actually contribute to the loss and are differentiated. The loss formulation is given by
\begin{equation}
    \mathcal{L}_w = \sum_{s=w+1}^m\mathcal{L}_2(\tilde{u}^{t+s},g_\theta^s(\tilde u^t))
\end{equation}
Another strategy to stabilize unrolling consists of truncating the backpropagation gradients \cite{sutskever2013}. It was reported to be beneficial in long unrolled physical systems \cite{list_chen_thuerey_2022}. In this approach, the forward process of the unrolling is not touched at all. However, the backpropagation paths are only evaluated in subsections of the full unrolled trajectory, effectively limiting the maximum length of a backpropagation path. The gradient propagation for unrolling horizons $m$ split into $v$ subsections reads as
\begin{equation}
    \frac{\partial \mathcal{L}_\mathrm{v}^s}{\partial \theta} =
    \sum_{B=B_v}^s \bigg[
    \frac{\partial \mathcal{L}^s}{\partial g_\theta^s} 
    \bigg ( \prod_{b=1}^{s-B}
    \frac{\partial g_\theta^{s-b}}%
    {\partial g_\theta^{s-b-1}} 
    \bigg)
    \frac{\partial g_\theta^B}
    {\partial \theta}\bigg], \quad \mathrm{with} \quad B_v = \bigg(\floor*{\frac{s-1}{m/v}}\frac{m}{v}\bigg)+1.
\end{equation}
We evaluate these methods in comparison to unrolled models trained with and without unrolling.
\begin{figure}
\centering
\includegraphics[width=0.5\textwidth]{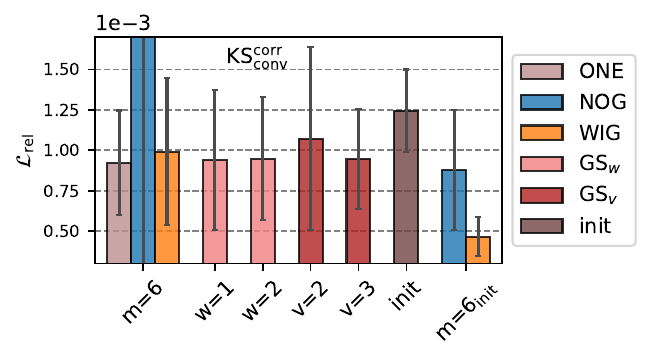}
\caption{Evaluation of gradient stopping techniques; $m=6$ shows performance of networks trained \textit{without} curriculum with 6 unrolling steps activated from the start; \textit{warm-up} steps with $w=1$ discarded steps at the beginning yields minor improvements; backpropagation splitting shows no improvements over the differentiable \gls{wig}; best performance is reached with unrolling and curriculum, starting with the \textit{init} network before unrolling 6 steps}
\label{fig:gradient_stop}
\end{figure}%
As shown in Figure \ref{fig:gradient_stop}, setting $w$=1 can yield mild improvements over training with the full chain if no curriculum is used, which is in line with the results in \cite{brandstetter2022message, prantl2022guaranteed}.
Dividing the backpropagation into subsections \citep{list_chen_thuerey_2022} likewise does not yield real improvements in our evaluation. 
We divided the gradient chain into two or three subsections, identified by the parameter $v$ in  Figure \ref{fig:gradient_stop}. 

The best performance is obtained with the full \wig{} setup and curriculum learning, where unrolled models are pre-trained with $m$=1 models. This can be attributed to the more accurate gradients of \wig{}, as all gradient-stopping variants above inevitably yield a mismatch between loss landscape and learning updates (question II).

\begin{figure}
\includegraphics[height=3.0cm]{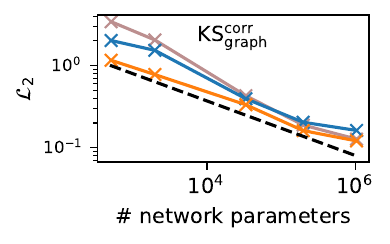} \hspace{-0.3cm}
\includegraphics[height=3.0cm]{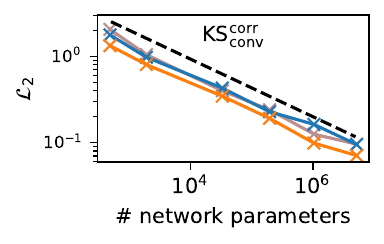} \hspace{-0.3cm}
\includegraphics[height=3.0cm]{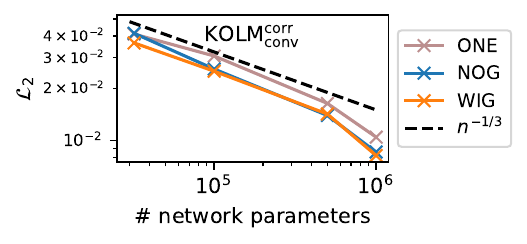} \hfill
\caption{Accuracy convergence over network size, correction networks converge with $n^{-1/3}$}
\label{fig:correction_convervence}
\end{figure}

\paragraph{Network Scaling}
Figure \ref{fig:corr_agnostic} shows clear, continuous improvements in accuracy for increasing network sizes. An obvious conclusion is that larger networks achieve better results. However, in the context of scientific computing, arbitrarily increasing the network size is not admissible. As neural approaches compete with established numerical methods, applying pure neural or correction approaches always entails accuracy, efficiency, and scaling considerations. The scaling of networks towards large engineering problems on physical systems has been an open question.
We compute a convergence rate between the test loss and the number of network parameters. To this end, we use an average test loss for each individual combination of network size and training setup. Here we use the $\mathcal{L}_2$ metric following the standard definition of convergence in numerical analysis \cite{holt1984numerical}. This metric provides consistency across timesteps. Relative errors result in unequally weighted timeframes when ground truth solution values vary across time. This is to be avoided for the medium-range test horizons investigated here (40 steps in \gls{ks}, 250 steps \gls{kolm}). The average is computed over the full set of random seeds used in our study, i.e., 8 to 20 individual training runs per variant and size depending on the physical system. The convergence rate estimates are based on the same test trajectories as previous evaluations like Figure \ref{fig:corr_agnostic}. For the correction setups, we estimate the convergence rate of the correction networks with respect to the parameter count $n$ to be $n^{-{1}/{3}}$, as shown in Figure \ref{fig:correction_convervence}.

Interestingly, the measured convergence rates persist across the tested physical system and the studied network architectures. This convergence rate is poor compared to classic numerical solvers. Consequently, neural networks are best applied for their intrinsic benefits which derive from the universal approximation theorem \cite{chen1995universal}. They possess appealing characteristics like data-driven fitting, reduced modeling biases, and flexible applications. In contrast, scaling to larger problems is more efficiently achieved by numerical approaches. In applications, it is thus advisable to combine both methods to render the benefits of both components. 
In this light, our correction results for small and medium-sized networks have more relevance to potential applications. 
Therefore, the two main conclusions from this subchapter are as follows: Firstly, the solver part in correction approaches scales more efficiently than the neural network part. Secondly, it becomes apparent how network size dominates the accuracy metrics (question III). Thus, parameter counts need to be equal when fairly comparing network architectures.

Another observation is the lack of overfitting: despite the largest \gls{ks} models being heavily overparameterized with five million parameters for a simulation with only 48 computational cells accuracy still improves. This finding aligns with recent insights on the behavior of modern large network architectures \citep{belkin2021fit}.
\subsection{Varied Learning Tasks}
To broaden the investigation of the unrolling variants,
we vary the learning task by removing the numerical solver from training and inference. This yields \textit{prediction} tasks where the networks directly infer the desired solutions. Apart from this increased difficulty, all other training modalities were kept constant, i.e., we likewise compare non-differentiable \nog{} models to full unrolling (\wig{}).%
\begin{figure}
\centering
\includegraphics[height=4.3cm]{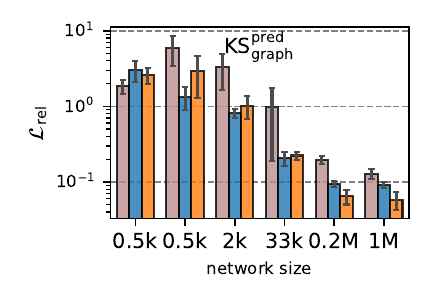}\hspace{-0.3cm}
\includegraphics[height=4.3cm]{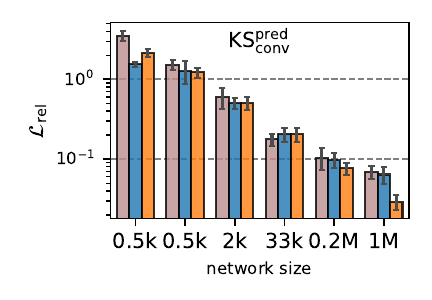} \hspace{-0.3cm}
\includegraphics[height=4.3cm]{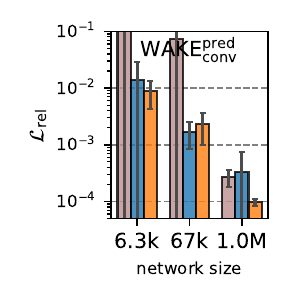}
\caption{Prediction setups with inference accuracy measured in terms of  $\mathcal{L}_\mathrm{rel}$ on \gls{ks} and \gls{wake}; \nog{} performs well for small network sizes, while \wig{} is show advantages for larger ones}
\label{fig:prediction_size}
\end{figure}

\begin{figure}
\centering
\includegraphics[width=.549\textwidth]{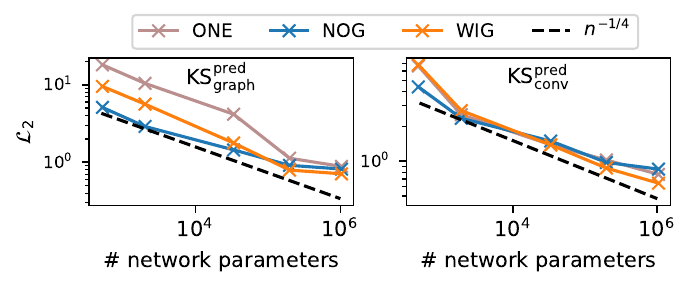}
\caption{Accuracy convergence over network size, prediction converges with $n^{-1/4}$}
\label{fig:prediction_convergence}
\end{figure}

\paragraph{Prediction}
The parameter count of models still dominates the accuracy but in contrast to before, 
the \nog{} setup performs better than both alternatives for smaller network sizes.
This is shown in Figure \ref{fig:prediction_size}. The \gls{ks} prediction setups are comparable to \cite{stachenfeld2021learned}, where convolutional ResNets of size 0.2M were trained on prediction tasks, albeit their networks feature an additional "dilation" ratio, specifying a sparsely populated filter. The \gls{wake} predictions are similar to \cite{morimoto2022generalization}.
The inference errors for the prediction setup are roughly an order of magnitude larger than the respective correction errors. 
This indicates that pure predictions more quickly diverge from the reference trajectory, especially for small network sizes.
For the setups in Figure \ref{fig:prediction_size}, \nog{} training achieves improvements of 41\%, and \wig{} improves on this by a further 30\%. In general, unrolling maintains strong benefits in prediction setups, but differentiability is less beneficial than in correction setups. Similarly to the correction task before, we use these test trajectories to estimate a convergence rate with respect to increasing parameter counts. Results are visualized in Figure \ref{fig:prediction_convergence}. The prediction convergence rate is slightly worse but comparable to the correction setups.

\begin{figure}
\centering
\includegraphics[height=3.7cm]{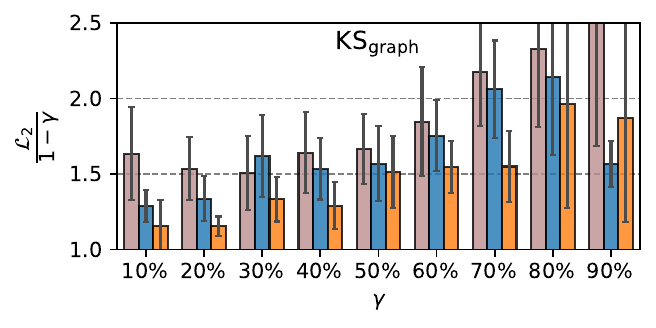}
\includegraphics[height=3.6cm]{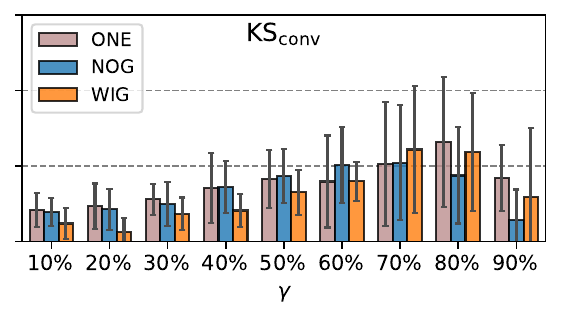} 
\vspace{-.5cm}
\caption{Transition from prediction to correction; a smooth transition in modeling difficulty is achieved by varying the timestep of the numerical solver embedded in the correction step $\gamma = \Delta t / \Delta \bar t$, with the solver timestep $\Delta t$ and the ground truth timestep $\Delta \bar t$; results are normalized by the correction difficulty $1-\gamma$}
\label{fig:corr_pred_transition}
\end{figure}

\paragraph{Smooth Transition}
While the learning task is typically mandated by the application, our setup allows us to investigate the effects of unrolling for a smooth transition from prediction to increasingly simple correction tasks, in line with Figure \ref{fig:flowchart}.
This is achieved by introducing a parameterized timestep $\Delta t =\gamma \Delta\bar t$ to the correction setup, where $\Delta\bar t$ represents the ground truth timestep. In contrast to the previous correction setup for the \gls{ks} equation, we use the ground truth solver $\Bar{\mathcal{S}}$ for this setup. Furthermore, as in the regular \gls{ks} setup, the physical state does not experience any reduction, such that $R(\bar{\mathbf{u}})=\bar{\mathbf{u}}$. The only source of error between the ground truth and the correction setup stems from the different time step sizes and thus depends on the value of $\gamma$. We formalize the setup as
\begin{equation}
    \mathbf{u}^{t+1} = f_\theta(\Bar{\mathcal{S}} (\mathbf{u}^t; \gamma\Delta\bar t)).
\end{equation}
With this setup a prediction task without any  numerical solver is constructed by setting $\gamma=0$, while $\gamma=1$ recovers the ground truth system where no correction is necessary. It is evident that the latter does not constitute a meaningful learning task, but for all values in between, the difficulty of correction tasks increases with decreasing $\gamma$. In other words,
we transition away from pure predictions by providing the network with improving inputs by increasing the time step of the reference solver for correction tasks.
Since the total error naturally decreases with simpler tasks, Figure \ref{fig:corr_pred_transition} shows the performance normalized by the task difficulty $1-\gamma$. The positive effects of \nog{} and especially \wig{} training carry over across the full range of tasks. Interestingly, the \nog{} version performs best for very simple tasks on the right sides of each graph. This is most likely caused by a relatively small mismatch of gradients and energy landscape.
As the task difficulty was changed by artificially varying the numerical solver's accuracy, this mimics the effects of basing the correction setups on different numerical schemes. Unrolling manifests a stable performance improvement across all numerical accuracies, from very crude approximations (10\%) to highly accurate ones that almost fit the ground truth on their own (90\%). We also observe these effects of unrolling for different architectures in convolutional and graph-based networks. Thus, unrolling promises benefits for many other correction setups utilizing numerical solvers of with different accuracy and various neural architectures.
\begin{figure}
\centering
\includegraphics[height=3.88cm]{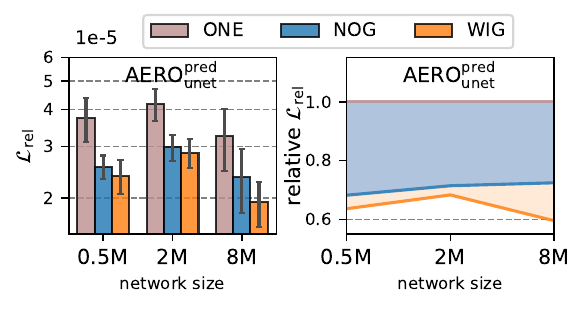}  
\caption{Left: Prediction of the \gls{aero} system, $\mathcal{L}_\mathrm{rel}$ errors of \one{}, \nog{}, and \wig{}; Right: Relative errors with respect to \one{}}
\label{fig:aero_results}
\end{figure}

\paragraph{Changing the Physical System}
As a complex, large-scale test scenario, our learning setups for the \gls{aero} system implement the recommendations we derived from the previous results. 
This dataset comprises aerofoil flows in the transonic regime and features shocks, fast-changing vorticial structures, and diverse samples around the critical $\text{Ma}=0.8$. As such, we deploy a modern attention-based U-Net \citep{oktay2018attention}.
Figure \ref{fig:aero_results} shows the inference evaluation of our trained models. Once again, unrolling increases accuracy at inference time. 
On average, unrolling improves the inference loss by up to 41\% in the \gls{aero} case. The results reflect our research targets: Network size has the largest impact, and choosing the right size balances performance and accuracy (III). Additionally, our models were trained with a learning rate scheduled curriculum to achieve the best results, details of which are found in \ref{app:aero_case}. Training on the \gls{aero} system resembles the behavior of lower-dimensional problems, e.g. the \gls{ks} system (IV). Similar to our previous prediction tasks, long-term gradients are less crucial but still deliver the best models (II). However, unrolling itself is essential for stable networks (I).

\begin{figure}
    \centering
    \includegraphics[height=3.8cm]{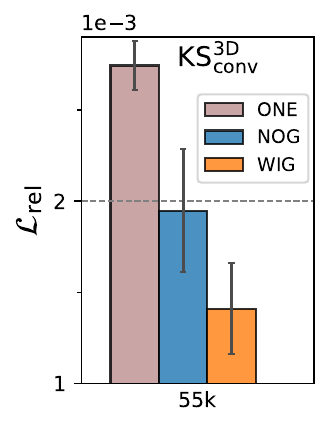} \ 
    \includegraphics[height=3.5cm]{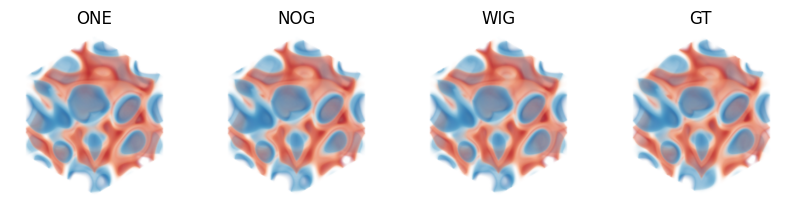}
    \caption{Comparison of the training variants on the three-dimensional \gls{ks} system, set up as a correction tasks with a convolutional ResNet; $\mathcal{L}_\mathrm{rel}$ evaluation on the left; visualization of the system state after 20 steps of simulation on the right}
    \label{fig:3d_ks}
\end{figure}

\paragraph{Extension to three dimensions}
To demonstrate that unrolling is effective regardless of the system's dimensionality, we conduct experiments on the three-dimensional \gls{ks} equation. 
We trained a convolutional ResNet to correct a lower-order solver, details as previously described in Section \ref{sec:ks3d}. 
The results are visualized in Figure \ref{fig:3d_ks}. A familiar pattern emerges as unrolled variants outperform the \one{} baseline. Once again, fully differentiable training with \wig{} performs best, while non-differentiable \nog{} unrolling comes in second.
While this test series focuses on a single, fixed network size, we can still observe that unrolling behaves similarly to the previous cases in higher dimensions. 

\subsection{The Potential of No-Gradient Unrolling}
Existing studies focus on fixed tasks, i.e. pure prediction tasks when no differentiable solver is available, and correction tasks when employing a differentiable solver. Given the large and complex implementations of solvers used in current scientific projects, often with decade-long histories, approaches that do not require the differentiability of these numerical solvers are especially attractive. Our experiments allow for quantifying the expected benefits when switching from a prediction task to a correction task with a non-differentiable solver.

\begin{figure}
\centering
\includegraphics[height=3.5cm]{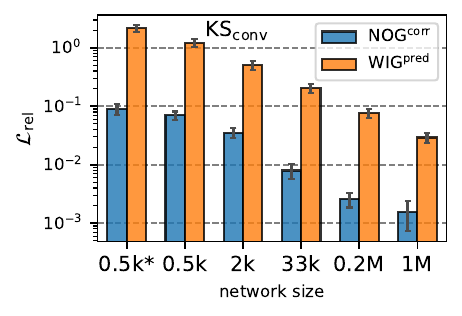}
\includegraphics[height=3.5cm]{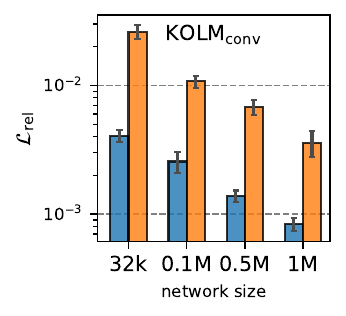}
\caption{Comparison of \wig{} prediction and \nog{} correction setups on the \gls{ks} system (left) and the \gls{kolm} flow (right)}
\label{fig:nog_corr_wig_pred}
\end{figure}

As illustrated in Figure \ref{fig:nog_corr_wig_pred}, this enables an average 13.3-fold accuracy improvement with the same network architecture and size. All architectures can benefit from training with \nog{} utilizing a non-differentiable numerical solver. The models with 0.5k parameters for the \gls{ks} perform 23.8 and 17.5 times better when supported by a low-fidelity solver, where the second value originates from a deeper variant of the 0.5k parameterization. The factor remains relatively constant for the remaining sizes, which all deploy deep architectures. Medium-sized networks with 2k and 33k parameters see a slight decrease in these improvements with values of  14.3 and 25.7 respectively, while the largest networks' accuracies improve by a factor of 30.0 and 18.9 once trained with \nog{} in a correction setup.
In the \gls{kolm} case, we observe smaller relative gains from introducing a numerical solver. Here, the factor is 4.9 on average. This relative factor is largest for small architectures, where we observe a value of 6.4 for the 32k network. Similarly to the \gls{ks} case, the benefits of using a numerical solver decrease for larger architectures. For our \gls{kolm} experiments the values are 4.19, 4.89, and 4.27 respectively. It has to be noted that these factors likely are specific to our setups and the used numerical solvers. However, the purpose of this evaluation is to illustrate the potential improvements made possible by correction unrolling.
Crucially, only an interface between the solver and the network is sufficient for \nog{} correction training, the core of the solver can remain untouched. These experiments indicate that there could be large benefits if scientific codes were interfaced with machine learning (questions I, III). We believe that this training modality could be a new common ground for machine learning and the computational sciences. In the past, machine learning focused works often focused on purely predictive tasks. As our results indicate, an interface with existing numerical architectures is not only necessary to deploy correction models in a scientific computing environment but additionally unlocks further accuracy improvements when models are trained with \nog{} unrolling. \\

\subsection{Computational Performance}
We measure the computational effort for both training and inference with our trained models. The \one{}, \nog{} and \wig{} variants have different computational complexity when it comes to calculating a network update. \one{} is the cheapest, as only one solver-network step needs to be computed. Both \nog{} and \wig{} are $m$-times more expensive in the forward pass. However, they differ in the backpropagation of gradients. The \nog{} setup does not differentiate solver operations, which makes the backpropagation pass cheaper than that of \wig{}. In principle, the gradient backpropagation of \nog{} setups is parallelizable across the unrolling horizon. While \wig{} requires a complete sequential backward pass through $m$ steps, \nog{} calculates \textit{independent} gradient contributions for each unrolled step. We report numbers for the training of one GCN of size 1M on the \gls{ks} system, as well as the CNN of size 0.1M on the \gls{kolm} system in Table \ref{tab:comp_perf_train}. Here, it seems that the TensorFlow implementation of the \gls{kolm} setup is capable of taking advantage of this parallelization, reducing the training time of \nog{} significantly when compared to \wig{}.

At inference time, \one{}, \nog{}, and \wig{} training variants all pose the same computational burden since their call signature is entirely identical. The computational time per inference step is listed in Table \ref{tab:comp_perf_inf} for the correction networks of \gls{kolm} setup. Since the numerical procedure used in the \gls{kolm} setup is relatively expensive, the inference time is dominated by the numerical solver. Switching to deeper architectures only results in small additional computational costs.
\begin{table}
    \begin{subtable}{.45\linewidth}
    \centering
    \begin{tabular}{c|c|c|c}
        Training time & 
        \hspace{0.5em} $\text{KS}_\text{graph}^\text{corr}$\hspace{0.5em} &
        \hspace{0.5em} $\text{KS}_\text{graph}^\text{pred}$\hspace{0.5em} &
        $\text{KOLM}_\text{conv}^\text{corr}$ \\         \hline
         \one{} & 34 min. & 28 min. & 3.6 h\\
         \nog{} & 84 min. & 64 min. & 10 h\\
         \wig{} & 87 min. & 67 min. & 16 h
    \end{tabular}
    \caption{}
    \label{tab:comp_perf_inf}
    \end{subtable}%
    \hfill
    \begin{subtable}{.45\linewidth}
    \centering
    \begin{tabular}{c|c}
         & Inference time (s) \\
         \hline
         32k & 0.0377 \\
         0.1M & 0.0389\\
         0.5M & 0.0391\\
         1M & 0.0401
    \end{tabular}
    \caption{}
    \label{tab:comp_perf_train}
    \end{subtable}
    \caption{(a): Computational time required for training a single graph model for \gls{ks} of size 1M for correction and prediction, and a convolutional model for \gls{kolm} of size 0.1M trained for a correction task; (b): Inference time for a single timestep of the \gls{kolm} model in a correction setting, averaged across 100 steps}
    \label{tab:com_perf}
\end{table}

\subsection{Discussion}
Our work has taken a task- and architecture-agnostic stance. Combined with the large scale of evaluations across more than three thousand models, this allows for extracting general trends. In line with the research targets from Section \ref{sec:introduction}, we arrive at the following set of recommendations:

\paragraph{(I) Non-differentiable unrolling as an alternative}
The mitigating effects on data shift introduced by unrolling can boost model performance. An unrolled but non-differentiable chain can increase performance by 33\% on average.
In addition, non-differentiable neural correction based simulators do not require a new implementation of existing numerical solvers and thus represent an attractive alternative for fields with large traditional code bases. 
When no differentiable implementations are available, previous work typically resorts to pure prediction models.
Instead, our results strongly encourage interfacing non-differentiable numerical implementations with machine learning, as doing so yields substantial accuracy improvements over prediction setups for identical network architectures and sizes.

\paragraph{(II) Differentiable unrolling delivers the best accuracy}
When other hyperparameters are fixed, a differentiable unrolling strategy consistently outperforms other setups.
At the cost of increased software engineering efforts, differentiable unrolling outperforms its one-step counterpart by 92\% on average. Long-term gradients prove to be especially important for correction setups where neural networks correct numerical solvers.

\paragraph{(III) correction approaches yield improved scaling:}
Our study across network sizes finds them predominant in terms of accuracy on inter- and extrapolative tests. However, we observe suboptimal convergence of the inference error when increasing the parameter count. These convergence rates compare poorly to the ones usually observed for numerical solvers. Most neural \gls{pde} simulators directly compete with numerical solvers. Thus, scaling to larger problems is critical for neural simulators. Unrolling can effectively train correction based simulators even when integrated with a non-differentiable solver. These combine the benefits of neural networks with the scaling properties of numerical solvers.

\paragraph{(IV) Low-dimensional problems match}
Our results translate between physical systems. Provided the systems are from a similar domain, i.e. we primarily consider turbulent chaotic systems,
high-level training behaviors translate between physical systems.
This validates the approach common in other papers, where broad studies were conducted on low-dimensional systems and only tested on the high-dimensional target.
\newline 

While conducting these experiments, we noticed how training with multiple unrolled setups is non-trivial and sensitive to hyperparameters. Special care is thus required to successfully train with long unrolling. Reliable training setups utilize a curriculum where the unrolled number of steps slowly increases. 
The learning rate needs to be adjusted to keep the amplitude of the gradient feedback at a stable level.

\section{Conclusion}
We have conducted in-depth empirical investigations of unrolling strategies for training neural \gls{pde} simulators. The inherent properties of unrolling were deduced from an extensive test suite spanning multiple physical systems, learning setups, network architectures, and network sizes. Our findings rendered four best practices for training autoregressive neural simulators via unrolling. Most importantly, beneficial effects of unrolling at training time were present regardless of the differentiability of the neural simulator. We analyzed the effective unrolling horizons of the differentiable \wig{} and non-differentiable \nog{} setups, and brought these into context with the characteristic timescale of the physical system and the number of simulation steps. At the same time, we have shown how the unrolling horizons in the \wig{} cannot be derived from the chaotic timescale of the target dynamics alone. The target dynamics are inadequate to measure gradient behavior in the early stages of the network training, which motivated the curriculum approach. This behavior becomes even more complicated in the \nog{} case, where the temporal discretization affects the gradient stability. Deriving a general rule to choose ideal unrolling horizons in the \wig{} and \nog{} setups could be an interesting avenue for future research.

Additionally, our test sets and differentiable solvers are meant to serve as a benchmark: The broad range of network sizes for popular architectures as well as the selection of common physical systems as test cases yield a flexible baseline for future experiments with correction and prediction tasks.
We have identified non-differentiable unrolling as an attractive alternative to simplistic one-step training and implementations of differentiable solvers. Our results lay the foundation for assessing the  gains that can be expected when deploying this method.
Lastly, we measured convergence rates for neural networks and saw how these architectures converge poorly compared to numerical solvers. We concluded that deploying neural networks may thus be beneficial when a particular problem can benefit from their inherent properties, while grid scaling is still more efficient with classical approaches. 

In the future, problem-tailored network architectures could potentially improve the scaling of neural networks. However, at the current point of research, it is advisable to combine neural and numerical approaches in correction setups to benefit from both. One of the applications that could benefit from this approach is training closure models for filtered or averaged \gls{pde}s, although non-differentiable unrolling in this context is restricted to correction-style formulations.
At the same time, there are limitations to our scope. Our physical systems live in the domain of non-linear chaos and are primarily connected to fluid mechanics. 
Consequently, there is no inherent guarantee that our results will straightforwardly transfer to different domains.
Additionally, we mentioned the relevance of our results to the scientific computing community. Our recommendations can help prioritize implementation efforts when designing unrolled training setups. 
However, we have only studied a subset of the relevant hyperparameters. In the context of unrolling, variations such as network width and depth, advanced gradient-stopping techniques, and irregularly spaced curriculums could be impactful. 
These topics are promising avenues for future research to further our understanding of autoregressive neural networks for scientific applications.

\subsubsection*{Acknowledgements}
This work was supported by the ERC Consolidator Grant \textit{SpaTe} (CoG-2019-863850).
\subsubsection*{Code Availablility}
The code for the project will be made available at  \url{https://github.com/tum-pbs/unrolling}.

\newpage
\bibliographystyle{plainnat}
\bibliography{bibliography}

\clearpage 
\begin{centering} \Large APPENDIX \end{centering}
\appendix
\section{Unrolling on Attractors}\label{app:unrolling}
Here we provide a proof for equation \eqref{eq:unroll_attractor_fit} for a one-dimensional system. It is based on the definitions and remarks shown in Chapter 4 of \cite{broer2011dynamical}. The discrete equivalent of an $\omega$-limit set defined therein is 
\begin{equation}
    \omega(u) = \{y \in \mathbb{R}|\forall T \in \mathbb{N} \hspace{0.5em} 
    \exists t>T :  g^t(u)=y\}.
\end{equation}
The $\omega$-limit set of an initial $u$ consists of the accumulation states that the trajectory converges to for large time horizons. In the following, we will use the $\omega$-limit set to show how unrolling converges to the learned dynamics attractor. Furthermore, it is shown that $w(u)$ is an attractor of the dynamics $A_g$ if there is an arbitrarily small neighborhood $\mathcal{N}_u$ such that
\begin{equation}\label{eq:intersection_neighborhood}
    g^t(\mathcal{N}_u)\subset \omega(u) \hspace{0.5em} \forall t \hspace{0.5em} \text{and} \hspace{0.5em} \bigcap_{t\geq0} g^t(\mathcal{N}_u)=\omega(u)=A_g.
\end{equation}

\paragraph{Lemma}
Consider a learned simulator $g:\mathbb{R}\xrightarrow{}\mathbb{R}$ being autoregressively unrolled at training for $m$ steps. Suppose that an attractor $A_g$ exists for the dynamics of $g$, and that the system is unrolled from states $u$ in a dataset $U$. The unrolling then converges to this attractor for large $m$ and sufficiently many initial states in the basin $u\in \mathcal{B}(A_g)$.

\noindent\textit{Proof}: 
An unrolled trajectory consists of $\mathrm{traj}(u)=\{g^s(u) | s\in[0,m]\}$. As $m\xrightarrow{}\infty$, the trajectory includes the $\omega$-limit set such that $\omega(u)\subset\mathrm{traj}(u)$. Then, with sufficiently many initial states in the basin $u\in \mathcal{B}(A_g)$ and using \eqref{eq:intersection_neighborhood}
\begin{equation}
    \bigcap_{u}\mathrm{traj}(u)\supset\omega(u)=A_g
\end{equation}
\qed
\section{Gradient Calculation}
\label{app:gradients}
\subsection{Correction}
The network parameter optimization was defined in Equation \ref{eq:unrolling_loss}. For one learning iteration, a loss was accumulated over an unrolled trajectory. We can write this total loss over the unrolled trajectory as a function of $g(\mathbf{u}^t)=f_\theta(\mathcal{S}(\mathbf{u}^t))$ such that
\begin{equation}
    \mathcal{L} = \sum_{t=1}^s  \mathcal{L}_2(\tilde{\mathbf{u}}^{t+\tau s}, g^s(\mathbf{u}^t)) = \sum_{t=1}^s  \mathcal{L}^s,
\end{equation}
where $g^s$ represents the recurrent application of multiple machine learning augmented simulation steps, and $\mathcal{L}^s$ represents the loss evaluated after that step. 
Note that the full backpropagation through this unrolled chain requires a differentiable solver for the correction setup. To test the effect of using a differentiable solver, we introduce two different strategies for propagating the gradients $\frac{\partial \mathcal{L}_u}{\partial \theta}$. The differentiable setup can calculate the full optimization gradients by propagating gradients through the solver. The gradients are thus evaluated as 
\begin{equation}
    \frac{\partial \mathcal{L}^s}{\partial \theta} = \sum_{B=1}^s \bigg[
    \frac{\partial \mathcal{L}^s}{\partial g^s} 
    \bigg ( \prod_{b=1}^{s-B}
    \frac{\partial g^{s-b}}{\partial f_\theta^{s-b}} 
    \frac{\partial f_\theta^{s-b}}{\partial \mathcal{S}^{s-b}}
    \frac{\partial \mathcal{S}^{s-b}}{\partial g^{s-b-1}} 
    \bigg)
    \frac{\partial g^B}{\partial f_\theta^B}
    \frac{\partial f_\theta^B}{\partial \theta}\bigg].
\end{equation}
We refer to this fully differentiable setup as \wig{}.
In contrast, if no differentiable solver is available, optimization gradients can only propagate to the network application, not through the solver. The gradients are thus evaluated as 
\begin{equation}
    \frac{\partial \mathcal{L}_s}{\partial \theta} = 
    \frac{\partial \mathcal{L}_s}{\partial g^s} 
    \frac{\partial g^s}{\partial f_\theta^s} 
    \frac{\partial f_\theta^s}{\partial \theta}.
\end{equation}
This setup is referred to as \nog{}. Most existing code bases in engineering and science are not fully differentiable. Consequentially, this \nog{} setup is particularly relevance, as it could be implemented using existing traditional numerical solvers. 

\subsection{Prediction}
Prediction operates on the same network parameter optimization from Equation \ref{eq:unrolling_loss}. The total loss over the unrolled trajectory is 
\begin{equation}
    \mathcal{L} = \sum_{i=t}^s  \mathcal{L}_2(\tilde{\mathbf{u}}^{t+\tau s}, f_\theta^s(\mathbf{u}^t)) = \sum_{t=1}^s  \mathcal{L}^s,
\end{equation}
where $f_\theta^s$ represents the recurrent application of the network and $\mathcal{L}^s$ represents the loss evaluated after that step. 
To test the effect of using long-term gradients, we introduce two different strategies for propagating the gradients $\frac{\partial \mathcal{L}_u}{\partial \theta}$. The differentiable setup can calculate the full optimization gradients by propagating gradients through the solver. The gradients are thus evaluated as 
\begin{equation}
    \frac{\partial \mathcal{L}^s}{\partial \theta} = \sum_{B=1}^s \bigg[
    \frac{\partial \mathcal{L}^s}{\partial f_\theta^s} 
    \bigg ( \prod_{b=1}^{s-B}
    \frac{\partial f_\theta^{s-b}}{\partial f_\theta^{s-b-1}} 
    \bigg)
    \frac{\partial f_\theta^B}{\partial \theta}\bigg].
\end{equation}
We refer to this fully differentiable setup as \wig{}.
In contrast, if no differentiable solver is available, optimization gradients can only propagate to the network application, not through the solver. The gradients are thus evaluated as 
\begin{equation}
    \frac{\partial \mathcal{L}_s}{\partial \theta} = 
    \frac{\partial \mathcal{L}_s}{\partial f_\theta^s} 
    \frac{\partial f_\theta^s}{\partial \theta}.
\end{equation}
This setup is referred to as \nog{}. Most existing code bases in engineering and science are not fully differentiable. Consequentially, this \nog{} setup is particularly relevance, as it could be implemented using existing traditional numerical solvers. 

\section{AERO case}
\label{app:aero_case}
\paragraph{Numerical data:}
The governing equations for the transonic flow over a NACA0012 airfoil are non-dimensionalized with the freestream variables (i.e., the density $\rho_{\infty}$, speed of sound $a_{\infty}$, and the chord length of the airfoil $c$), and can be expressed in tensor notation as
\begin{align}
    \begin{split}
\frac{\partial \rho u_\mathrm{i}}{\partial x_\mathrm{i}}=0 \\
\frac{\partial \rho u_\mathrm{i} u_\mathrm{j}}{\partial x_\mathrm{j}}=-\frac{\partial p}{\partial x_\mathrm{i}} + \frac{\partial \tau_{x_{\mathrm{i}} x_{\mathrm{j}}}}{\partial x_\mathrm{j}} \\
\frac{\partial (\rho E + p)u_\mathrm{i}}{\partial x_\mathrm{i}} = \frac{\partial (-q_\mathrm{i} +u_\mathrm{j} \tau_{x_{\mathrm{i}} x_{\mathrm{j}}} )}{\partial x_\mathrm{i}}
    \end{split}
\end{align}%
where the shear stress (with Stokes’ hypothesis) and heat flux terms are defined as
\begin{equation*}
\tau_{x_{\mathrm{i}} x_{\mathrm{j}}}=\mu\frac{M_{\infty}}{Re_{\infty}}[(\frac{\partial u_\mathrm{i}}{\partial x_\mathrm{j}}+\frac{\partial u_\mathrm{j}}{\partial x_\mathrm{i}})-\frac{2}{3}\frac{\partial u_{\mathrm{k}}}{\partial x_{\mathrm{k}}}\delta_{\mathrm{i}\mathrm{j}}]
\end{equation*} 
and 
\begin{equation*}
q_{x_{\mathrm{i}}}=-\frac{\mu}{Pr}\frac{M_{\infty}}{Re_{\infty}(\gamma-1)}\frac{\partial \Theta}{\partial x_{\mathrm{i}}}.
\end{equation*}
Here, Reynolds number is defined as $Re_{\infty}=\rho_{\infty}\sqrt{u_{\infty}^2+v_{\infty}^2}c/\mu_{\infty}$; $\gamma$ is the ratio of specific heats, 1.4 for air; the laminar viscosity $\mu$ is obtained by Sutherland's law (the function of temperature), and the turbulent viscosity $\mu_T$ is determined by turbulence models; laminar Prandtl number is constant, i.e.
$Pr=0.72$.
The relation between pressure $p$ and total energy $E$ is given by
\begin{equation*}
p=(\gamma-1)\bigl[\rho E-\frac{1}{2}\rho u_{\mathrm{i}}u_{\mathrm{i}}\bigr].
\end{equation*}
Note also that from the equation of state for a perfect gas, we have $p=\rho a^2/\gamma$ and temperature $\Theta=a^2$.
As we perform 2D high-resolution quasi-direct numerical simulations, no turbulence model is employed. 

The finite-volume method numerically solves the equations using the open-source code CFL3D. The mesh resolution $1024\times256$ is kept the same for cases, i.e. 256 grid cells in the wall-normal direction, 320 grid cells in the wake, and 384 grid cells around the airfoil surface. 
The convective terms are discretized with third-order upwind scheme, and viscous terms with a second-order central difference.  

An inflow/outflow boundary based on one-dimensional Riemann invariant is imposed at about 50\emph{c} away from the airfoil in the (\emph{x}, \emph{y}) plane. The grid stretching is employed to provide higher resolution near the surface and in the wake region, and the minimal wall-normal grid spacing is $6\times10^{-4}$ to ensure $y^+_n<1.0$. A no-slip adiabatic wall boundary condition is applied on the airfoil surface. 
The non-dimensional time step is $0.008c/{U_{\infty}}$. 

In the transonic regime, airflow behavior becomes more complex due to the formation of shock waves and supersonic and subsonic flow areas on the airfoil surfaces. When the Mach number is below $M=0.77$, the unsteadiness in the flowfield is mainly caused by the high-frequency vortex shedding. At around $M=0.8$, a series of compression waves coalesce to form a strong shock wave, and the flow structures are dominated by alternately moving shock waves along the upper and lower sides of the airfoil. For $M>0.88$, the strong shock waves become stationary on both surfaces.
To cover all possible flow regimes, the samples in the training dataset are generated at $M=[0.75, 0.8, 0.825, 0.88, 0.9]$, and the test samples are performed at $M=0.725$ and $M=0.775$.

We evaluate the vortex shedding frequency to identify the characteristic timescale of the \gls{aero} simulations. The frequency $f_sc/U_\infty$ was individually evaluated for every hyper-parameterization in the training dataset and averaged to obtain a characteristic timescale $t_{c,\text{AERO}}=1/{f_s} = 0.455$. This timescale is covered by roughly 15 prediction steps of the neural simulator.

The snapshots are saved at every four simulation steps (i.e., $dt_{sampling}=0.032c/U_{\infty}$) and spatially downsampled by 4x and 2x in the circumferential and wall-normal directions, respectively. In the case of $M=0.85$, there are 1000 snapshots; for other cases, there are 500 snapshots.

\paragraph{Neural Network Architectures:}
We implemented Attention U-Net \citep{oktay2018attention}. It consists of three encoder blocks, each progressively capturing features from the input image through convolution and downsampling. Following the encoder blocks, there's a bottleneck block for information compression with a higher number of channels. Subsequently, the architecture includes three decoder blocks, which use skip connections to integrate features from both the bottleneck and corresponding encoder blocks during the upscaling process. These decoder blocks gradually reduce the number of channels, culminating in a 1x1 convolutional output layer that generates pixel-wise predictions. We train networks of varying sizes by adjusting the number of features, as indicated in Table \ref{tab-app:aero-arch}. The training was performed in PyTorch \citep{NEURIPS2019_9015}.
\begin{table}[tbh]
    \caption{AERO Network Architectures and Parameter Counts}
    \label{tab-app:aero-arch}
    \centering
    \begin{tabular}{l| r c l }
        \toprule
        \textit{Architecture} &   \textit{\# Parameters}  & \textit{Features in encoder  \& bottleneck blocks} \\
        \midrule
        UNet, 0.5m &  511332  & [16, 32, 64] [128] \\ 
        UNet, 2.0m & 2037956  & [32, 64, 128] [256] \\
        UNet, 8.1m & 8137092  & [64, 128, 256] [512] \\
        \bottomrule
    \end{tabular}
\end{table}
\paragraph{Network Training and Evaluation} The model is trained with Adam and a mini-batch size of 5, with training noise, for up to 500k iterations. A learning rate of $6\times10^{-4}$ is used for the first 250k iterations and then decays exponentially to $6\times10^{-5}$.
The training curriculum increases the number of unrolled steps from $m=1$ to $m=4$ and then $m=9$.

$\mathcal{L}_\mathrm{rel}$ errors are computed and accumulated over 13 prediction steps (equivalent to 52 simulator steps) of simulation for two test cases with previously unseen Mach numbers: $M=0.725$ and $M=0.775$, corresponding to shock-free case and near-critical condition case.

\section{Additional Results}
\label{sec:appendix_additional_results}
\subsection{Metric Comparison}
\begin{figure}
\centering
    \begin{tabular}{cc}
          \includegraphics[height=4cm]{ks_corr_gcn_model_size_h20.pdf} &   
          \includegraphics[height=4cm]{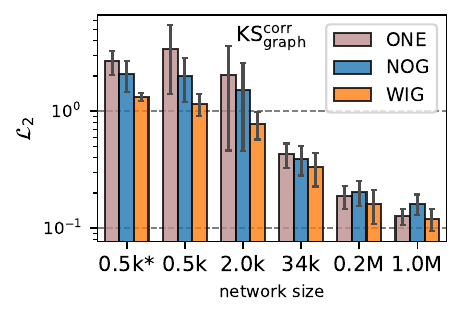} \\
        
        \includegraphics[height=4cm]{ks_pred_cnn_model_size_h20.pdf} &
        \includegraphics[height=4cm]{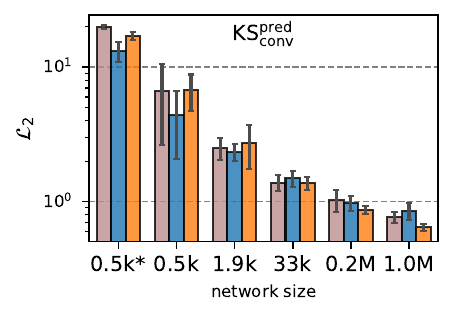}\\
    \end{tabular}
\caption{Metric comparison between relative and absolute error metrics}
\label{fig:metric_comparison}
\end{figure}

The choice of metric can affect the interpretation of results. In the main evaluations of the paper, we showcase the results in terms of a relative error metric, which is normalized by ground truth data. Relative error values approaching 1 indicate that the compared trajectory de-correlates from the ground truth. This effect is especially pronounced in chaotic dynamics and motivates shorter evaluation horizons (e.g., 12 steps for \gls{ks}). As a long-term stability measure, we used the divergence-time in figure \ref{fig:ks_corr_cnn_divergence}. By putting a threshold on the $\mathcal{L}_2$ error, this metric evaluates the number of steps for which the solution remains stable. However, it does not contain any information about the accuracy of a solution with respect to the ground truth trajectory. The absolute $\mathcal{L}_2$ metric in figure \ref{fig:metric_comparison} can be seen as an intermediate between relative and divergence-time metrics. It depicts both the accuracy of a solution compared to the ground truth and the growing energies in divergent trajectories. In the present evaluation of the \gls{ks} system, we have thus used intermediate-length horizons (40 steps).
\subsection{Prediction on the KOLM system}
\begin{figure}[h]
\centering
\includegraphics[width=0.3\textwidth]{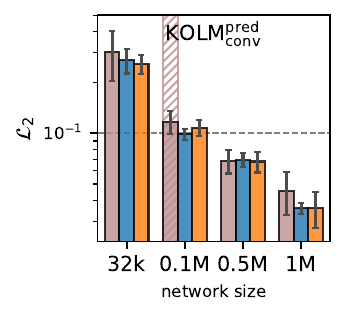}
\caption{Comparison of $\mathcal{L}_2$ errors for CNN predictions on the \gls{kolm} flow}
\label{fig:kolm_cnn_pred}
\end{figure}

The networks used for the correction of the \gls{kolm} system were also trained for prediction. The results are visualized in Figure \ref{fig:kolm_cnn_pred}. Unsurprisingly, the overall error values are larger than for the respective correction setups, where an additional numerical prior supports the network. Like the other prediction and correction setups, the errors decrease with increasing network size. Unrolled training modalities perform best at all network sizes. The quantitative values of the loss metric are also tabulated in \ref{tab:kolm_cnn_pred_size}.

\subsection{Interpolation and Extrapolation Tests}

\begin{figure}[h]
\centering
\includegraphics[width=0.7\textwidth]{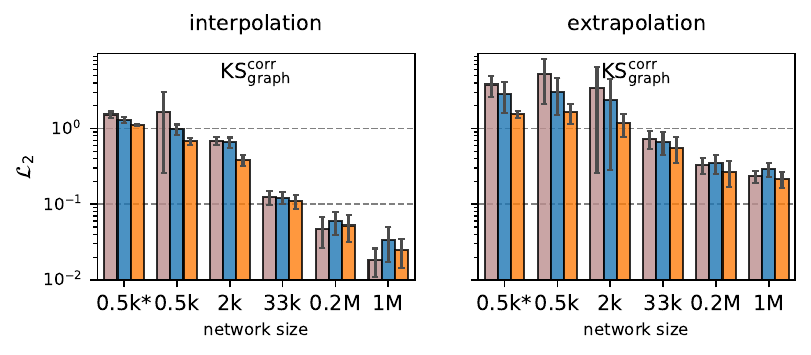}
\includegraphics[width=0.7\textwidth]{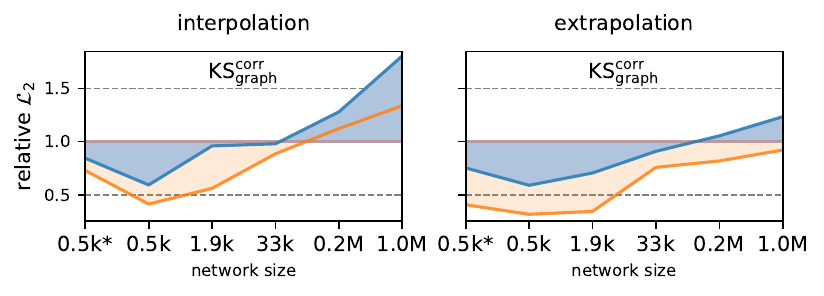}
\caption{Comparison of $\mathcal{L}_2$ errors on interpolative and extrapolation test sets for GCNs on \gls{ks}}
\label{fig:ks_gcn_interp_extrap}
\end{figure}

\begin{figure}[h]
\centering
\includegraphics[width=0.8\textwidth]{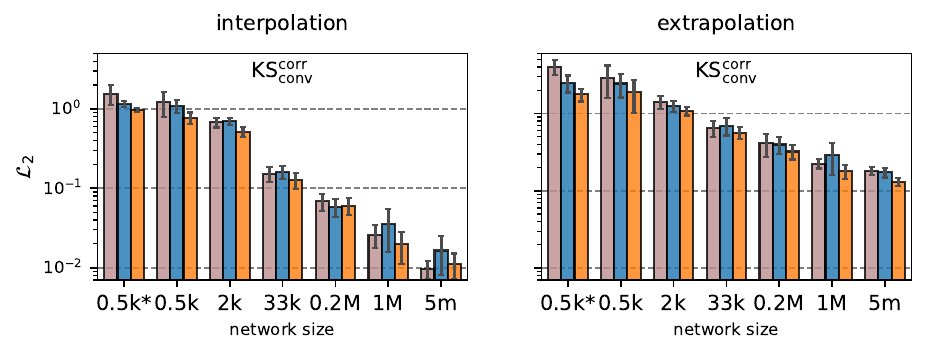}
\includegraphics[width=0.8\textwidth]{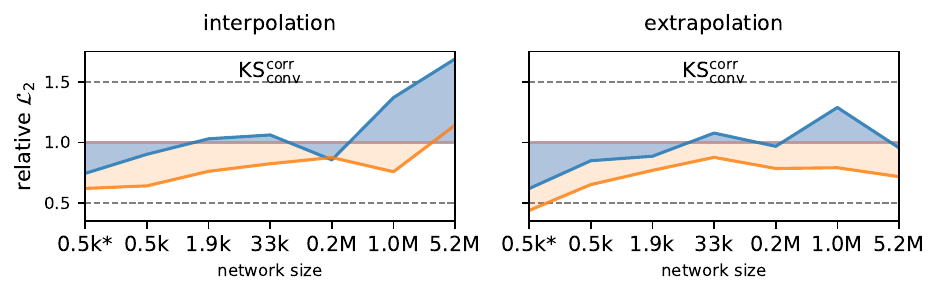}
\caption{Comparison of $\mathcal{L}_2$ errors on interpolative and extrapolation test sets for CNNs on \gls{ks}}
\label{fig:ks_cnn_interp_extrap}
\end{figure}

\begin{figure}[h]
\centering
\includegraphics[width=0.49\textwidth]{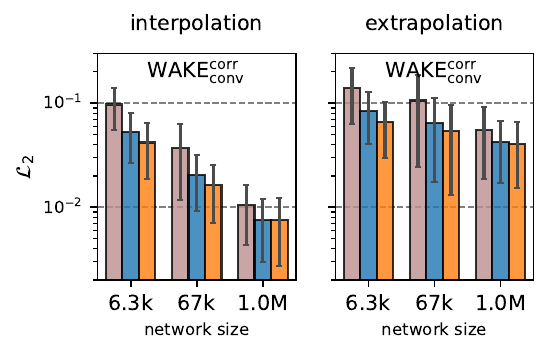}
\includegraphics[width=0.49\textwidth]{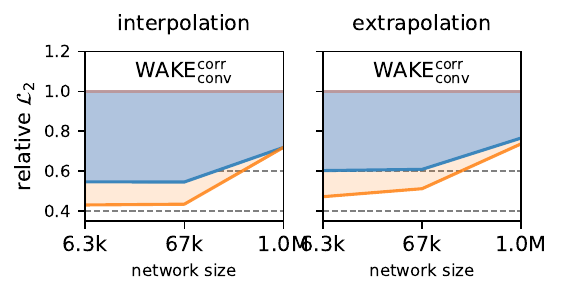}
\caption{Comparison of $\mathcal{L}_2$ errors on interpolative and extrapolation test sets for \gls{wake}}
\label{fig:wake_interp_extrap}
\end{figure}

\begin{figure}[h]
\centering
\includegraphics[width=0.6\textwidth]{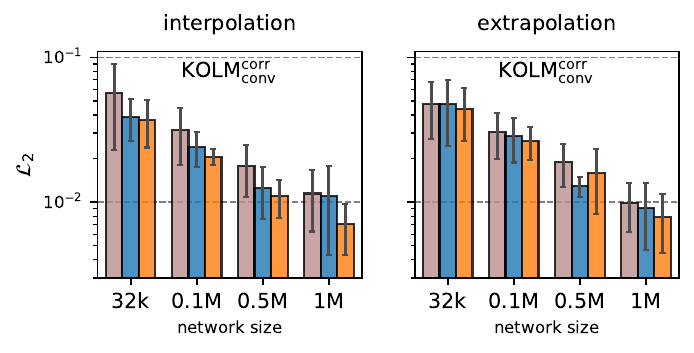}
\includegraphics[width=0.6\textwidth]{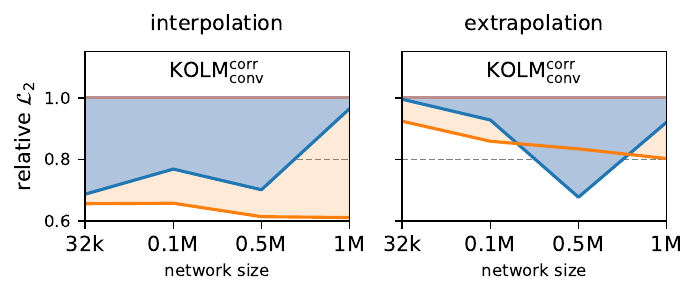}
\caption{Comparison of $\mathcal{L}_2$ errors on interpolative and extrapolation test sets for \gls{kolm}}
\label{fig:kolm_interp_extrap}
\end{figure}

Interpreting physical hyperparameters is necessary for models to generalize to extrapolative test cases. We want to test whether extrapolation to new physical hyperparameters benefits from unrolling or long-term gradients. In the main sections, all evaluations accumulated their results from interpolative and extrapolative test cases. In this section, we explicitly differentiate between interpolation and extrapolation tests. Figures \ref{fig:ks_gcn_interp_extrap}, \ref{fig:ks_cnn_interp_extrap}, \ref{fig:wake_interp_extrap}, and \ref{fig:kolm_interp_extrap} compare the model performance on interpolative and extrapolative test cases with respect to the physical parameters and depicts performance relative to the ONE baseline for various model sizes. Overall accuracy is worse on the harder extrapolative test cases. 
Similarly to the combined tests from the main section \wig{} performs best on average, both on interpolative or extrapolative data. However, the long-term gradients introduced by this method do not seem to explicitly favor inter- or extrapolation.
Nonetheless, the positive aspects of \wig{} training are not constrained to interpolation, but successfully carry over to extrapolation cases.

\subsection{Dataset Sizes}
\begin{figure}[h]
\centering
\includegraphics[width=0.5\textwidth]{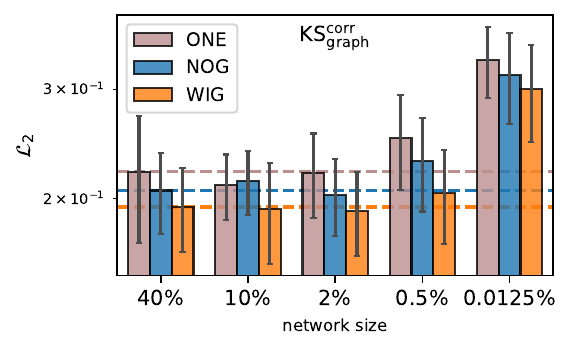}
\caption{Dataset size variations in percent of the full training set}
\label{fig:dataset_size}
\end{figure}

Unrolling connects multiple dataset frames in one trajectory. This connectedness encodes the physical relation between subsequent frames that ultimately has to be learned by the neural network. Given that unrolled setups observe more of these physical connections, one might expect that unrolling decreases the necessary dataset size. To test this hypothesis, we incrementally decreased the amount of training data. The number of training iterations was kept constant throughout the process. The models are then evaluated on our full test sets. Figure \ref{fig:dataset_size} compares the $\mathcal{L}_2$ for variations in the training dataset size. A measurable difference in inference accuracy only appears for dataset sizes smaller than $X\%$ of the original dataset. While the accuracy of \wig{} does indeed deteriorate slightly later than \nog{} and especially \one{}, this transition is confined to a small section of $Y\%$ dataset size. In practice, training a setup in this narrow dataset margin is unlikely and thus not mentioned as a strong benefit of unrolling.  

\subsection{Best performing models}
\begin{figure}[h]
\centering
\includegraphics[width=0.715\textwidth]{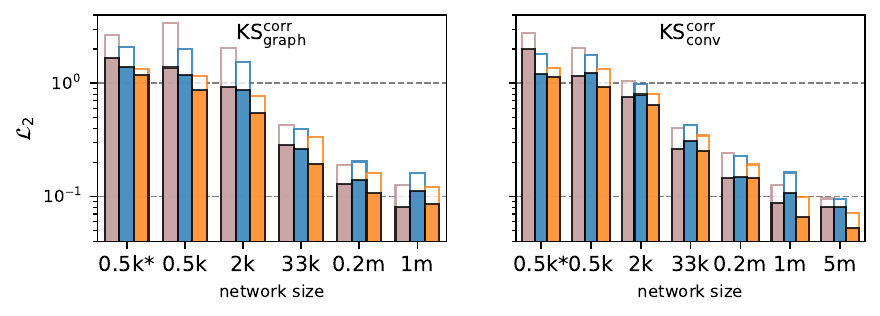}
\includegraphics[width=0.275\textwidth]{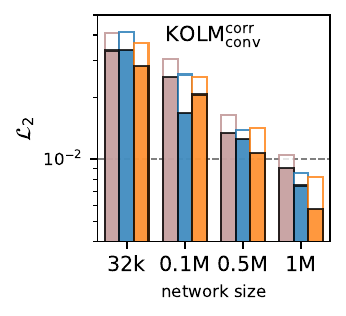}
\caption{$\mathcal{L}_2$ errors of the best performing models; the outline in the background represent the average accuracy}
\label{fig:bestmodels}
\end{figure}

Our error measurements relied on statistical evaluations, which in turn were based on multiple randomized training runs. \nog{} and especially \wig{} showed clear benefits in these statistical evaluations.
However, we must also consider the computational cost of training (differentiable) unrolled setups. Unrolling $m$ steps with the \nog{} setup increases the computational cost of one training iteration $m$-fold. Differentiable training in the \wig{} setup adds even more computations due to the backpropagation through the solver. 
Consequently, training multiple models and selecting the best might be a viable approach for \one{} models, where training costs are lower. Thus, we evaluate the best-performing models in a separate analysis. Figure \ref{fig:bestmodels} depicts the best inference $\mathcal{L}_2$ achieved by a given model size, architecture, and learning setup. The best \one{} models are on par with \nog{} or \wig{} for some network sizes and architectures. However, in many other cases, the average unrolled \wig{} setup still performs better or similar to the best \one{} model. In light of the fact that these best models were selected out of 8 (\gls{kolm}) or 20 (\gls{ks}) training runs, training with unrolling is ultimately more resource-efficient if best performance is sought. Thus, our recommendation of training with \nog{} or \wig{} approaches persists.

\subsection{Additional Architecture}
\label{app:additional_architecture}
\begin{figure}
\vspace{-0.5cm}
\centering
\includegraphics[height=4cm]{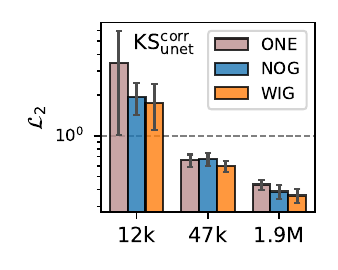} 
\caption{Comparison of $\mathcal{L}_2$ errors for the U-Net architecture on the KS system}
\label{fig:ks_corr_unet}
\end{figure}
We tested the generalization of our findings toward attention-gated U-nets \citep{oktay2018attention}. The implementation is a 1D version of the network used in the \gls{aero} cases introduced in Appendix \ref{app:aero_case}. Crucially, it features attention gates in its skip connections allowing the network to efficiently mix global and local structures in the output. The $\mathcal{L}_2$ errors of this architecture on the \gls{ks} system are visualized in Figure \ref{fig:ks_corr_unet}. The behavior of \one{}, \nog{}, and \wig{} matches well with other architectures (i.e. conv and graph nets) studied in the main section. We still observe the best performance with fully differentiable \wig{} unrolling, while \nog{} offers slimmer benefits over \one{}. We can conclude that 
our observations regarding the positive effects of unrolling with and without gradients transfer to attention-based networks.

\FloatBarrier
\section{Inference Visualizations}
This section visualizes typical inference trajectories on which our evaluations were based. Due to the vast amount of trained models, only a very small subset of all models are visualized. The goal is to contextualize the results from the main paper and previous appendix evaluations. The following figures show various training modalities for a fixed initialization, which was randomly selected from our trained models.\newline

\begin{figure}[h]
    \centering
    \includegraphics[width=0.95\textwidth]{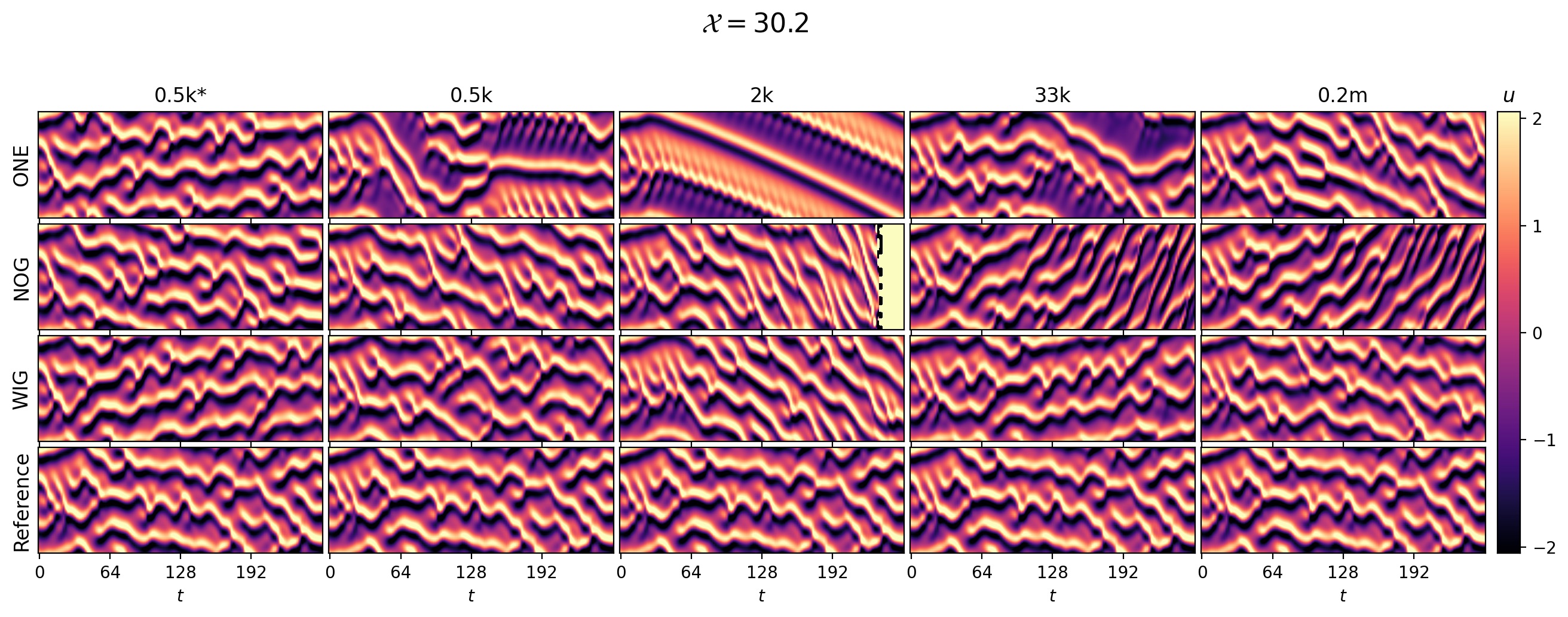}
    \caption{Inference error trajectory visualization of the CNN correction on the \gls{ks} system for extrapolation on low $\mathcal{X}$}
    \label{fig:vis_ks_cnn_evo_30_2}
\end{figure}
\begin{figure}[h]
    \centering
    \includegraphics[width=0.95\textwidth]{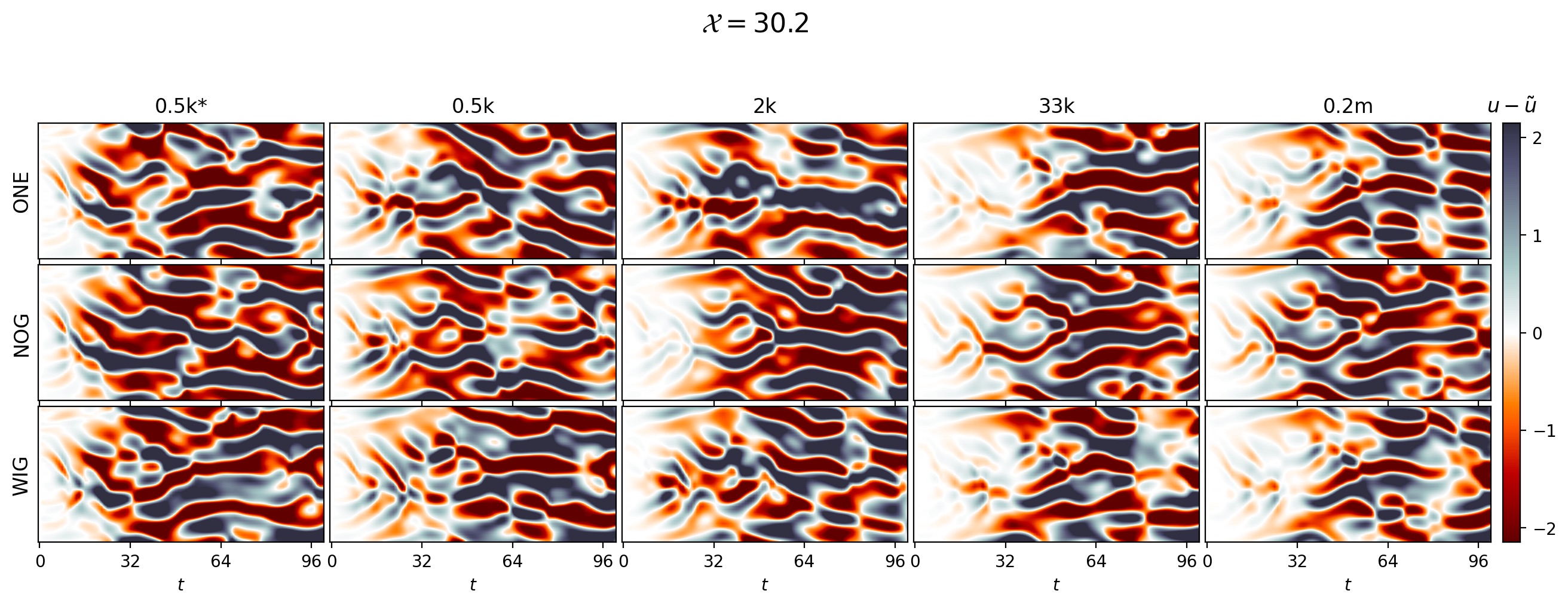}
    \caption{Inference error trajectory visualization of the CNN correction on the \gls{ks} system for extrapolation on high $\mathcal{X}$}
    \label{fig:vis_ks_cnn_err_30_2}
\end{figure}%

\begin{figure}[h]
    \centering
    \includegraphics[width=0.95\textwidth]{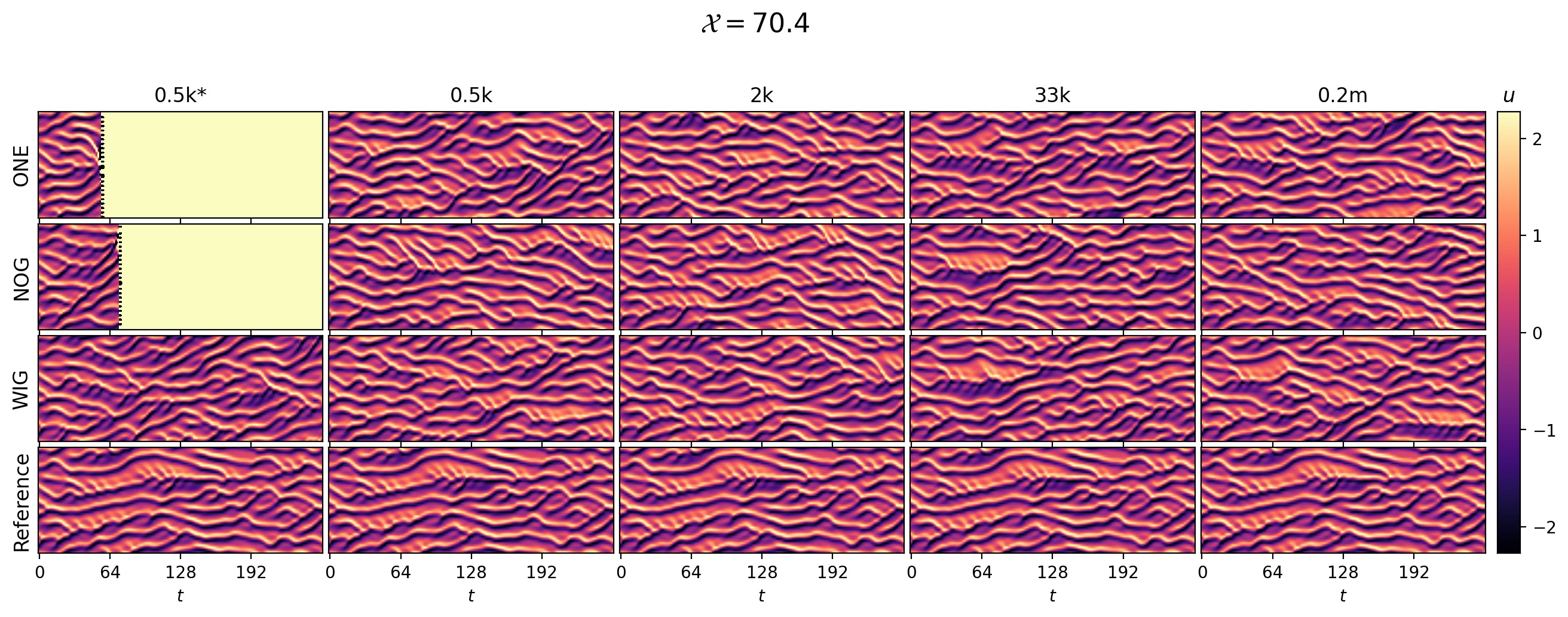}
    \caption{Inference error trajectory visualization of the GCN correction on the \gls{ks} system for extrapolation on low $\mathcal{X}$}
    \label{fig:vis_ks_cnn_evo_70_4}
\end{figure}
\begin{figure}[h]
    \centering
    \includegraphics[width=0.95\textwidth]{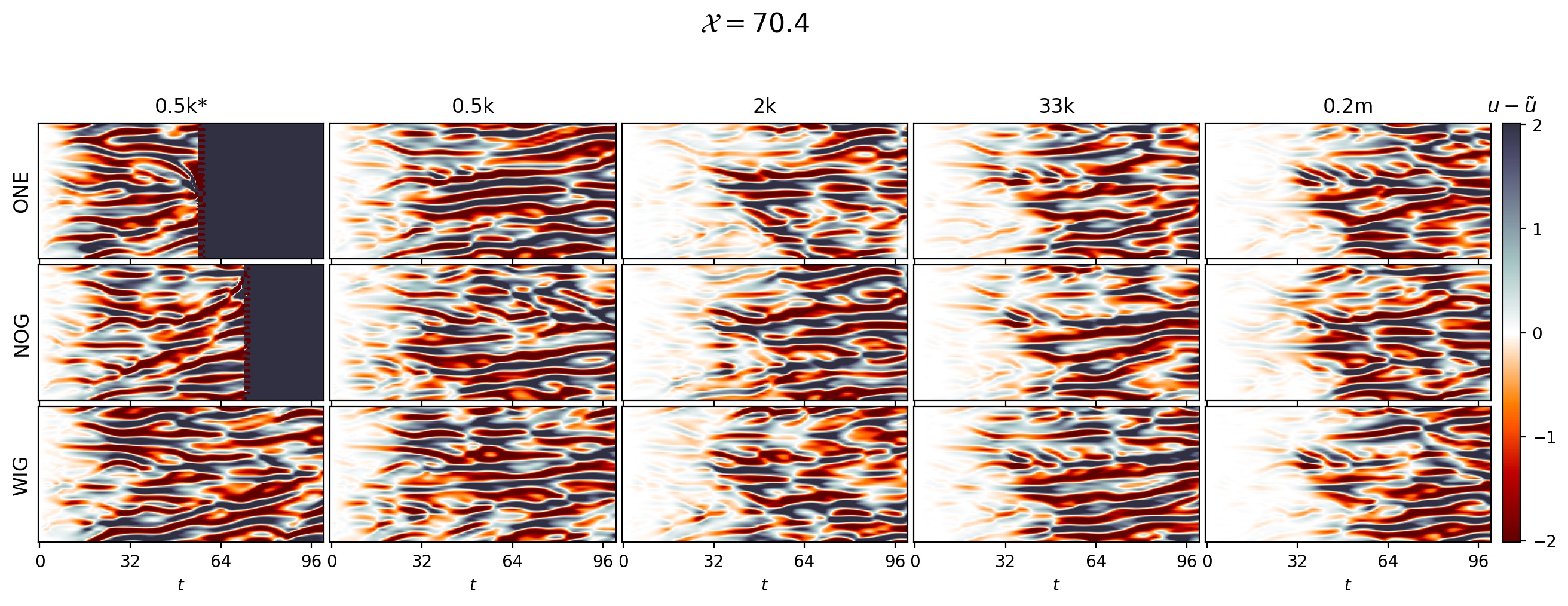}
    \caption{Inference error trajectory visualization of the GCN correction on the \gls{ks} system for extrapolation on high $\mathcal{X}$}
    \label{fig:vis_ks_cnn_err_70_4}
\end{figure}%

Figure \ref{fig:vis_ks_cnn_evo_30_2} displays a sample inference trajectory for convolutional correction models trained with the three unrolling variants. The inference case is selected from the extrapolation dataset, in this case for the small domain size of $\mathcal{X}=30.2$, which lies outside of the training range. The difference to the reference is further visualized in Figure \ref{fig:vis_ks_cnn_err_30_2}. Here, the distance to the target state as $u-\tilde u$ is used. This error was evaluated for the full range of model sizes as introduced in Section \ref{sec:systems_architectures} and listed in Table \ref{tab-app:ks-arch}. The same visualization procedure was also done for the other extrapolation test on large domains with $\mathcal{X}=70.4$. Figures \ref{fig:vis_ks_cnn_evo_70_4} and \ref{fig:vis_ks_cnn_err_70_4} similarly display the state trajectories and errors for a sample network. A clear trend of longer correlation horizons is visible in the error figures. As already expanded on in the main paper, larger architectures stay in proximity to the target evolution for longer inference rollouts. At the same time, unrolling reduces errors for a fixed size since NOG and WIG have smaller amplitudes. On the large domain sizes in Figures \ref{fig:vis_ks_cnn_evo_70_4} and \ref{fig:vis_ks_cnn_err_70_4}, the networks also show instabilities when trained with \one{} or \nog{}.

\begin{figure}[h]
    \centering
    \includegraphics[width=0.97\textwidth]{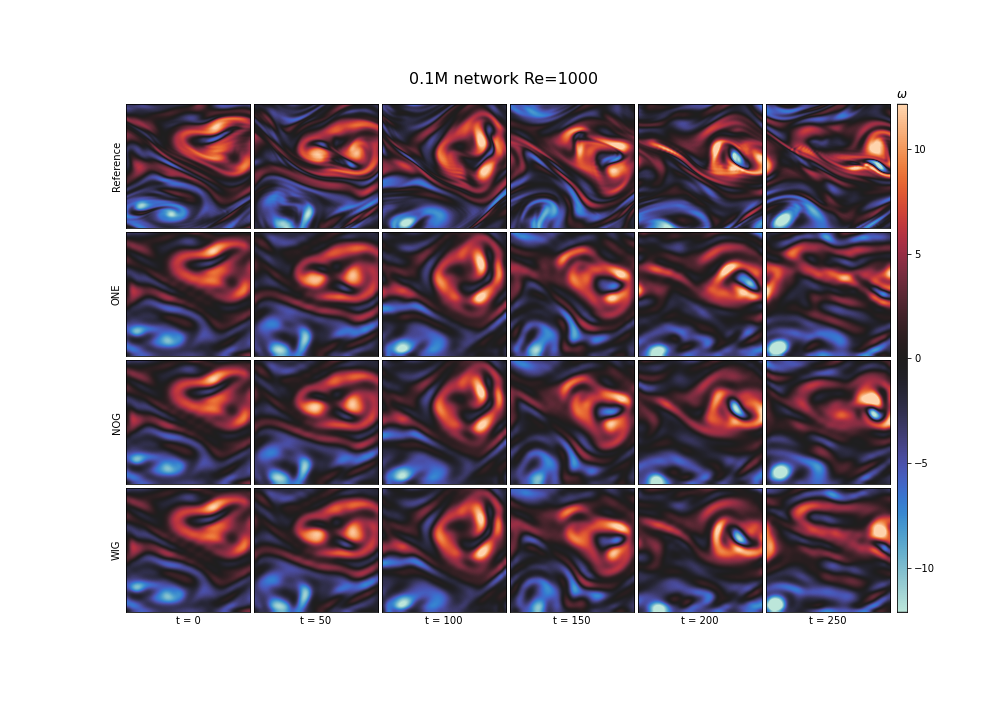}
     \caption{Inference vorticity after 250 steps of the CNN correction on the \gls{kolm} system for interpolation on intermediate Re=600, reference data is shown in the left column}
    \label{fig:vis_kolm_vort_cnn_evo_time_1000}
\end{figure}

\begin{figure}[h]
    \centering
    \includegraphics[width=0.60\textwidth]{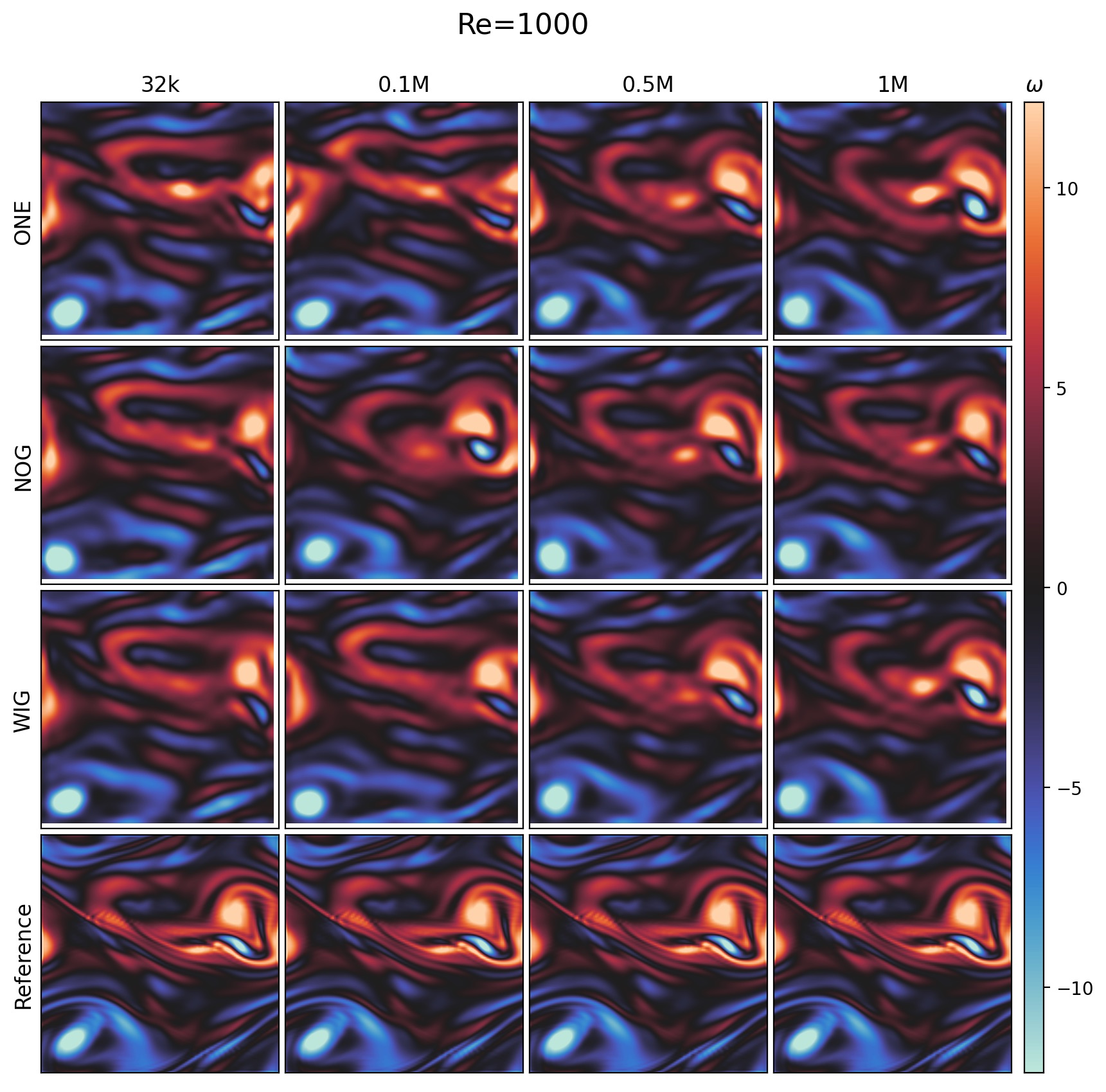}
     \caption{Inference vorticity after 250 steps of the CNN correction on the \gls{kolm} system for extrapolation on high Re=1000, reference data is shown in the left column}
    \label{fig:vis_kolm_vort_cnn_evo_1000}
\end{figure}

\begin{figure}[h]
    \centering
    \includegraphics[width=0.60\textwidth]{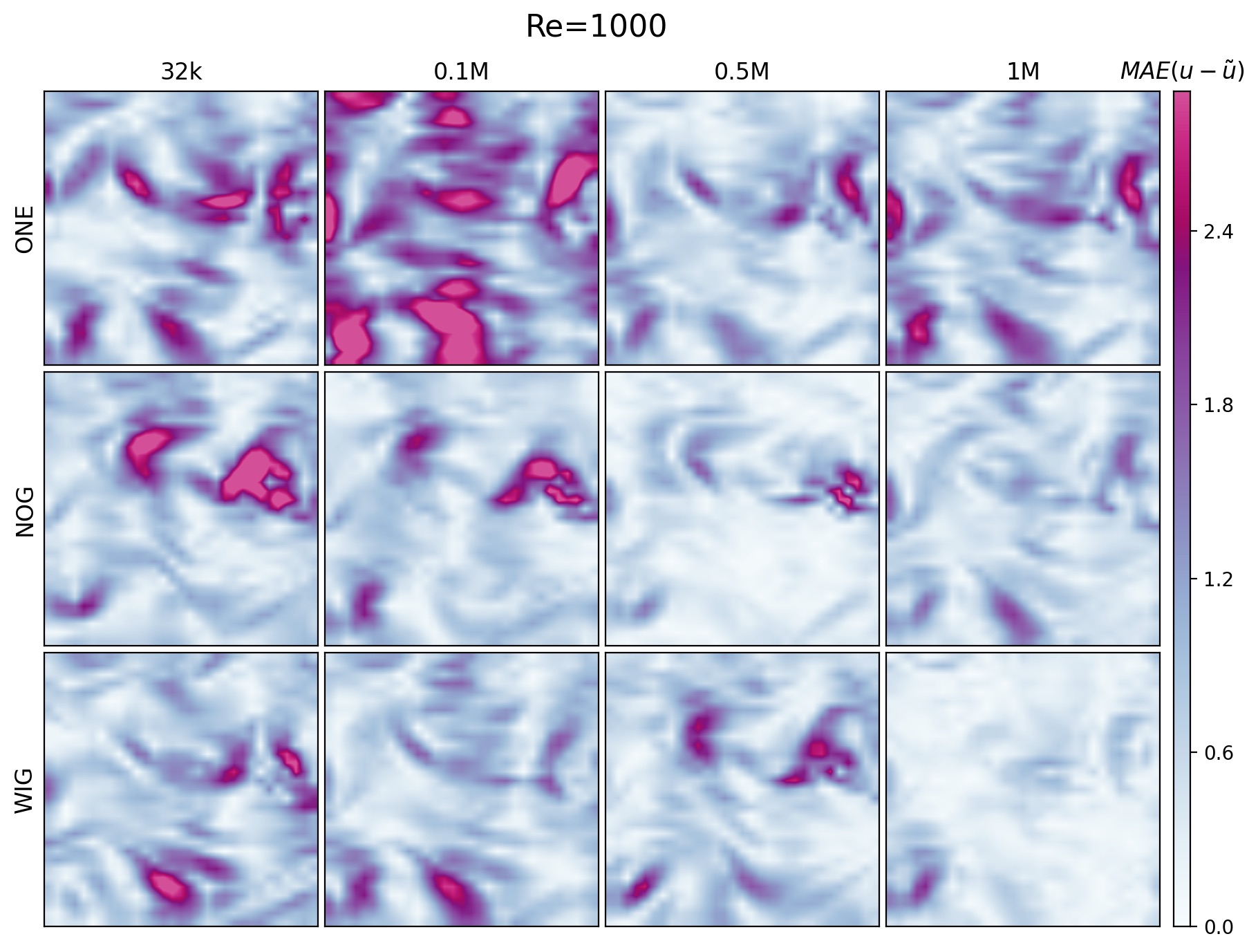}
     \caption{Inference vorticity after 250 steps of the CNN correction on the \gls{kolm} system for interpolation on intermediate Re=600, reference data is shown in the left column}
    \label{fig:vis_kolm_vort_cnn_err_1000}
\end{figure}

\FloatBarrier
For the \gls{kolm} system, we visualize sample rollout trajectories for the extrapolation case of Re=1000 in Figure \ref{fig:vis_kolm_vort_cnn_evo_time_1000}. The displayed horizon spans 250 timesteps. The final step of this inference rollout is further visualized in Figure \ref{fig:vis_kolm_vort_cnn_evo_1000} for the whole range of training setups, including unrolling variants and network sizes. Finally, a visualization of the solution error is provided in Figure \ref{fig:vis_kolm_vort_cnn_err_1000}. The visualizations show how those networks that were trained with \one{} and \nog{} unrolling tend to misinterpret points of high vorticity and generally produce larger errors. As pointed out in the main sections, network size is, along with training modality, a decisive contributor to inference accuracy. This observation can also be made in the error visualizations in \ref{fig:vis_kolm_vort_cnn_err_1000}. 


\FloatBarrier
\newpage
\section{Evaluation Data}
\label{app:evaluation_data}
Herein, we provide the statistical data used to generate the figures in the main paper. The error data shown is always gathered from evaluations on a combined test set from interpolative and extrapolative cases, as described in Appendix \ref{sec:systems_architectures}. For each initial frame in the test set, a trajectory is computed and errors with respect to the ground truth are accumulated, yielding one scalar error value per trajectory. Means and standard deviations are now calculated across these trajectories on a per-model-size and per-training-modality basis. Thus, the tables show these statistics for various model sizes, which are listed for the \gls{ks} system in Table \ref{tab-app:ks-arch}, for the wake system in Section \ref{sec:systems_wake}, for the \gls{kolm} case in Table \ref{tab-app:kolm-arch}, and for the \gls{aero} case in Table \ref{tab-app:aero-arch}. We denote the mean with an overline, e.g. as $\overline{\mathrm{ONE}}$, and the standard deviation as $\sigma$. Additionally, we perform a one-sided Welch's t-test to the statistical significance of the empirical tests and their resulting distributions. For distributions with different means and standard deviations, \cite{welch1947generalization} defines a t-test as 
\begin{equation}
    t = \frac{|\overline{X}_1-\overline{X}_2|}{\sqrt{\dfrac{\sigma_1^2}{M_1}+\dfrac{\sigma_2^2}{M_2}}}.
\end{equation}
This holds for two distributions with mean $\overline{X}_i$, standard deviation $\sigma_i$, and sample sizes $M_i$. The p-value for a one-sided significance test is then computed as 
\begin{equation}
    p = Pr(T\leq t|H_0),
\end{equation}
giving us the probability that the null-hypotheses (i.e. both distributions are identical) is true.

The differences in the studied training modalities are statistically significant for smaller network setups. As model sizes increase, the distribution of trained models becomes more similar. Our heavily overparameterized networks (e.g. 1.0M parameters for 48 degree of freedom KS system) all show highly accurate and stable predictions (see divergence time metric). Based on the theoretical considerations above, this means that the attractor of the learned dynamics is closely aligned with the ground truth system. These setups thus display less data shift and the benefits of unrolling are reduced. When these conditions in the overparameterized regime are met, NOG models are at a disadvantage due to their mismatch between gradients and loss landscape. However, these heavily oversized architectures are not practically relevant for scientific computing due to their weak scaling compared to numerical approaches. This evaluation confirms that for relevant, small to medium-sized networks our results are statistically significant and hence conclusions can be drawn.

The tables relate to the figures as follows:
\begin{itemize}
    \item Figure \ref{fig:corr_agnostic} visualizes Tables \ref{tab:ks_gcn_corr_size}, \ref{tab:wake_cnn_corr_size}, \ref{tab:kolm_corr_size}
    \item Figure \ref{fig:unrolling_horizon} visualizes Tables \ref{tab:ks_cnn_corr_unrol}, \ref{tab:kolm_corr_cnn_curriculum}, \ref{tab:kolm_corr_cnn_schedule}
    \item Figure \ref{fig:gradient_stop} visualizes Table \ref{tab:ks_cnn_corr_gradstop}
    \item Figure \ref{fig:prediction_size} visualizes Tables \ref{tab:ks_gcn_pred_size}, \ref{tab:ks_cnn_pred_size}, \ref{tab:wake_cnn_pred_size}
    \item Figure \ref{fig:corr_pred_transition} visualizes Tables \ref{tab:ks_gcn_transition}, \ref{tab:ks_cnn_transition}constant
    \item Figure \ref{fig:aero_results} visualizes Table \ref{tab:aero_cnn_pred_size}    
\end{itemize}
\begin{table}[h]
    \centering
    \caption{Correction GCN $\mathcal{L}_\mathrm{rel}$ errors on the \gls{ks} system}
    \begin{tabular}{lrrrrrrrr}
\toprule
{} &  $\overline{\mathrm{ONE}}$ &  $\overline{\mathrm{NOG}}$ &  $\overline{\mathrm{WIG}}$ &  $\sigma({\mathrm{ONE}})$ &  $\sigma({\mathrm{NOG}})$ &  $\sigma({\mathrm{WIG}})$ &  $p^\mathrm{NOG}_\mathrm{ONE}$ &  $p^\mathrm{WIG}_\mathrm{ONE}$ \\
\midrule
0.5k* &                    0.13866 &                    0.12531 &                    0.07415 &                   0.03573 &                   0.04180 &                   0.00884 &                        0.14228 &                        0.00000 \\
0.5k  &                    0.12246 &                    0.08551 &                    0.04125 &                   0.04741 &                   0.02651 &                   0.00716 &                        0.00212 &                        0.00000 \\
2k    &                    0.25568 &                    0.06132 &                    0.02138 &                   0.80377 &                   0.08010 &                   0.00746 &                        0.14434 &                        0.10011 \\
33k   &                    0.00913 &                    0.00856 &                    0.00563 &                   0.00292 &                   0.00418 &                   0.00309 &                        0.31063 &                        0.00036 \\
0.2M  &                    0.00232 &                    0.00270 &                    0.00173 &                   0.00065 &                   0.00073 &                   0.00082 &                        0.04417 &                        0.00788 \\
1M    &                    0.00130 &                    0.00196 &                    0.00132 &                   0.00039 &                   0.00048 &                   0.00030 &                        0.00001 &                        0.40292 \\
\bottomrule
\end{tabular}

    \label{tab:ks_gcn_corr_size}
\end{table}%
\begin{table}[h]
    \centering
    \caption{Correction CNN $\mathcal{L}_\mathrm{rel}$ errors on the \gls{ks} system}
    \begin{tabular}{lrrrrrrrr}
\toprule
{} &  $\overline{\mathrm{ONE}}$ &  $\overline{\mathrm{NOG}}$ &  $\overline{\mathrm{WIG}}$ &  $\sigma({\mathrm{ONE}})$ &  $\sigma({\mathrm{NOG}})$ &  $\sigma({\mathrm{WIG}})$ &  $p^\mathrm{NOG}_\mathrm{ONE}$ &  $p^\mathrm{WIG}_\mathrm{ONE}$ \\
\midrule
0.5k* &                    0.11646 &                    0.09048 &                    0.05381 &                   0.02157 &                   0.01912 &                   0.00893 &                        0.00013 &                        0.00000 \\
0.5k  &                    0.07308 &                    0.06987 &                    0.04797 &                   0.01431 &                   0.01148 &                   0.01033 &                        0.21954 &                        0.00000 \\
2k    &                    0.04174 &                    0.03547 &                    0.02306 &                   0.01018 &                   0.00640 &                   0.00432 &                        0.01252 &                        0.00000 \\
33k   &                    0.00728 &                    0.00797 &                    0.00582 &                   0.00227 &                   0.00227 &                   0.00205 &                        0.17330 &                        0.01953 \\
0.2M  &                    0.00318 &                    0.00255 &                    0.00204 &                   0.00178 &                   0.00069 &                   0.00045 &                        0.07378 &                        0.00422 \\
1M    &                    0.00122 &                    0.00155 &                    0.00080 &                   0.00020 &                   0.00082 &                   0.00016 &                        0.04649 &                        0.00000 \\
2M    &                    0.00078 &                    0.00079 &                    0.00052 &                   0.00015 &                   0.00014 &                   0.00010 &                        0.42027 &                        0.00000 \\
\bottomrule
\end{tabular}

    \label{tab:ks_cnn_corr_size}
\end{table}%
\begin{table}[h]
    \centering
    \caption{Correction CNN $\mathcal{L}_\mathrm{rel}$ errors on the \gls{wake} system}
    \begin{tabular}{lrrrrrrrr}
\toprule
{} &  $\overline{\mathrm{ONE}}$ &  $\overline{\mathrm{NOG}}$ &  $\overline{\mathrm{WIG}}$ &  $\sigma({\mathrm{ONE}})$ &  $\sigma({\mathrm{NOG}})$ &  $\sigma({\mathrm{WIG}})$ &  $p^\mathrm{NOG}_\mathrm{ONE}$ &  $p^\mathrm{WIG}_\mathrm{ONE}$ \\
\midrule
6.3k &                    0.00078 &                    0.00042 &                    0.00026 &                   0.00019 &                   0.00010 &                   0.00005 &                        0.00000 &                        0.00000 \\
67k  &                    0.00020 &                    0.00013 &                    0.00010 &                   0.00008 &                   0.00005 &                   0.00004 &                        0.00075 &                        0.00000 \\
1.0M &                    0.00013 &                    0.00009 &                    0.00008 &                   0.00005 &                   0.00004 &                   0.00004 &                        0.01031 &                        0.00122 \\
\bottomrule
\end{tabular}

    \label{tab:wake_cnn_corr_size}
\end{table}%
\begin{table}[h]
    \centering
    \caption{Correction CNN $\mathcal{L}_\mathrm{rel}$ errors on the \gls{kolm} system}
    \begin{tabular}{lrrrrrrrr}
\toprule
{} &  $\overline{\mathrm{ONE}}$ &  $\overline{\mathrm{NOG}}$ &  $\overline{\mathrm{WIG}}$ &  $\sigma({\mathrm{ONE}})$ &  $\sigma({\mathrm{NOG}})$ &  $\sigma({\mathrm{WIG}})$ &  $p^\mathrm{NOG}_\mathrm{ONE}$ &  $p^\mathrm{WIG}_\mathrm{ONE}$ \\
\midrule
32k  &                    0.00406 &                    0.00407 &                    0.00361 &                   0.00081 &                   0.00045 &                   0.00037 &                        0.48124 &                        0.01460 \\
0.1M &                    0.00302 &                    0.00257 &                    0.00249 &                   0.00030 &                   0.00049 &                   0.00022 &                        0.00058 &                        0.00000 \\
0.5M &                    0.00163 &                    0.00139 &                    0.00141 &                   0.00016 &                   0.00014 &                   0.00019 &                        0.00001 &                        0.00015 \\
1M   &                    0.00102 &                    0.00084 &                    0.00080 &                   0.00008 &                   0.00010 &                   0.00011 &                        0.00000 &                        0.00000 \\
\bottomrule
\end{tabular}

    \label{tab:kolm_corr_size}
\end{table}%
\begin{table}[h]
    \centering
    \caption{Prediction GCN $\mathcal{L}_\mathrm{rel}$ errors on the \gls{ks} system}
    \begin{tabular}{lrrrrrrrr}
\toprule
{} &  $\overline{\mathrm{ONE}}$ &  $\overline{\mathrm{NOG}}$ &  $\overline{\mathrm{WIG}}$ &  $\sigma({\mathrm{ONE}})$ &  $\sigma({\mathrm{NOG}})$ &  $\sigma({\mathrm{WIG}})$ &  $p^\mathrm{NOG}_\mathrm{ONE}$ &  $p^\mathrm{WIG}_\mathrm{ONE}$ \\
\midrule
0.5k* &                    1.84653 &                    3.05853 &                    2.59502 &                   0.39942 &                   0.95652 &                   0.64054 &                        0.00000 &                        0.00004 \\
0.5k  &                    5.96483 &                    1.34493 &                    2.95659 &                   2.59318 &                   0.45774 &                   1.65345 &                        0.00000 &                        0.00005 \\
2k    &                    3.30063 &                    0.81323 &                    1.02441 &                   1.64430 &                   0.10700 &                   0.33664 &                        0.00000 &                        0.00000 \\
33k   &                    0.97240 &                    0.20468 &                    0.22368 &                   0.78061 &                   0.04296 &                   0.02868 &                        0.00004 &                        0.00006 \\
0.2M  &                    0.19699 &                    0.09459 &                    0.06454 &                   0.02300 &                   0.00875 &                   0.01415 &                        0.00000 &                        0.00000 \\
1M    &                    0.12911 &                    0.09182 &                    0.05792 &                   0.01819 &                   0.00936 &                   0.01514 &                        0.00000 &                        0.00000 \\
\bottomrule
\end{tabular}

    \label{tab:ks_gcn_pred_size}
\end{table}%
\begin{table}[h]
    \centering
    \caption{Prediction CNN $\mathcal{L}_\mathrm{rel}$ errors on the \gls{ks} system}
    \begin{tabular}{lrrrrrrrr}
\toprule
{} &  $\overline{\mathrm{ONE}}$ &  $\overline{\mathrm{NOG}}$ &  $\overline{\mathrm{WIG}}$ &  $\sigma({\mathrm{ONE}})$ &  $\sigma({\mathrm{NOG}})$ &  $\sigma({\mathrm{WIG}})$ &  $p^\mathrm{NOG}_\mathrm{ONE}$ &  $p^\mathrm{WIG}_\mathrm{ONE}$ \\
\midrule
0.5k* &                    3.48794 &                    1.54234 &                    2.15773 &                   0.49023 &                   0.11446 &                   0.26813 &                        0.00000 &                        0.00000 \\
0.5k  &                    1.52333 &                    1.27744 &                    1.22393 &                   0.21687 &                   0.40071 &                   0.17834 &                        0.01037 &                        0.00001 \\
2k    &                    0.59395 &                    0.50286 &                    0.50700 &                   0.17407 &                   0.08076 &                   0.09128 &                        0.02017 &                        0.02758 \\
33k   &                    0.17768 &                    0.20454 &                    0.20458 &                   0.03064 &                   0.04219 &                   0.03852 &                        0.01340 &                        0.00964 \\
0.2M  &                    0.10441 &                    0.09865 &                    0.07651 &                   0.03165 &                   0.02102 &                   0.01385 &                        0.25063 &                        0.00044 \\
1M    &                    0.06933 &                    0.06417 &                    0.02924 &                   0.01334 &                   0.01582 &                   0.00596 &                        0.13601 &                        0.00000 \\
\bottomrule
\end{tabular}

    \label{tab:ks_cnn_pred_size}
\end{table}%
\begin{table}[h]
    \centering
    \caption{Prediction CNN $\mathcal{L}_\mathrm{rel}$ errors on the \gls{wake} system}
    \begin{tabular}{lrrrrrrrr}
\toprule
{} &  $\overline{\mathrm{ONE}}$ &  $\overline{\mathrm{NOG}}$ &  $\overline{\mathrm{WIG}}$ &  $\sigma({\mathrm{ONE}})$ &  $\sigma({\mathrm{NOG}})$ &  $\sigma({\mathrm{WIG}})$ &  $p^\mathrm{NOG}_\mathrm{ONE}$ &  $p^\mathrm{WIG}_\mathrm{ONE}$ \\
\midrule
6.3k &            408020369.94856 &                    0.01375 &                    0.00886 &           816040738.02572 &                   0.01530 &                   0.00462 &                        0.01565 &                        0.01565 \\
67k  &                    0.07258 &                    0.00167 &                    0.00230 &                   0.12260 &                   0.00082 &                   0.00131 &                        0.00682 &                        0.00722 \\
1.0M &                    0.00027 &                    0.00033 &                    0.00010 &                   0.00009 &                   0.00042 &                   0.00001 &                        0.28444 &                        0.00000 \\
\bottomrule
\end{tabular}

    \label{tab:wake_cnn_pred_size}
\end{table}%
\begin{table}[h]
    \centering
    \caption{Prediction CNN $\mathcal{L}_2$ errors on the \gls{kolm} system}
    \begin{tabular}{lrrrrrrrr}
\toprule
{} &  $\overline{\mathrm{ONE}}$ &  $\overline{\mathrm{NOG}}$ &  $\overline{\mathrm{WIG}}$ &  $\sigma({\mathrm{ONE}})$ &  $\sigma({\mathrm{NOG}})$ &  $\sigma({\mathrm{WIG}})$ &  $p^\mathrm{NOG}_\mathrm{ONE}$ &  $p^\mathrm{WIG}_\mathrm{ONE}$ \\
\midrule
32k   &                    0.30455 &                    0.27062 &                    0.25849 &                   0.09936 &                   0.04430 &                   0.03258 &                        0.19624 &                        0.11662 \\
0.1M  &                    0.11714 &                    0.09850 &                    0.10791 &                   0.01839 &                   0.00787 &                   0.01187 &                        0.00978 &                        0.12649 \\
0.5Mk &                    0.06879 &                    0.06942 &                    0.06790 &                   0.01117 &                   0.00656 &                   0.00948 &                        0.44579 &                        0.43303 \\
1M    &                    0.04561 &                    0.03582 &                    0.03595 &                   0.01311 &                   0.00272 &                   0.00856 &                        0.02881 &                        0.05153 \\
\bottomrule
\end{tabular}

    \label{tab:kolm_cnn_pred_size}
\end{table}%
\begin{table}[h]
    \centering
    \caption{Prediction Unet $\mathcal{L}_2$ errors on the \gls{aero} system}
    \begin{tabular}{lrrrrrr}
\toprule
{} &  $\overline{\mathrm{ONE}}$ &  $\overline{\mathrm{NOG}}$ &  $\overline{\mathrm{WIG}}$ &  $\sigma({\mathrm{ONE}})$ &  $\sigma({\mathrm{NOG}})$ &  $\sigma({\mathrm{WIG}})$ \\
\midrule
0.5M &                    0.05046 &                    0.03143 &                    0.02975 &                   0.01302 &                   0.00420 &                   0.00221 \\
2M   &                    0.03302 &                    0.02815 &                    0.02829 &                   0.00251 &                   0.00090 &                   0.00106 \\
8M   &                    0.02892 &                    0.02392 &                    0.02388 &                   0.00505 &                   0.00153 &                   0.00204 \\
\bottomrule
\end{tabular}

    \label{tab:aero_cnn_pred_size}
\end{table}%
\begin{table}[h]
    \centering
    \caption{Correction CNN $\mathcal{L}_2$ errors on \gls{ks} system for multiple unrollings}
    \begin{tabular}{lrrrr}
\toprule
{} &  $\overline{\mathrm{NOG}}$ &  $\overline{\mathrm{WIG}}$ &  $\sigma({\mathrm{NOG}})$ &  $\sigma({\mathrm{WIG}})$ \\
\midrule
m=2  &                    0.34026 &                    0.33344 &                   0.06799 &                   0.07425 \\
m=3  &                    0.33358 &                    0.32817 &                   0.07953 &                   0.05533 \\
m=4  &                    0.32845 &                    0.31876 &                   0.06794 &                   0.05780 \\
m=5  &                    0.36300 &                    0.29802 &                   0.06534 &                   0.06446 \\
m=6  &                    0.34506 &                    0.29201 &                   0.05682 &                   0.06101 \\
m=8  &                    0.36041 &                    0.28614 &                   0.04762 &                   0.04553 \\
m=10 &                    0.39730 &                    0.28358 &                   0.07861 &                   0.06345 \\
m=12 &                    3.30768 &                    0.26729 &                   8.65391 &                   0.07201 \\
m=14 &                        inf &                    0.25408 &                   0.00000 &                   0.06418 \\
m=16 &                        inf &                    0.24387 &                   0.00000 &                   0.05274 \\
m=18 &                        inf &                    0.25749 &                   0.00000 &                   0.05718 \\
m=20 &                        inf &                    3.46048 &                   0.00000 &                   9.12454 \\
\bottomrule
\end{tabular}

    \label{tab:ks_cnn_corr_unrol}
\end{table}%
\begin{table}[h]
    \centering
    \caption{Correction CNN $\mathcal{L}_2$ errors on \gls{kolm} system for multiple curriculums}
    \begin{tabular}{lrrrr}
\toprule
{} &  $\overline{\mathrm{NOG}}$ &  $\overline{\mathrm{WIG}}$ &  $\sigma({\mathrm{NOG}})$ &  $\sigma({\mathrm{WIG}})$ \\
\midrule
1-2-4 &                    0.03064 &                    0.01635 &                   0.00344 &                   0.00172 \\
2-2-4 &                    0.01046 &                    0.04140 &                   0.00084 &                   0.00506 \\
4-4-4 &                    0.02576 &                    0.01394 &                   0.00511 &                   0.00144 \\
\bottomrule
\end{tabular}

    \label{tab:kolm_corr_cnn_curriculum}
\end{table}%
\begin{table}[h]
    \centering
    \caption{Correction CNN $\mathcal{L}_2$ errors on \gls{kolm} system when trained without learning rate schedules}
    \begin{tabular}{lrrrrrr}
\toprule
{} &  $\overline{\mathrm{ONE}}$ &  $\overline{\mathrm{NOG}}$ &  $\overline{\mathrm{WIG}}$ &  $\sigma({\mathrm{ONE}})$ &  $\sigma({\mathrm{NOG}})$ &  $\sigma({\mathrm{WIG}})$ \\
\midrule
32k  &                    0.04860 &                    0.04563 &                    0.05165 &                   0.00651 &                   0.00528 &                   0.00626 \\
0.1M &                    0.03478 &                    0.03701 &                    0.03708 &                   0.00271 &                   0.00689 &                   0.00433 \\
0.5M &                    0.03090 &                    0.02827 &                    0.03057 &                   0.00267 &                   0.00320 &                   0.00332 \\
1M   &                    0.02935 &                    0.02244 &                    0.02635 &                   0.00965 &                   0.00398 &                   0.00394 \\
\bottomrule
\end{tabular}

    \label{tab:kolm_corr_cnn_schedule}
\end{table}%
\begin{table}[h]
    \centering
    \caption{Correction CNN $\mathcal{L}_2$ errors on \gls{ks} system for gradient stopping variants}

    \begin{tabular}{lrrrrrr}
\toprule
{} &       m=6 &  m=6$_\mathrm{init}$ &       w=1 &       w=2 &       v=1 &       v=2 \\
\midrule
$\overline{\mathrm{WIG}}$ &  0.388526 &             0.281911 &  0.374733 &  0.396608 &  0.441374 &  0.407062 \\
$\sigma({\mathrm{WIG}})$  &  0.092057 &             0.050417 &  0.097563 &  0.102014 &  0.156538 &  0.101093 \\
\bottomrule
\end{tabular}

    \label{tab:ks_cnn_corr_gradstop}
\end{table}%
\begin{table}[h]
    \centering
    \caption{GCN $\mathcal{L}_2$ errors on \gls{ks} system when transitioning from prediction to correction}
    \begin{tabular}{lrrrrrr}
\toprule
{} &  $\overline{\mathrm{ONE}}$ &  $\overline{\mathrm{NOG}}$ &  $\overline{\mathrm{WIG}}$ &  $\sigma({\mathrm{ONE}})$ &  $\sigma({\mathrm{NOG}})$ &  $\sigma({\mathrm{WIG}})$ \\
\midrule
10\% &                    1.46963 &                    1.15836 &                    1.04210 &                   0.27693 &                   0.09666 &                   0.15044 \\
20\% &                    1.22752 &                    1.06870 &                    0.92173 &                   0.16561 &                   0.11952 &                   0.05283 \\
30\% &                    1.05448 &                    1.13224 &                    0.93338 &                   0.17234 &                   0.18949 &                   0.10398 \\
40\% &                    0.98373 &                    0.92035 &                    0.77333 &                   0.16078 &                   0.12198 &                   0.09403 \\
50\% &                    0.83278 &                    0.78407 &                    0.75527 &                   0.11489 &                   0.12352 &                   0.11875 \\
60\% &                    0.73871 &                    0.70152 &                    0.61906 &                   0.14414 &                   0.09365 &                   0.06889 \\
70\% &                    0.65280 &                    0.61808 &                    0.46474 &                   0.10726 &                   0.09699 &                   0.07094 \\
80\% &                    0.46519 &                    0.42852 &                    0.39318 &                   0.10353 &                   0.10320 &                   0.13866 \\
90\% &                    0.26651 &                    0.15659 &                    0.18686 &                   0.09774 &                   0.01506 &                   0.06874 \\
\bottomrule
\end{tabular}

    \label{tab:ks_gcn_transition}
\end{table}%
\begin{table}[h]
    \centering
    \caption{CNN $\mathcal{L}_2$ errors on \gls{ks} system when transitioning from prediction to correction}
    \begin{tabular}{lrrrrrr}
\toprule
{} &  $\overline{\mathrm{ONE}}$ &  $\overline{\mathrm{NOG}}$ &  $\overline{\mathrm{WIG}}$ &  $\sigma({\mathrm{ONE}})$ &  $\sigma({\mathrm{NOG}})$ &  $\sigma({\mathrm{WIG}})$ \\
\midrule
10\% &                    1.08743 &                    1.07423 &                    1.00705 &                   0.10005 &                   0.08438 &                   0.09433 \\
20\% &                    0.98619 &                    0.97025 &                    0.85021 &                   0.12132 &                   0.10936 &                   0.07485 \\
30\% &                    0.89545 &                    0.87275 &                    0.82945 &                   0.07255 &                   0.10187 &                   0.07454 \\
40\% &                    0.81295 &                    0.81708 &                    0.72305 &                   0.13808 &                   0.10260 &                   0.06438 \\
50\% &                    0.70735 &                    0.71738 &                    0.66283 &                   0.09532 &                   0.08907 &                   0.07525 \\
60\% &                    0.55892 &                    0.60302 &                    0.55926 &                   0.12181 &                   0.10013 &                   0.05164 \\
70\% &                    0.45414 &                    0.45675 &                    0.48273 &                   0.12301 &                   0.11392 &                   0.12604 \\
80\% &                    0.33173 &                    0.28746 &                    0.31830 &                   0.08661 &                   0.06363 &                   0.07834 \\
90\% &                    0.14203 &                    0.11449 &                    0.12929 &                   0.02205 &                   0.01993 &                   0.04566 \\
\bottomrule
\end{tabular}

    \label{tab:ks_cnn_transition}
\end{table}%

\end{document}